\documentclass[twocolumn,10pt]{article}

\usepackage[version=3]{mhchem}
\usepackage[margin=2cm]{geometry}
\usepackage{graphicx}
\usepackage{amsmath,amssymb}
\usepackage[utf8]{inputenc}
\usepackage[T1]{fontenc}
\usepackage{authblk}
\usepackage[numbers,sort&compress]{natbib}
\usepackage{abstract} 
\usepackage[usenames,dvipsnames]{xcolor}
\usepackage{hyperref}
\usepackage{xr-hyper}

\definecolor{burntorange}{rgb}{0.85,0.2,0}
 \hypersetup{
     colorlinks=true,
     linkcolor=blue,
     filecolor=magenta,      
     urlcolor=cyan,
     citecolor=burntorange,
     pdftitle={Overleaf Example},
     pdfpagemode=FullScreen,
     }

\usepackage{xspace}

\newcommand{\cacmin}{\smash{$d\,^{\rm Ca-C}_{\rm 6,min}$}\xspace}
\newcommand{\ccamin}{\smash{$d\,^{\rm C-Ca}_{\rm 6,min}$}\xspace}
\newcommand{\carc}{\smash{Ca$^{\rm CI}$}\xspace}
\newcommand{\crc}{\smash{C$^{\rm CI}$}\xspace}

\newcommand{\nionrc}{\smash{$N_{\rm ion}^{\rm CI-C}$}\xspace}
\newcommand{\nionc}{\smash{$N_{\rm ion}^{\rm C}$}\xspace}
\newcommand{\nion}{\smash{$N_{\rm ion}$}\xspace}
\newcommand{\conc}{\smash{$\chi_{\rm ion/w}$}\xspace}
\newcommand{\ncl}{\smash{$N_{\rm cl}$}\xspace}
\newcommand{\kt}{$k_{\rm B}T$\xspace}
\newcommand{\rg}{$R_g^{\rm C}$\xspace}

\newcommand{\la}{$\lambda_A$\xspace}
\newcommand{\lb}{$\lambda_B$\xspace}
\newcommand{\kab}{$k_{AB}$\xspace}
\newcommand{\flux}{$\Phi_{A}$\xspace}

\newcommand{\caion}{\smash{$\rm Ca^{2+}$}\xspace}
\newcommand{\cion}{\smash{$\rm CO_{3}^{2-}$}\xspace}
\newcommand{\ow}{\smash{O$_{\rm w}$}\xspace}

\newcommand{\caco}{\smash{CaCO$_3$}\xspace}

\newcommand{\SI}{the Supplementary Information\xspace}
\newcommand{\mainp}{the main paper\xspace}
\newcommand{\infretis}{$\infty$RETIS\xspace}
\newcommand{\pcross}{$P_{\rm cross}$\xspace}

\begin{document}

\title{\textbf{Organisation and aggregation of calcium carbonate ions in water via molecular simulation}}

\author[1]{Elizaveta Sidler\thanks{Corresponding author: \href{mailto:elizaveta.sidler@ntnu.no}{\texttt{elizaveta.sidler@ntnu.no}}}}
\author[1]{Rita S. Dias}
\author[1]{Raffaela Cabriolu}

\affil[1]{\small Department of Physics, Norwegian University of Science and Technology (NTNU), Trondheim 7491, Norway}

\date{} 

    \twocolumn[
    \maketitle
    \begin{onecolabstract}
      \noindent Crystallisation of calcium carbonate (\caco) in aqueous solutions is essential for carbon dioxide sequestration and the sustainable production of calcium carbonate materials, which are widely used in industry and increasingly explored for novel applications. While crystallisation is often analysed within the framework of classical nucleation theory, \caco and other ionic crystals often form through amorphous precursor phases, motivating the need for alternative descriptions. In this work, we investigate the relationship between the formation of the first \caco coordination sphere, a fundamental unit of local ordering, and ionic aggregation using two complementary simulation approaches: on-the-fly probability-enhanced sampling (OPES) and asynchronous and infinite-swap replica exchange transition interface sampling (\infretis). OPES was used to explore free-energy landscapes, whereas \infretis was used to determine reaction kinetics. We studied systems at two levels of supersaturation, between which the formation of the first coordination sphere exhibited distinct kinetic behaviour. These differences were found to be linked to concentration-dependent ionic aggregation. At lower concentration, aggregation remains limited, whereas higher concentration promotes extensive spinodal-like aggregation. For both concentrations, the first coordination sphere forms within the pre-formed clusters through one of two distinct pathways: either within a cluster close to the minimum size required for its formation, or within larger amorphous aggregates whose size and structure depend on concentration. The coexistence of these pathways at both levels of supersaturation reveals the multifaceted nature of early-stage \caco crystallisation and highlights the importance of differentiating ionic aggregation and organisation.
      \vspace{1.5em}
    \end{onecolabstract}
    ]
\saythanks

\section{Introduction}

Ionic nucleation in water has long been recognized as a topic of considerable scientific and practical importance. Calcium carbonate (\caco), one of the most abundant minerals on Earth, has been been the subject of extensive research aimed at elucidating nucleation mechanisms. The precipitation of calcium carbonate in the ocean is a crucial component in the global carbon cycle and biomineralisation.\cite{carbon_cycle, carbon_cycle_ocean, soil_sequestration, carbon_cycle_climate,bacteria_biomin, biomin_paths, biomin_evolution, biomin_coccoliths} The diversity of structural architectures and high compatibility with living organisms endow \caco with exceptional potential for biomedical applications, such as tissue engineering, drug delivery, and anticancer-specific therapy.\cite{appl_comb, appl_biomed, biomed_nanoparticles, biomed_nanoparticles_new}

The broad functionality of \caco is closely associated with the remarkable polymorphism of its forms. Its three main crystalline polymorphs that precipitate from solution are calcite, aragonite, and vaterite.\cite{CAV} Calcite structure is the most stable under ambient conditions. Aragonite is a stable high-pressure phase of CaCO$_3$ and is metastable under ambient conditions.\cite{caco3_pd} Vaterite is widely regarded as one of the possible metastable precursors of calcite. Despite extensive investigations, its crystal structure remains poorly understood and has yet to be unequivocally determined.\cite{vaterite_str} The polymorphism of \caco is also manifested through temperature- and pressure-dependent structural phase transitions.\cite{str_trans, Bayarjargal_phase_diagr_VII}

The diversity of crystalline forms of precipitated \caco is generally associated with the presence of distinct crystallisation pathways governed by the intricate interplay between thermodynamic and kinetic factors. Within the framework of classical nucleation theory (CNT), crystallisation from solution proceeds via monomer-by-monomer association leading to the formation of a critical nucleus. Alternatively, non-classical nucleation theories (NCNTs) suggest that crystallisation may proceed through intermediate metastable phases or amorphous precursors, which subsequently transform into crystalline phases. Moreover, in the formulations of some NCNTs, these metastable intermediates arise from the aggregation of stable prenucleation species. The multistep crystallisation of \caco proceeding via amorphous calcium carbonate (ACC) is well-reported.\cite{paths_coexistence, interface_ACC_MD, ACC_to_calcite_via_vaterite} However, the extent to which \caco crystallisation follows classical or non‑classical pathways remains an open question.\cite{PNC_NCNT, CNT_NCNT_review, PNC_no_1, CNT_NCNT_both_review}

Several metastable intermediates and precursors beyond ACC have been proposed for crystallisation of \caco, including prenucleation clusters (PNCs) and dense liquid phase (DLP). The existence of PNCs was first suggested by Gebauer et al. on the basis of titration experiments.\cite{PNC_first_exp} Subsequent experimental studies provided further evidence supporting the presence of such species in solution.\cite{PNC_second_exp, PNC_exp_new} Molecular dynamics (MD) simulations carried out by Demichelis et al. characterised PNCs as dynamically ordered liquid-like oxyanion polymers (DOLLOPs).\cite{PNC_MD_yes} However, later studies questioned the stability of PNCs and proposed that, although present, they may be statistically and kinetically insignificant in the nucleation process.\cite{DLP_intro_PNC_MD_no, PNC_no_1, PNC_no_2} Owing to their small size and high dynamical nature, PNCs remain challenging for direct experimental observations, and their precise role in \caco crystallisation therefore remains unresolved. 

Liquid–liquid phase separation (LLPS) of highly saturated \caco, resulting in the formation of DLP, was first proposed by Faatz et al. as a pathway to the formation of spherical ACC particles.\cite{LLPS_ACC_Faatz} Later, Wolf et al. provided experimental observations of liquid-like \caco nanoparticles with characteristic sizes of 100–300 nm in acoustically levitated droplets.\cite{LLPS_1} Further molecular simulation research suggested that LLPS may proceed through either spinodal or binodal pathways, depending on \caco concentration, and provided structural analysis of the resulting DLP.\cite{LLPS_MD_1, PNC_to_ACC_LLPS, DLP_intro_PNC_MD_no} Whether LLPS should be interpreted within NCNTs or can be explained in terms of CNT is debated in different experimental and computational studies.\cite{DLP_intro_PNC_MD_no, LLPS_2019, PNC_to_ACC_LLPS}
Several works further indicate that DLP can evolve into ACC or crystalline \caco under appropriate conditions.\cite{DLP_to_ACC_new, LLPS_MD_1, DLP_into_ACC_withMg}

ACC is a metastable, non-crystalline phase of \caco that plays an important role in biomineralisation as a precursor to crystalline biogenic calcium carbonate.\cite{ACC_bio_stable, ACC_shells} Structurally, ACC exhibits strong polyamorphism: as a precursor for different crystalline phases of \caco, ACC may adopt a proto-calcite, proto-aragonite, or proto-vaterite short-range order; in addition, ACC exists in hydrous and anhydrous forms.\cite{ACC_forms_review, proto_aragonite} Several molecular simulation studies conducted diverse structural and energy analysis for anhydrous and hydrated ACC with varying levels of hydration.\cite{water_is_a_key, ACC_hyd_SIM, ACC_hydr_2010, ACC_dehydration_sim, nanodomains_ACC}.

The transition of ACC into stable crystalline phases has been reported to proceed through a combination of dissolution-recrystallisation and solid-state transformation processes.\cite{dissol_recryst, interface_ACC_MD, ACC_dehydration_sim, ACC_to_calcite_via_vaterite, dissol_repr_new} The transformation mechanism and the resulting polymorph have been reported to depend on various conditions, such as particle size, ACC formation rate, pH, and temperature.\cite{dissol_recr_part_time, dissol_recr_part_size, ACC_stab_add, polymorph_T_phase_diag, ACC_stab_add_ions} 

As demonstrated by the extensive literature cited above, molecular simulations, and MD in particular, have shown to be a powerful tool for investigating \caco species and phases, as well as their evolution and transformations. However, direct simulation of \caco crystallisation using conventional MD remains challenging, as the relevant time scales typically exceed those accessible for this method. To address this limitation, a range of advanced simulation techniques can be applied.\cite{methods_review}

Transition interface sampling, and its extension, replica exchange transition interface sampling (RETIS), provide a framework for calculating rate constants and analysing reaction kinetics through Monte Carlo (MC)-based sampling of unbiased MD trajectories.\cite{TIS, RETIS} More recently,  by Zhang et al. developed a parallelised variant of RETIS, termed \infretis, which utilises advanced MC algorithms combining subtrajectory moves and asynchronous replica exchange.\cite{Infretis} \infretis enables efficient sampling of rare events and facilitates computationally demanding rate calculations in large complex systems.

Metadynamics is a widely used enhanced sampling method that facilitates the exploration of configurational space across free-energy barriers.\cite{metad} The method relies on constructing a history-dependent bias potential on the fly, and is used to reconstruct free-energy surfaces (FESs) that may be difficult to obtain with conventional MD. More recently, Invernizzi et al. introduced on-the-fly probability-enhanced sampling (OPES), a related bias‑driven approach designed to accelerate convergence of free energy estimates.\cite{OPES}

Here, we investigate the early stages of calcium carbonate crystallisation in aqueous solution using atomistic molecular simulations. We consider crystallisation through two related processes: ionic aggregation and organisation. These processes may be closely coupled, with ions adopting ordered arrangements as they aggregate. For \caco, however, aggregation and structural organisation often exhibit a more complex relationship, as large amorphous aggregates and prenucleation species can form prior to the emergence of ordered structures.

To investigate ionic aggregation and the influence of concentration, we consider aqueous \caco systems at two high levels of supersaturation, each represented by two different system sizes. Enhanced sampling methods are employed to characterise local ionic organisation through the formation of the first coordination sphere between calcium and carbonate ions. To address both the kinetics and thermodynamics of this process, we apply \infretis and OPES. \infretis is used to calculate rate constants of the formation of the first coordination sphere, while OPES is employed to explore the FESs underlying this reaction. Ionic aggregation occurs concurrently through the formation and growth of ionic clusters. 

Both organisation and aggregation, as well as the relationship between them, are examined in the post-simulation analysis. This approach extends the understanding of the formation and evolution of amorphous precursor clusters beyond the aggregation of calcium and carbonate ions, which has been the main focus of many previous studies. In particular, we investigate the development of local ionic organisation within these clusters and identify the cluster types that mediate the coordination reaction. Furthermore, we investigate whether coordination-sphere formation depends on the choice of the central ion by comparing the formation of coordination spheres centred around calcium ions (Ca--C) and carbonate ions (C--Ca).

\section{Model and Methods}\label{sec:method}

\subsection{Molecular dynamics}\label{sec:method_MD}

All MD simulations were performed using the full-atomistic interaction potential for \caco developed by Raiteri et al., together with the SPC/Fw water model.\cite{Raiteri_2015, water_pot} This force field was parametrised on the basis of solubility and hydration free energy data to accurately reproduce the early stages of the association of calcium and carbonate ions in water, making it well-suited for the present study. All simulations were carried out with periodic boundary conditions. 

In this work, the composition of molecular systems is expressed as the fraction of the number of ions (\caion and \cion) to the number of water molecules (H$_2$O): 

\begin{equation}
\chi_{\rm ion/w} = \frac{N_{\text{Ca}^{2+}} + N_{\text{CO}_3^{2-}}}{N_{\text{H}_2\text{O}}}
\label{Eq_TIS}
\end{equation}

Using this notation, hereafter referred to as concentration for practical reasons, the effect of the concentration of calcium and carbonate ions in water was evaluated for \conc~= 0.01 ($0.28$ M for each ionic species) and \conc~= 0.02 (0.56 M for each ionic species). To examine the influence of the total number of ions (\nion) in the simulation, we performed simulations with \nion~= 72 (36 \caion ions and 36 \cion ions) and \nion~= 108 (54 \caion ions and 54 \cion ions) for each \conc. In total, four system compositions were considered: (\conc, \nion)~= (0.02, 72), (0.01, 72), (0.02, 108), (0.01, 108).

Each of the four systems was created by randomly placing the specific number of water molecules, and \caion and \cion ions in the simulation box. The systems were subjected to a potential energy minimisation, allowing the simulation box to relax to a target pressure of 0 bar. Subsequently, the four systems were heated to $T=300$ K and equilibrated at this temperature for 10 ns using MD in the NPT ensemble with a pressure maintained at 0 bar. The equilibrated systems were used as starting configurations for \infretis and OPES. We employed \textsc{lammps} MD engine \cite{LAMMPS} for all MD simulations.

\subsection{\infretis}\label{sec:method_infretis}

\infretis provides a framework for determining the rate constant \kab for the rare transition between states A and B separated by energy barriers. An order parameter (OP) $\lambda$ is employed to distinguish the two states and track progress between them. 
In the \infretis method, $n+1$ interfaces $\lambda_i$ ($i=0,\ldots,n$) are placed along the OP between states A and B. The first and last interfaces (\la~$ = \lambda_0$ and \lb~$ = \lambda_n$) define the transition region just outside the basins of states A and B. In our study, the reaction proceeds from larger to smaller values of $\lambda$, such that state A is defined by $\lambda>$~\la and state B by $\lambda<$~\lb. The intermediate interfaces ($\lambda_i$, where $i=1,\ldots,n-1$) are positioned sequentially between \la and \lb across the barriers in the transition region.

\begin{figure*}[h!]
    \centering
    \includegraphics{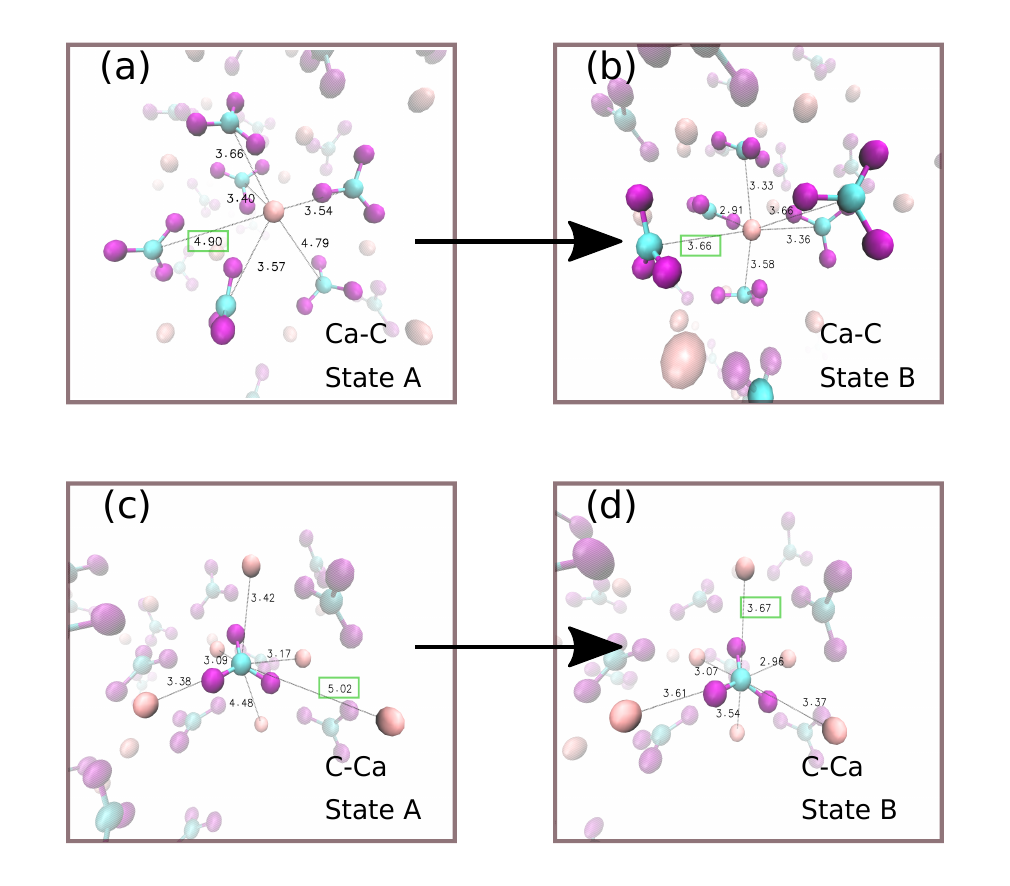}
    \caption{Simulation snapshots of the system with \conc~= 0.01 and \nion~= 72 when (a) \cacmin~$\geq$~\la ($\approx 4.90$ \AA{}), (b) \cacmin~$\leq$~\lb ($\approx 3.66$ \AA{}), (c) \ccamin~$\geq$~\la ($\approx 5.02$ \AA{}), (d) \ccamin~$\leq$~\lb ($\approx 3.67$ \AA{}). Only calcium and carbonate ions are shown; water is omitted. Ions outside the corresponding coordination sphere are semitransparent. The distances between the central atom (Ca or C) and its 6 nearest neighbours (C or Ca) are indicated, with the corresponding 6$_{\rm th}$ minimum distance enclosed in the green frame. Calcium atoms are shown in light pink, carbon atoms are in turquoise, and oxygen atoms are in magenta.}
    \label{fig:snapshots_OP}
\end{figure*}

After choosing an OP and placing the interfaces, MD trajectories are sampled following the \infretis-specific MC-based path-sampling protocol. All sampled trajectories that start in state A and cross interface $\lambda_i$ at least once constitute the path ensemble $[i^+]$. In addition, all trajectories in the $[i^+]$ ensembles have reached either state A or state B. The $[i^+]$ ensembles are introduced to facilitate sampling of the transition region and the paths connecting states A and B, which are termed reactive paths. 

To estimate the flux \flux across the first interface, trajectories are sampled in the path ensemble $[0^-]$. Except for their first and last frames -- which are located between \la and $\lambda_1$ -- these trajectories remain confined within state A ($\lambda >$~\la). \flux is subsequently calculated from the average path lengths $\langle t^{[0^-]} \rangle$ and $\langle t^{[0^+]} \rangle$ of the ensembles $[0^-]$ and $[0^+]$, respectively, using the formula:\cite{10.1063/1.4989844}

\begin{equation}
\Phi_{A} = \frac{1}{\langle t^{[0^+]} \rangle + \langle t^{[0^-]} \rangle }
\label{Eq_flux}
\end{equation}

The trajectories in an $[i^+]$ ensemble can be used to evaluate the conditional crossing probability $P_A(\lambda_{i+1}|\lambda_{i})$ -- defined as the probability that a trajectory that has crossed $\lambda_i$ reaches $\lambda_{i+1}$ before recrossing \la. The conditional probabilities from all $[i^+]$ ensembles then enable the computation of $P_A(\lambda_B|\lambda_A)$, the probability that a trajectory that started in state A reaches state B before returning to state A (also referred to as \pcross), according to the formula: 
\begin{equation}
P_A(\lambda_B|\lambda_A) = \prod_{i=0}^{n-1}P_A(\lambda_{i+1}|\lambda_{i}),
\label{Eq_P_cross}
\end{equation}
After calculating $P_A(\lambda_B|\lambda_A)$ and $\Phi_{A}$, \kab is obtained using the formula:
\begin{equation}
k_{AB} = \Phi_{A}P_A(\lambda_B|\lambda_A)\\
\label{Eq_k_AB}
\end{equation}
%
%
To study the formation of the first calcium carbonate coordination sphere with \infretis, we need to define an appropriate OP. In the two-component model of \caco used here, the first coordination sphere forms either around \caion with \cion neighbours (Ca--C), or around \cion with \caion neighbours (C--Ca). Because a carbon atom is located at the centre of a carbonate ion, we explicitly consider only distances between calcium and carbon atoms, excluding the oxygen atoms of \cion from the analysis; hence, Ca--C and C--Ca notations. In all three relevant polymorphs of \caco, the first coordination sphere with respect to calcium and carbon atoms contains 6 neighbours. Consequently, we used two OPs -- the Ca--C and C--Ca 6$_{\rm th}$ minimum distances (\cacmin and \ccamin, respectively) -- each suitable for studying the formation of the corresponding first coordination sphere of \caco. 

To calculate \cacmin, we first determine the distance between each Ca atom and its sixth nearest C atom. We then select the minimum of these distances across all Ca atoms. Thus, \cacmin indicates the first occurrence of sixfold coordination between a calcium ion and carbonate ions. The C--Ca 6$_{\rm th}$ min distance (\ccamin) is obtained using the same approach, with the roles of Ca and C interchanged.


Next, we place \la and \lb interfaces. In previous studies of a two-ion system containing one \caion and one \cion in water, Raiteri et al. identified the most stable state for the neutral \caion--\cion pair at a Ca--C distance $\simeq$ 3.5 \AA{}, corresponding to the contact ion-pair configuration.\cite{water_is_a_key} In the same study, this stable state was found to be separated from the first water-shared state by a free energy barrier located between $\simeq$ 3.5 \AA{} and $\simeq$ 5 \AA{}. The minima at larger Ca--C distances were found to be separated by lower free energy barriers. Consequently, the present work focuses on transitions across the barrier between the stable contact state and the first water-shared state. Additionally, according to the Ca-C radial distribution functions (RDFs) for crystalline calcium carbonate, the first coordination sphere lies within a distance of approximately 4 \AA{} for all three main polymorphs. Using these distances as a reference, we conducted a preliminary study (detailed in Section~\ref{sec:Interfaces} in the Supplementary Information) to determine the positions of \la and \lb with respect to the \cacmin OP. Based on this analysis, \lb was placed at 3.7 \AA{} and \la at 4.9 \AA{}. Thus, state B, defined by \cacmin $\leq \lambda_B$, features configurations with at least one fully formed sixfold Ca--C coordination sphere. The chosen \lb ensures that state B encompasses the first coordination sphere of each of the three relevant polymorphs, thereby avoiding polymorph-specific bias. Meanwhile, state A, with \cacmin $\geq$ \la, encompasses the accessible region of OP space preceding the free energy barrier of interest. The indicated values of \la$= 4.9$ \AA{} and \lb$ = 3.7$ \AA{} were employed in all \infretis simulations -- for all combinations of \conc and \nion and for both \cacmin and \ccamin OPs. Simulation snapshots with the molecular systems in states A and B are presented in Figure~\ref{fig:snapshots_OP}.

The formation of the first coordination sphere, when represented by the conventional OP of the coordination number (CN), is often characterised by multiple energy barriers of comparable height preceding state B (see Section~\ref{sec:local_cn}). In contrast, when represented by the 6$_{\rm th}$ minimum distance OP, the process can be described as a transition across a single energy barrier. This simplifies the placement of \la and \lb. Moreover, unlike CN, the 6$_{\rm th}$ minimum distance is naturally continuous and does not require a cut-off radius, which would influence the shape of the projected energy landscapes and, consequently, the placement of interfaces. Notably, the choice of OP affects the definitions of states A and B and the interface placement, while the underlying physics of the process remains unchanged.

\cacmin and \ccamin depend only on the positions of two dynamically selected atoms. Their indices are determined at each timestep and may vary throughout the simulation. For each OP, the atom at the centre of the corresponding coordination sphere, i.e. the atom with the minimum sixth-neighbour distance, is referred to as the central ion (CI). The central ions are denoted \carc and \crc for \cacmin and \ccamin, respectively.

\infretis simulations were performed using the \cacmin OP for four systems with the following compositions: (\conc, \nion) = (0.02, 72), (0.01, 72), (0.02, 108), (0.01, 108), to study the formation of the first Ca--C coordination sphere. 
In addition, \infretis simulations were carried out for two systems with (\conc, \nion) = (0.02, 72) and (0.01, 72) using \ccamin OP to investigate the difference between the formation of the Ca--C and C--Ca first coordination spheres. 


While the same values of \la and \lb were used in all \infretis simulations, the number and positions of the intermediate interfaces differed between systems. The optimal interface placement was determined iteratively using the \texttt{$\infty$tools} python package. The exact protocol for the \infretis simulations is presented in Section~\ref{sec:SI_protocol} of \SI.

$P_A(\lambda_{i+1}|\lambda_{i})$ (and, consequently, $P_A(\lambda_{B}|\lambda_{A})$) were estimated using the weighted histogram analysis method provided by the \texttt{$\infty$tools} python package.\cite{WHAM} In addition, conditional free energy (CFE) profiles were computed along different coordinates in configuration space using \texttt{$\infty$tools}. CFEs are history-dependent, with the lowest value corresponding to the most frequently visited state, which, in \infretis evaluations of \kab, is always state A. Consequently, CFE landscapes do not yield standard unconditional free energies across the entire visited configuration space. However, unlike standard free energy landscapes, which are determined solely by equilibrium thermodynamics, CFE surfaces generated through path reweighting are influenced by both thermodynamics and kinetics. As a result, CFE landscapes reveal energy barriers and reaction pathways that depend not only on energy differences but also on kinetic factors, such as diffusion. More information on this topic can be found in the work by van Erp et. al.\cite{CFE}

Here, we used the $\infty$\textsc{retis} python library developed by Zhang et al., coupled with \textsc{lammps} MD engine, to perform \infretis simulations.\cite{Infretis, LAMMPS} To enable parallel path sampling, \infretis employs a system of `workers'. We used 4 to 6 workers, each of which employed 16 to 32 CPUs. The supercomputer LUMI, hosted by CSC (Finland) was used for all \infretis simulations. For the largest system with \conc $=0.01$ and \nion $=108$, the production simulation required 200,000 CPU-hours while using 6 workers, each allocated 32 CPUs to simulate 32500 \infretis steps.

\subsection{OPES}

OPES explores FESs in a space defined by a set of collective variables (CVs).\cite{OPES} CVs are low-dimensional functions of atomic coordinates ($\mathbf{s}({\mathbf{R}})$) selected to capture the essential degrees of freedom of a process of interest. In OPES, the unbiased probability distribution $P_n(\mathbf{s})$ is estimated on the fly from a sum of Gaussian kernels $K(\mathbf{s},\mathbf{s}_k)$ deposited along the CV space according to:

\begin{equation}
P_n(\mathbf{s}) = \frac{\sum_k^nw_k K(\mathbf{s},\mathbf{s}_k)}{\sum_k^nw_k},
\label{eq:OPES}
\end{equation}

\noindent where $n$ denotes the current iteration step, index $k$ acts as a discretized time variable ($k = 1, \ldots, n$), and the weights are given by $w_k=e^{\beta V_{k-1}(\mathbf{s}_k)}$. This estimate is then used to construct the time-dependent adaptive bias $V_n(\mathbf{s})$, which targets the well-tempered probability distribution according to:

\begin{equation}
V_n(\mathbf{s})=(1-1/\gamma)\frac{1}{\beta}\log
\left( \frac{P_n(\mathbf{s})}{Z_n} + \epsilon \right),
\label{eq:OPES}
\end{equation}

\noindent for $n_\text{{th}}$ iteration, where $\gamma$ and $\epsilon$ are the bias factor and the regularization term, respectively. Both $\gamma$ and $\epsilon$ depend on a specified barrier-height parameter $\Delta E$, which limits the maximum value of $V_n(\mathbf{s})$ and thus represents the largest free energy barrier that can be overcome (see \SI in ref. \cite{OPES} for the exact definition of $\gamma$ and $\epsilon$). $\beta = (k_{\rm B}T)^{-1}$, where $k_{\rm B}$ is the Boltzmann constant, and $Z_n$ is a normalization factor over the explored CV space. Conceptually, the applied bias enhances sampling and facilitates the crossing of free-energy barriers. After completion of the simulation, the free-energy landscapes are reconstructed via reweighting using a weighted kernel density estimation as a probability estimator. The free energies obtained with OPES are fundamentally functions of the biased CVs. However, through reweighting, FESs can also be reconstructed along additional unbiased CVs.

\begin{table*}[h!]
\small
\caption{Results of \infretis simulation with the OP corresponding to the indicated type of the coordination sphere (Ca--C or C--Ca for \cacmin or \ccamin, respectively), for systems with the specified \conc and \nion.}
\centering
\begin{tabular*}{\textwidth}{@{\extracolsep{\fill}}llllll}
\hline
$d_{6, \rm min}$ & \conc & \nion & $P_{\rm cross}$ (*$10^{-4}$) & $\Phi_A$ (ns$^{-1}$) & $k_{AB}$ (ns$^{-1}$)\\
\hline
Ca--C & 0.02  &  72  &  0.7 $\pm$ 0.3  &  42 $\pm$ 7    &  0.003 $\pm$ 0.001  \\
Ca--C & 0.02  &  108 &  1.4 $\pm$ 0.3  &  240 $\pm$ 45  &  0.033 $\pm$ 0.008  \\
Ca--C & 0.01  &  72  &  1.2 $\pm$ 0.5  &  154 $\pm$ 26  &  0.017 $\pm$ 0.006  \\
Ca--C & 0.01  &  108 &  1.2 $\pm$ 0.6  &  106 $\pm$ 26  &  0.013 $\pm$ 0.006  \\
\hline
\hline
C--Ca & 0.02  &  72  &  6  $\pm$ 2  &  32 $\pm$ 6   &  0.020 $\pm$ 0.006  \\
C--Ca & 0.01  &  72  &  8  $\pm$ 2  &  50 $\pm$ 16  &  0.040 $\pm$ 0.011  \\
\hline
\end{tabular*}
\label{tab:rates}
\end{table*}

OPES simulations were performed for the four system compositions described in Section~\ref{sec:method_MD}. Each OPES simulation applied bias simultaneously to the two CVs defined by \cacmin and \ccamin (see Section~\ref{sec:method_infretis}). The final FESs were evaluated along the biased CVs and several auxiliary unbiased CVs (\nionrc, CNs, etc.; see Section~\ref{sec:analysis}). Because the \infretis OPs and OPES CVs correspond to the same physical variables, we hereafter use OP and CV interchangeably.

OPES calculations were carried out in the NPT ensemble at $T=300$ K and $P=0$ bar for simulation times of 350--550 ns, using a barrier-height parameter $\Delta E$ = 30 \kt. For efficient sampling of the region of interest, \cacmin and \ccamin were restrained by an upper harmonic wall potential placed at 10 \AA{}. Gaussian kernels were deposited every 500 MD steps; to limit compressibility of the kernels due to the adaptive Gaussian width $\sigma$, a minimum kernel width $\sigma_{min}=0.03$ was imposed for both CVs. All OPES calculations were carried out on the SCARF computing HPC cluster, with access provided by the STFC Computing Department staff based at Daresbury. The OPES simulation of the largest system with \conc $=0.01$ and \nion $=108$ required 309,000 CPU-hours, using an allocation of 128 CPUs to integrate 350 ns of OPES MD. OPES was employed through \textsc{plumed}2.9 \cite{PLUMED} coupled with \textsc{lammps} MD engine \cite{LAMMPS}.

\subsection{Analysis}\label{sec:analysis}\indent

While the OPs of \cacmin and \ccamin monitor the formation of the corresponding first coordination sphere, they do not explicitly monitor ionic aggregation or capture the size, connectivity, or overall structure of the calcium carbonate ionic aggregates. As such, in order to analyse aggregation of calcium and carbonate ions, we performed cluster analysis based on distances between the Ca and C atoms. The cluster containing the central ion associated with the corresponding OP (CI, \carc for \cacmin OP and \crc for \ccamin OP, see Section~\ref{sec:method_infretis}) is termed the CI-cluster (central-ion-cluster). Further details on the cluster analysis and connectivity are provided in \SI, Section~\ref{sec:SI_cluster_analysis}.

From the cluster analysis, several properties were obtained: the size of any clusters (\nionc) and the size of the CI-cluster (\nionrc, CI-C stands for CI-cluster), both expressed as the number of ions; the total number of clusters (\ncl); and the radius of gyration of any cluster (\rg).

Since the OPES bias was applied simultaneously along \cacmin and \ccamin, two \nionrc values were computed at each simulation step, corresponding to the clusters containing the respective reference CIs (\carc and \crc). When the two reference CIs belonged to the same ionic aggregate, the resulting CI-clusters were identical and had the same \nionrc. When the reference CIs belonged to different aggregates, the resulting CI-clusters were distinct and generally had different \nionrc values, although aggregates of equal size could occur by chance.

The structure of the CI-cluster was analysed with radial distribution functions (RDFs, $g(r)$), computed for all \caco atoms or between the specific atom types. Additionally, we computed CNs associated with the different atom types. In order to analyse the shape of the CI-cluster, we employed the $K_1$ and $K_2$ shape parameters, derived from the eigenvalues of the gyration tensor. Details on the calculation of radial distribution functions, CNs, and shape parameters are provided in Sections~\ref{sec:SI_RDF}, \ref{sec:SI_CN_meth}, and \ref{sec:SI_K1_K2_meth} of \SI.

\section{Results and Discussion}\label{sec:results}

\subsection{Rate constants from \infretis}\label{sec:results_ca_c}

The rate constants \kab, fluxes \flux, and crossing probabilities \pcross for the formation of the first coordination sphere evaluated from the \infretis simulations are summarized in Table~\ref{tab:rates} for all considered systems and investigated OPs.

\begin{figure*}[h]
    \centering
    \includegraphics{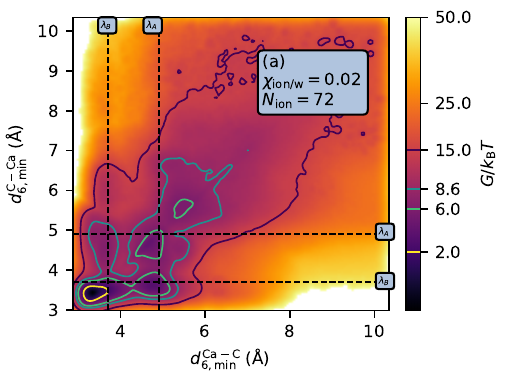}
    \includegraphics{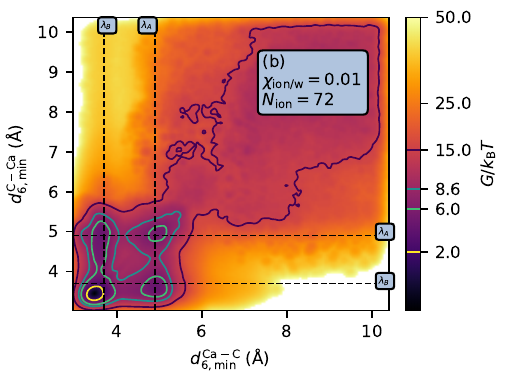}
    \includegraphics{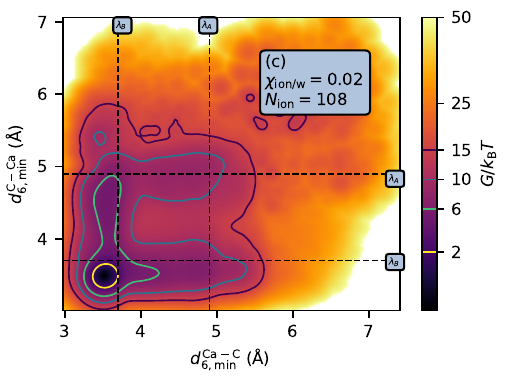}
    \includegraphics{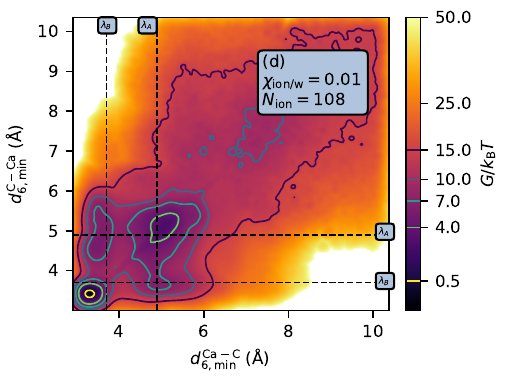}
    \caption{FESs projected onto the \cacmin and \ccamin for the systems with compositions indicated in the legends. LFERs are outlined by the free energy contour lines.}
    \label{fig:6min_cac_cca_OPES}
\end{figure*}

To examine the effect of composition, we first compare the rate constants obtained using \cacmin OP. For \conc~= 0.01, the systems with \nion~= 72 and 108 exhibit statistically indistinguishable values of \kab and identical \pcross values. The flux for the smaller system (\nion~= 72) is higher than that for the larger system (\nion~= 108), although the two values remain broadly comparable. 
In contrast, at the higher \conc~= 0.02, the rate constant for the system with \nion~= 72 is found to be approximately an order of magnitude smaller than that for the system with \nion~= 108. This difference is primarily driven by the lower flux through \la in the smaller system, whereas the reduction in the crossing probability, although present, is comparatively modest. Therefore, the observed lower \kab is primarily associated with a reduced flux from  state A rather than with a substantial reduction in the crossing probability.

For the larger systems (\nion~= 108), which are less affected by finite-size effects and therefore closer to the bulk limit, \kab is higher for \conc~= 0.02 than for \conc~= 0.01, a difference that is again driven by the increased flux through \la. Combined with the different dependence of \kab on \nion, this observation suggests that the formation of the first \caco coordination sphere exhibits different kinetic behaviour at the two \conc.

Next, we compare the rate constants for the formation of the Ca--C coordination sphere with that of the C--Ca coordination sphere calculated with \infretis for two systems with \nion~=72, and \conc~=0.02 and 0.01 (see Table~\ref{tab:rates}).

For both \conc, the crossing probabilities for the formation of the first C--Ca coordination sphere are higher than those for the formation of the first Ca--C coordination sphere. However, at a given concentration, the two \pcross values remain within the same order of magnitude, indicating only subtle kinetic variations in the transition region between the two reactions. The flux difference between the two reactions is negligible for the system with \conc~= 0.02, whereas for \conc~= 0.01, the flux associated with the C--Ca OP is lower than that associated with the Ca--C OP. Nevertheless, differences in \pcross lead to consistently larger \kab values for the formation of the first C--Ca coordination sphere, exceeding those for the first Ca--C coordination sphere by factors of 6.7 and 2.4 at \conc = 0.02 and 0.01, respectively. These differences in \kab suggest broadly comparable kinetics with only a modest preference for the formation of the C--Ca coordination sphere. The evolution of $\ln(P_{\rm cross})$ along the \cacmin and \ccamin OPs is discussed in Section~\ref{sec:SI_ln_P} in \SI. 

\subsection{Energy barriers from OPES}\label{sec:results_barr_OPES}

The \infretis study of the kinetics of the formation of Ca--C and C--Ca coordination spheres was complemented by the OPES calculations of the free energy $G$ along the corresponding CVs. While \infretis provides a dynamical description of the process and an estimate of \kab, OPES enables a broader exploration of the configurational space, including state B, and allowed us to determine the positions and relative depths of the free energy minima.

The FESs projected onto \cacmin and \ccamin for the four studied systems are shown in Figure~\ref{fig:6min_cac_cca_OPES}. Across all systems, the global minimum of free energy is located where both \cacmin and \ccamin values correspond to state B (both \cacmin and \ccamin $\leq$~\lb). Moreover, as expected for an OPES simulation, a broader range of \cacmin and \ccamin values was sampled compared to \infretis because the bias potential promotes exploration of high-free-energy regions.

The FESs for the systems with \nion~= 72 and \conc~=0.01 and 0.02, and with \nion~= 108 at \conc~= 0.01 contain a qualitatively similar set of low-free-energy routes (LFERs) for the transition from large \cacmin and \ccamin values to the global minimum. Specifically, in these three systems, as the \cacmin and \ccamin decrease from their largest values, the LFERs first reach a distinct local minimum (AA), located at \cacmin~$\simeq$~\ccamin~$\simeq 5 \AA{}$ for the two systems with \conc~=0.01 and \cacmin~$\simeq$~\ccamin~$\simeq 4.5 \AA$ for the system with \conc~= 0.02. Although the latter minimum falls within the region $\lambda \leq$~\la, we classify the AA minimum in all three systems as belonging to state A. In general, the precise locations of states A and B in the OPES FESs may differ slightly from the \la and \lb definitions used in \infretis. Since these differences do not affect the qualitative features of the FESs, states A and B are considered to be separated at \cacmin~= \ccamin~= 4~\AA{}, near the maximum of the free energy barrier, in the analysis of all OPES FESs.

Minimum (AA) serves as a branching point. From it, transitions to three additional local minima are possible: minimum (BA), where \cacmin is in state B and \ccamin is in state A; minimum (AB), where \cacmin is in state A and \ccamin is in state B; and the global minimum (BB), where both OPs are in state B. Consequently, minimum (AA) gives rise to three possible LFERs to the global minimum: (LFER i) \cacmin A~$\rightarrow$~B (minimum AA~$\rightarrow$~BA) followed by \ccamin A~$\rightarrow$~B (minimum BA~$\rightarrow$~BB) ; (LFER ii) \ccamin A~$\rightarrow$~B (minimum AA~$\rightarrow$~AB) followed by \cacmin A~$\rightarrow$~B (minimum AB~$\rightarrow$~BB); and (LFER iii) where both \cacmin and \ccamin A~$\rightarrow$~B transition simultaneously (minimum AA~$\rightarrow$~BB). The LFER (iii) exhibits a substantially higher free energy barrier and a narrower bottleneck than the other two across all three systems. Among the other two LFERs, the FESs show that the transition of \ccamin into state B before the transition of \cacmin into state B (AA~$\rightarrow$~AB along LFER ii) involves a lower free energy barrier than the opposite order (AA~$\rightarrow$~BA along LFER i), and is therefore kinetically more accessible. This observation is consistent with the \infretis results, where the formation of the first C--Ca coordination sphere exhibits higher \pcross than the formation of the first Ca--C coordination sphere.

However, when considering the full transition towards the global minimum (where both OPs have reached state B), only the systems with \nion = 72 unambiguously favour LFER (ii). In this possible route after the initial transition of \ccamin into state B (AA~$\rightarrow$~AB), the subsequent transition of \cacmin into state B (AB~$\rightarrow$~BB) encounters a lower free energy barrier than the competing transition in which \ccamin returns to state A (AB~$\rightarrow$~AA). For \conc~=0.01 and \nion~=108, the transition of \cacmin into state B (AB~$\rightarrow$~BB) following the initial transition of \cacmin into state B (AA~$\rightarrow$~AB along LFEC (ii) encounters a higher free energy barrier than the competing transition in which \ccamin returns to state A (AB~$\rightarrow$~AA). This makes the progression toward global minimum kinetically less favourable along this route for this system. Importantly, all discussed differences between the free-energy barrier heights associated with LFERs (i) and (ii) are relatively small in all three FESs, indicating that both LFERs remain kinetically accessible in these systems.

For the system with \nion~=108 and \conc~=0.02 (see Figure~\ref{fig:6min_cac_cca_OPES}(c)), no distinct free-energy minima are observed apart from the global minimum (BB). The FES suggests that LFER (i) (AA~$\rightarrow$~AB~$\rightarrow$~BB) is the most kinetically accessible route, characterized by a nearly barrierless transition. This absence of a significant activation barrier is even more apparent by the one-dimensional free-energy profiles along each CV that exhibit barriers of at most 1~\kt (see Figure~\ref{fig:SI_FE_1d} in \SI). Consequently, whether the transition proceeds first along \cacmin or \ccamin is of limited significance for this system. In contrast, the pronounced reduction in the free energy barrier with increasing \nion for the system with \conc~= 0.02 is notable and consistent with the corresponding increase in \kab observed in the \infretis simulations. For \conc~= 0.01, the FESs exhibit a different trend with increasing \nion: the barrier remains and even slightly increases between \nion~= 72 and =108. Therefore, the OPES results support the conclusion drawn from the \infretis simulation, namely that the two concentrations exhibit different kinetic behaviour.

The presented analysis of the LFERs for the formation of the Ca--C and C--Ca coordination spheres is conducted explicitly within the framework of transition state theory which infers pathway preferences from free-energy barrier heights but does not provide an explicit dynamical description of the reaction process. All subsequent analyses of LFERs (based on FESs, not CFE landscapes) share the same limitations.

\begin{figure*}[h!]
    \centering
    \includegraphics{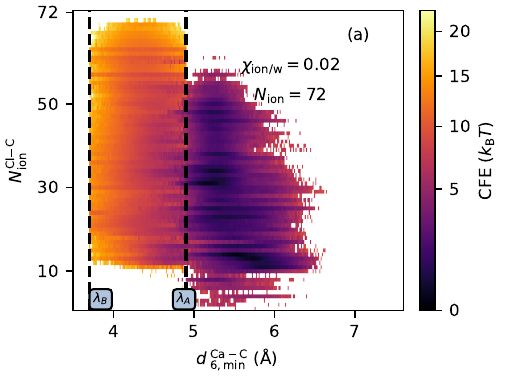}
    \includegraphics{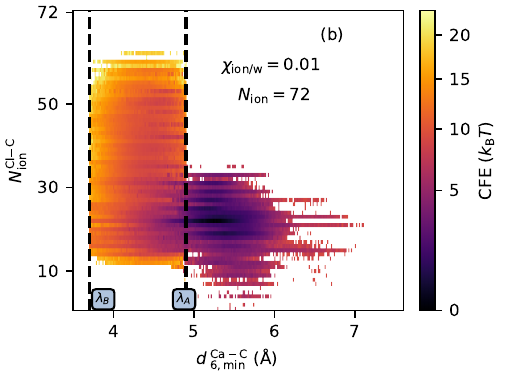}
    \includegraphics{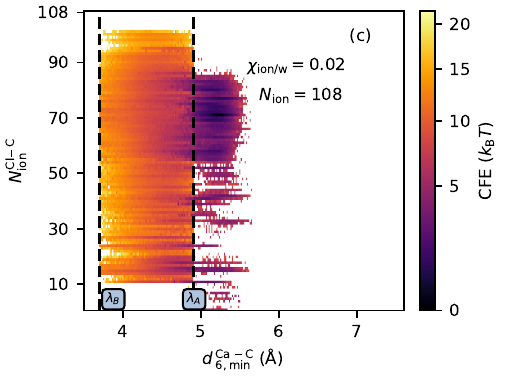}
    \includegraphics{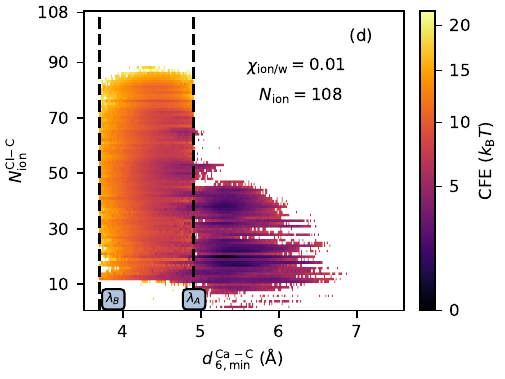}
    \caption{CFE surfaces as a function of the \nionrc and \cacmin for the systems with compositions indicated in the legends.}
    \label{fig:6min_cl_size}
\end{figure*} 

A numerical comparison of the free energy barriers computed with OPES and the CFE barriers derived from \infretis revealed that the former are lower for all systems. For example, for the system with (\conc, \nion)~=~0.01, 108, OPES shows a barrier of $\simeq$ 4.5 \kt at \cacmin~= 4 \AA{}, whereas \infretis yields a barrier of $\simeq$ 9 \kt at the same \cacmin. This discrepancy arises from the fundamental methodological differences between the two approaches. Free energy barriers obtained with OPES reflect the reconstructed equilibrium FES and do not explicitly account for the kinetic aspects of the transition. Moreover, OPES barriers are highly sensitive to how the transition is represented in the chosen CVs. By contrast, \infretis samples ensembles of reactive trajectories directly, thereby incorporating both thermodynamic and kinetic factors, while remaining far less sensitive to the choice of OPs.\cite{CFE}

An additional factor contributing to the difference in barrier heights is that OPES explores the multidimensional CV space using both \cacmin and \ccamin CVs. In contrast, \infretis defines interfaces along a single OP for a specific A-to-B reaction. Nevertheless, the sampled trajectories are not restricted to the chosen OP and evolve in the full phase space, allowing transitions associated with the other reaction to occur as well, as demonstrated in Section~\ref{sec:SI_CFE_OPs} in \SI. Despite the quantitative differences, both methods revealed the same qualitative trends. 

\subsection{Ionic aggregation and formation of the first coordination sphere}\label{sec:cluster_analysis}

To investigate the relationships between the aggregation of calcium and carbonate ions into the CI-cluster and the formation of the first coordination sphere, we performed a cluster analysis over the trajectories collected with \infretis and OPES.

\begin{figure*}[h!]
    \centering
    \includegraphics{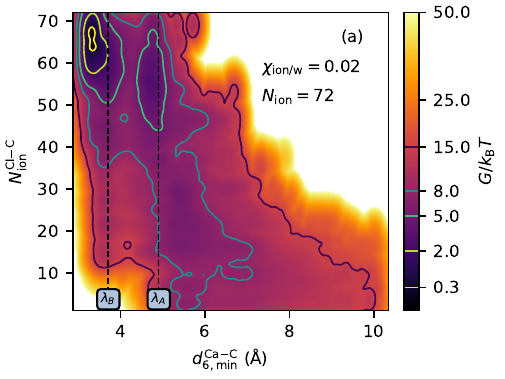}
    \includegraphics{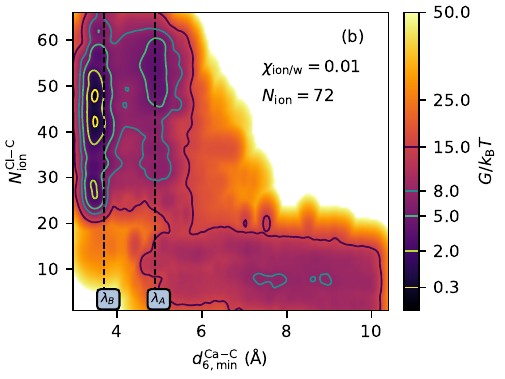}
    \includegraphics{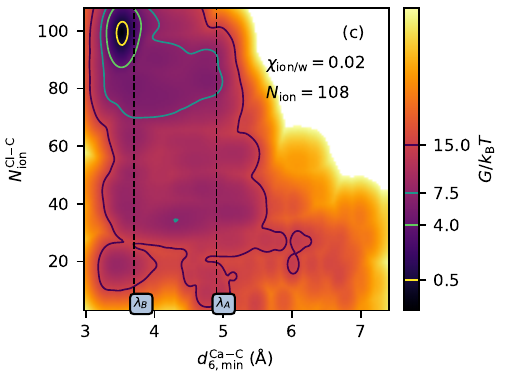}
    \includegraphics{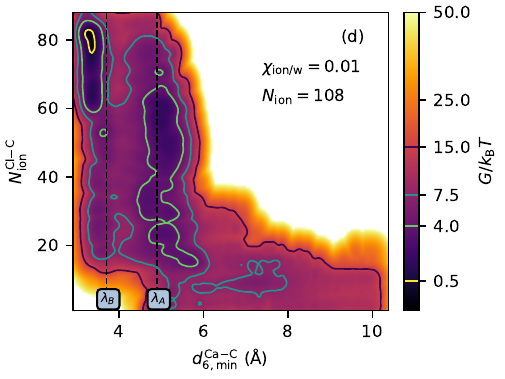}
    \caption{FESs projected onto \cacmin and the corresponding \nionrc for the systems with compositions indicated in the legends.}
    \label{fig:6min_cac_RC_size_OPES}
\end{figure*}

Figure~\ref{fig:6min_cl_size} shows the CFE surfaces, obtained from \infretis simulations, for the four systems as functions of the OP \cacmin and the size of the corresponding CI-cluster \nionrc. All energy minima in Figure~\ref{fig:6min_cl_size}, as well as in the other CFE landscapes presented herein, are located before \la, which is consistent with the CFE definition given in Section~\ref{sec:method_infretis}. The OPES FESs as functions of \cacmin and the corresponding \nionrc are shown in Figure~\ref{fig:6min_cac_RC_size_OPES} for the four systems. While \infretis samples transition region excessively, it does not sample state B thoroughly as this is not required for the calculation of \kab. By contrast, the OPES approach is well-suited to sampling energy minima, thus fulfilling this gap. Consistently with FESs in Figure~\ref{fig:6min_cac_cca_OPES}, the FESs shown in Figure~\ref{fig:6min_cac_RC_size_OPES} exhibit the global free energy minima in state B for all four systems.

Both systems with \conc~= 0.01 show specific aggregation behaviour. At \nion~=72 (Figure~\ref{fig:6min_cl_size}(b)), the basin of state A spans \nionrc values from 9 to 33 with a minimum located at \nionrc $\simeq 22$. This CFE valley reaches maximum values of $\approx 7-8$ \kt; similar maximum CFE values are observed for state A across all other CFE surfaces calculated in this work. At \nion~=108 (Figure~\ref{fig:6min_cl_size}(d)), the low-CFE region associated with state A extends up to \nionrc $\simeq 48$, and, while retaining a minimum at \nionrc $\simeq 22$, exhibits two new minima centred at \nionrc $\simeq 12$ and $\simeq 38$. Within the transition region, \nionrc reaches values significantly higher than those observed in state A for both \nion. This suggests ionic aggregation, limited before the reaction and reaching larger \nionrc during the reaction.

FESs computed with OPES reveal similar aggregation behaviour for the systems with \conc~= 0.01, although some differences relative to the \infretis results are observed. For \nion~= 72 (Figure~\ref{fig:6min_cac_RC_size_OPES}(b)), the free-energy basin in state A, defined here as the region with the free-energy values below $\leq 8$~\kt to facilitate comparison with the \infretis results, extends from \nionrc $\simeq 30$ to $\simeq 60$, reaching a minimum at \nionrc $\simeq 50$. These \nionrc values were not observed in the \infretis simulation, revealing a discrepancy between the two methods. This discrepancy is discussed in detail in Section~\ref{sec:SI_infretis_vs_OPES} in \SI, from which we conclude that the minima found with OPES may not correspond to a primary reaction pathway, despite being energetically feasible. Consequently, hereafter, we consider the \infretis representation of state A for this system in the context of cluster analysis.
In state B, the FES exhibits three competing minima of comparable depth located at \nionrc $\simeq 28, 42$, and $48$. Consequently, the \nionrc values associated with free-energy minima are smaller in state B than in state A. However, relative to the minimum in state A identified from the \infretis CFE surface, the free-energy minima in state B occur at larger \nionrc values.

For \nion~= 108 (Figure~\ref{fig:6min_cac_RC_size_OPES}(d)), the FES exhibits a low-free-energy region spanning \nionrc $\simeq 10$ to \nionrc $\simeq 70$ ($\leq 8$~\kt) in state A, which is in good agreement with the \infretis results. In state B, the free energy minimum forms an extended region from \nionrc $\simeq 60$ to $\simeq 90$, with its lowest free energy point at \nionrc $\simeq 80$. Thus, for this system, the FES suggests an increase in CI-cluster size in state B relative to state A. 

The limited \nionrc in state A and the increase in \nionrc during or after the transition are consistent across both \infretis and OPES results. Moreover, no characteristic change in the energy landscapes with increasing \nion is revealed. Although larger cluster sizes become accessible at \nion $=108$, state A remains characterized by clear energy minima at small cluster sizes (\nionrc $\leq 60$).

For the systems with \conc~= 0.02, we observe larger CI-clusters compared to the systems with \conc~= 0.01. For \nion~= 72 (Figure~\ref{fig:6min_cl_size}(a)), the CFE valley in state A spans \nionrc values from $\simeq 10$ to $\simeq 50$, with the lowest points located at \nionrc~$\simeq 14, 31$, and 34. For \nion~= 108 (Figure~\ref{fig:6min_cl_size}(c)), CFE landscape exhibits a single pronounced minimum in state A, centred at a much larger \nionrc~$\simeq 71$ and spanning the range \nionrc~$\simeq 50$--$80$. Small \nionrc values are only sparsely represented in state A, appearing as semi-isolated low energy configurations rather than a continuous free energy region. Thus, for this concentration, the stability basins corresponding to state A on the CFE landscapes for the two \nion occupy distinct, non-overlapping ranges of the \nionrc values. Nevertheless, both reach much larger \nionrc values than at \conc~= 0.01. In the transition region, the \nionrc of both systems span a wide range of values, same as at lower \conc. However, unlike at lower \conc, this broad distribution does not signify a substantial increase in \nionrc. Instead, only a modest expansion towards larger CI-clusters is observed, while at \nion~=108, the transition region is dominated by additional smaller CI-cluster sizes relative to state A.

Visual inspection of the CFE surfaces does not reveal clear reaction channels associated with specific cluster sizes for all four systems. To gain further insight on this topic, we performed a detailed analysis of the \infretis reactive trajectories, which presented in Section~\ref{sec:trans_region}. 

FESs calculated from the OPES simulation also show more extensive aggregation for \conc~=0.02 compared to \conc~=0.01. For \nion~= 72 (Figure~\ref{fig:6min_cac_RC_size_OPES}(a)), the FES exhibits a broad low free energy region ($\leq 8$~\kt) covering the entire range of \nionrc. This broad range of accessible \nionrc states only slightly exceeds that observed in the CFE landscape computed using \infretis. However, the free-energy minimum in state A is located at \nionrc~$\simeq 55$, differing from the positions of the CFE minima and suggesting that a large CI-cluster forms before the first coordination sphere is formed. In state B, the free energy minimum is centred around \nionrc $\simeq 65$, indicating only a modest increase in CI-cluster size after the reaction. 

As \nion increases to 108 (Figure~\ref{fig:6min_cac_RC_size_OPES}(c)), the FES no longer exhibits a distinct free energy minimum in state A. Instead, the only distinct free-energy minimum is located at \nionrc $\simeq 100$ in state B. This minimum lies within a valley of low free energy ($\leq 8$~\kt) that connects to state A, suggesting that the formation of such large CI-clusters may precede the formation of the first coordination sphere.

Taken together, both methods reveal that for \conc~= 0.02, calcium and carbonate ions form large aggregates (reaching 66\% of the maximum \nionrc for both \nion values in \infretis, and even larger \nionrc in OPES) already in state A. Moreover, for the larger systems (\nion~= 180), neither method revealed a distinct energy minimum for small \nionrc ($\leq 50$). Such energy landscapes, especially compared to those for \conc~= 0.01, suggest a spinodal-like mechanism of ionic aggregation for this \conc. Reaction progression through this mechanism appears to be limited in the smaller systems and prominent in the larger system. Thus, for \conc~= 0.02, both ionic association and the rate constants for coordination-sphere formation exhibit a dependence on \nion.

Importantly, the presented FESs are projected onto \nionrc, which is not the biased CV. Therefore, hidden barriers along this coordinate cannot be completely excluded. However, the broad range of \nionrc values observed in the FESs indicates extensive sampling of this variable. Furthermore, the path-sampling \infretis simulations -- where trajectories propagate without biasing forces -- also exhibit consistent spinodal-like aggregation. Finally, spinodal aggregation of \caion and \cion ions has previously been reported for this force field \cite{LLPS_MD_1, PNC_to_ACC_LLPS}.

It is interesting to note that, in OPES FES for all systems, the transition region is characterised by LFERs with $G \leq 8$~\kt. The \nionrc interval covered by these LFERs overlap with that associated with the free energy minima in state A. This could indicate that the CI-clusters do not undergo significant size changes during the formation of the first coordination sphere.

The FES and CFE landscapes calculated with respect to \ccamin and the corresponding \nionrc are reported in Section~\ref{sec:SI_FE_CFE} of \SI. For each composition, the CFE surfaces expressed in terms of \cacmin or \ccamin are qualitatively and quantitatively similar, exhibiting the same dependencies on \conc and \nion. 

\begin{figure*}[h!]
\includegraphics{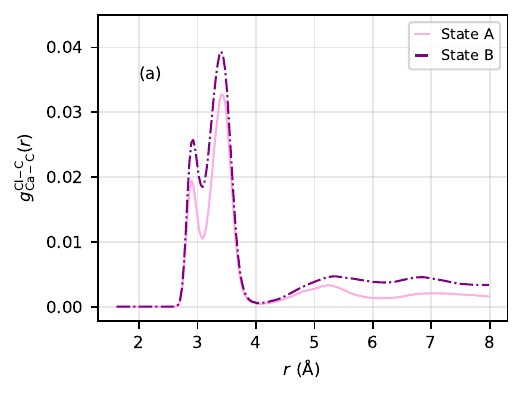}
\includegraphics{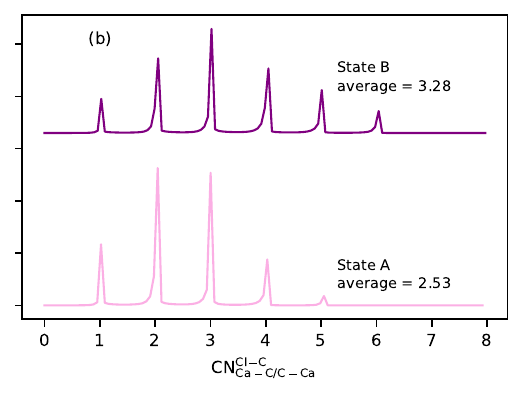}
\caption{(a) Ca--C RDFs in states A and B, as indicated in the legend; (b) CN$^{\rm CI-C}_{\rm Ca-C/C-Ca}$ histograms in states A and B, as indicated in the legend, including contributions from both Ca--C and C--Ca coordinations within the CI-cluster. All plots characterise the CI-cluster associated with \cacmin OP in the system with \conc $=0.01$ and \nion $=108$.}
\label{fig:struct}
\end{figure*}

In summary, both the \infretis and OPES methods reveal different trends in ionic aggregation for the two \conc. These trends exhibit different relationships with the formation of the first coordination sphere, with the latter serving as a measure of ionic organization for the two concentrations. For \conc~= 0.01, ionic aggregation in state A is limited, while further cluster growth occurs in the transition region during the formation of the first coordination sphere or in state B. We refer to this ionic aggregation-organization relationship as stepwise. For \conc $=0.02$, ionic aggregation proceeds directly towards larger clusters through a spinodal-like mechanism in state A, followed by the formation of the first coordination sphere. Accordingly, we refer to this ionic aggregation-organization relationship as aggregation-first. 

The dependence of ionic aggregation on solvent concentration is a known, entropically driven phenomenon. In contrast, the connection between system composition and the formation of the first coordination sphere -- which is a strictly local property -- is less well understood. To gain insight into the latter, we analyse whether ionic aggregation provides a link between system composition and local ionic organization.

In Section~\ref{sec:SI_flux} in \SI, we discuss the relationship between the flux through \la and the CFE surfaces shown in Figure~\ref{fig:6min_cl_size}. 
The flux is correlated with the extent of state A (size of the state-A region on the FES), which varies with system composition at least partly due to differences in ionic aggregation. In particular, the system with \conc $=0.02$ and \nion $=108$ exhibits a distinct narrow region of large \nionrc values in state A and, accordingly, a higher flux. Thus, the change in flux with system composition is partially attributable to the differences in ionic aggregation. Consequently, the different kinetic behaviours for the formation of the first coordination sphere, which arise primarily from differences in the flux, are also linked to these aggregation trends. Moreover, aggregation in state A determines the sizes of the CI-clusters from which the reaction originates (see Section~\ref{sec:trans_region} for further discussion).

\subsection{Structure of the CI-cluster}\label{sec:CI_struct}

The structure of amorphous and liquid-like \caco species simulated with the force field used in this work has been extensively characterized in previous studies \cite{LLPS_MD_1, ACC_hyd_SIM, PNC_MD_yes, PNC_to_ACC_LLPS}. In the context of these earlier observations, we describe the structural features of the CI-cluster.

Hereafter, state A is analysed using \infretis trajectories and state B using OPES trajectories. The choice of \infretis for state A is motivated by its unbiased sampling, the greater availability of state-A configurations relative to OPES, and the results of the cluster analysis in Section~\ref{sec:cluster_analysis}. We first calculated RDFs between different atom types within the CI-cluster. Figure~\ref{fig:struct}(a) shows the Ca--C RDFs for states A and B for the system with \conc~= 0.01 and \nion~= 108. The Ca--C RDFs computed for other compositions do not exhibit significant differences and are presented in Figure~\ref{fig:rdf_SI_1}(a,b) of \SI. Section~\ref{sec:SI_RDFs} of \SI further provides RDFs calculated between other atom types for all considered system compositions and OPs. 

For all systems in both states A and B, the overall profiles and peak positions of the computed RDFs closely match reported data for both DLP (reported in \cite{PNC_to_ACC_LLPS,DLP_to_ACC_new}) and ACC (reported in \cite{ACC_hyd_SIM, ACC_dehydration_sim}), which exhibit similar short-range structural features. Any differences in peak heights relative to published values likely arise from subtle differences in local ordering and force-field resolution. To distinguish between these two phases, we computed the Ca--C/C--Ca CN, including contributions from both Ca--C and C--Ca coordinations. Histograms of the Ca--C/C--Ca CN within the CI-cluster associated with \cacmin (CN$^{\rm CI-C}_{\rm Ca-C/C-Ca}$) in states A and B are shown in Figure~\ref{fig:struct}(b), for the system with \conc~= 0.01 and \nion~= 108. Corresponding histograms for all investigated systems and OPs are provided in Figure~\ref{fig:CN_hist_SI} in \SI. Across all systems, the overall average CN in state A lies between 2 and 3, consistent with values previously reported for DLP ionic aggregates simulated with the same force field.\cite{DLP_intro_PNC_MD_no, LLPS_MD_1, DLP_to_ACC_new}

In state B, although the RDFs do not change significantly in shape, their peak heights increase. The average CN also increases, reaching values above 3 across all systems and thereby exceeding the values reported for DLP clusters. This increase in ionic coordination likely reflects the displacement of first-shell water molecules by carbonate ions, and, consequently, progressive dehydration (see Section~\ref{sec:water} for further discussion). These structural characteristics are more consistent with hydrous ACC, which is more highly coordinated and less hydrated than DLP clusters. This interpretation is in line with previous studies reporting the dehydration-driven transformation of DLP into ACC.\cite{LLPS_MD_1, PNC_to_ACC_LLPS, LLPS_2019, DLP_to_ACC_new} 

The increase in the CN may alternatively reflect a slow ongoing transformation of the clusters toward local ordering characteristic of anhydrous crystalline calcium carbonate polymorphs, i.e. a direct DLP-to-crystalline transformation process that has been previously proposed.\cite{DLP_intro_PNC_MD_no, LLPS_1} Cautious evidence for this interpretation may be found the C--C RDFs (Figure~\ref{fig:rdf_SI_1}(f) in \SI) which, for several systems, exhibit two small maxima at $r \simeq 3$ \AA{} and $r \simeq 3.5$ \AA{} in state B that are not characteristic of ACC. In addition, the differences between the C--C RDFs computed for different systems suggest subtle variations in structural ordering across the systems.

We have also calculated the Ca--C/C--Ca CN as a function of CI-cluster size (Figure~\ref{fig:CN_per_cl_size_SI} in \SI). In state B (Figure~\ref{fig:CN_per_cl_size_SI}(b) and \ref{fig:CN_per_cl_size_SI}(d) in \SI), for all systems, the CN reaches maximum of $\gtrsim 3.4$ at the \nionrc corresponding to the free energy minimum of the respective system (see Figure~\ref{fig:6min_cac_RC_size_OPES}). In addition, the CN exhibits a second distinct maximum around \nionrc $=11-17$ in both state A and B for several systems, including all with \nion $=108$. As discussed below (Section~\ref{sec:trans_region}), this maximum reflects a reactive pathway rather than the free-energy minima.

Previous simulation studies have identified DOLLOPs as a distinct class of hydrated \caco aggregates.\cite{PNC_MD_yes} However, this structure is not consistent with the large CN values ($\geq 4$) observed in our CI-clusters (Figure~\ref{fig:CN_hist_SI} in \SI). Consequently, the CI-clusters obtained in this work do not exhibit a DOLLOP structure. Moreover, analysis presented in Section~\ref{sec:SI_GC} of \SI reveals that DOLLOP-structured clusters, if present, represent a negligible fraction of all clusters in the simulation box for all systems' compositions. Finally, in Section~\ref{sec:SI_K1_K2} of the \SI, we analyse the shapes of the CI-clusters, completing their structural characterization. We found that the lowest free-energy states are generally rod-like. The exception is the system with \conc~=0.02 and \nion~=108, whose state A is predominantly spherical according to the \infretis simulation. In addition, the free-energy minima are relatively broad, indicating shape fluctuations rather than rigid morphologies.

Overall, for all systems, the amorphous CI-clusters exhibit a structure most consistent with that of DLP in state A. The transition from state A to state B is associated with mild structural transformations within the CI-clusters, reflected in an increase in the CN.

Importantly, the OPs of \cacmin and \ccamin do not, by definition, track the transformation of an entire cluster. They only probe local structural changes within a cluster, which may potentially represent the initial steps toward the formation of a new stable phase (see screenshots in Section~\ref{sec:SI_snapshots} in \SI). Moreover, although CNs provide valuable insight and allow us to distinguish between DLP-like and ACC-like local environments, a definitive discrimination between a dynamic liquid-like DLP and a more rigid ACC would require dynamic descriptors, such as a mean square displacement. Consequently, the terms DLP and ACC are used here only as structural references rather than as definitive assignments of the CI-clusters to macroscopic phases. In addition, the present discussion of cluster structure is restricted to sufficiently large clusters with \nionrc $\gtrsim 20$. Based on the \rg values (Figure~\ref{fig:SI_av_rg} in \SI), these clusters exceed the length scale corresponding to the second RDF peak, making structural comparisons based on coordination environments more meaningful. It is also worth noting that, despite using \infretis trajectories for the analysis of state A, the same structural picture emerges from the structural analysis of the OPES trajectories. 

\subsection{Analysis of reactive paths}\label{sec:trans_region}

Next, we analysed the reactive paths obtained using \infretis. Reactive paths are defined as trajectories along which the OP successfully crosses from \la and \lb, thereby completing the reaction. As such, reactive paths allow us to analyse the reaction mechanisms and to identify the routes through which the transition proceeds.

The OPs of \ccamin and \cacmin describe the formation of the first coordination sphere through the distance between two atoms, as defined in Section~\ref{sec:method_infretis}. We evaluated the probability that the corresponding pair of atoms belongs to the same cluster along the reactive path, and found that this probability exceeds 98.8\% for all considered systems. This quantity was computed as the fraction of frames, across all reactive trajectories, in which the ions associated with the OPs at a given frame belonged to the same cluster. Therefore, the transition across the energy barrier between \la and \lb occurs predominantly through positional reorganisation of ions within the CI-cluster, rather than association of ions from different clusters.

\begin{figure}[h!]
    \includegraphics{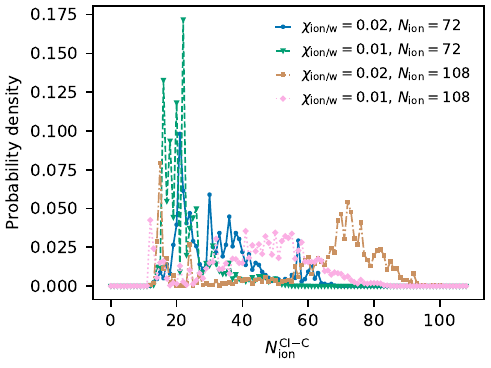}

    \caption{Probability density of \nionrc along the reactive paths, normalised by the corresponding trajectory length to prevent longer trajectories from contributing disproportionately. The plots were obtained from the \infretis simulations with the \cacmin OP. The system compositions are indicated in the legend.}
    \label{fig:cl_size_RP}
\end{figure}

The distribution of \nionrc along the reactive paths was computed for simulations with \cacmin OP, as shown in Figure~\ref{fig:cl_size_RP}. The maxima in these distributions identify the CI-cluster sizes that are most frequently associated with the reaction. 

For the larger systems (\nion~= 108), a clear bimodal distribution of \nionrc is observed, with two distinct peaks of different shapes separated by a minimum. The first peak, which is higher and narrower, is located at \nionrc values as small as 12 and 15 for \conc~= 0.01 and 0.02, respectively. The second peak, which is slightly lower and wider, appears at \nionrc~$\approx 72$ for \conc~= 0.02, and spans approximately from 32 to 58 for \conc~= 0.01. 

Such \nionrc distribution suggests the existence of two reaction pathways that differ in the characteristics of the cluster within which the reaction occurs. Along the first pathway, featuring small CI-clusters, \nionrc~$\approx 12-15$, formation of the first coordination sphere is mediated by the CI-cluster close to the minimum cluster size required to complete the coordination sphere. The minimum cluster comprises 12 ions: one central ion, six nearest neighbours, and five additional ions to maintain electroneutrality. Along the second pathway, featuring a larger CI-cluster, \nionrc~$\approx 32-80$ at different \conc, the reaction occurs within an amorphous cluster with the structure discussed in Section~\ref{sec:CI_struct}. The starting configurations of the CI-clusters along the two pathways are also characterized by slightly different CN values, with those associated with the first pathway having a slightly higher CNs (state A in Figure~\ref{fig:CN_per_cl_size_SI}(a) in \SI). In state B, the CN exhibits two maxima as a function of \nionrc, each associated with one of the two pathways (Figure~\ref{fig:CN_per_cl_size_SI}(b) in \SI). Moreover, more than half of the reactive trajectories (1482 out of 2143 for \conc~= 0.01, 1648 out of 2373 for \conc~= 0.02) remain within a single pathway throughout their evolution across the barrier, without transitioning to the alternative pathway. The trajectories associated with the two pathways are distinguished by their characteristic durations. The first pathway is considerably faster, trajectory lengths of $\approx$ 5 and 10 ps for \conc~= 0.02 and 0.01, respectively. By contrast, the amorphous-cluster-mediated pathway is substantially slower, with typical trajectory lengths of roughly 30 and 70 ps for \conc~= 0.02 and 0.01, respectively.

The existence of these two pathways is particularly intriguing because it may reflect the coexistence of two crystallisation mechanisms. The first pathway, involving formation of a compact $\sim 12$-ion cluster with local coordination environments reaching CN~= 6, is difficult to reconcile with amorphous prenucleation species such as DLP. Instead, it appears more consistent with CNT, where such cluster could represent an early stage in the growth toward a critical nucleus. In contrast, the second pathway, which proceeds within a large amorphous cluster, may be indicative of a multistep nucleation mechanism involving a hydrated amorphous precursor. The observation of both pathways at both concentrations, together with their markedly different timescales, makes this finding particularly noteworthy. Nevertheless, this interpretation remains speculative and requires further investigation.

The differences in \nionrc along the second pathway between the two \conc are consistent with the concentration-dependent aggregation behaviour discussed in Section~\ref{sec:cluster_analysis}. The results therefore again suggest a link between ionic aggregation and the first-coordination sphere formation: concentration-dependent aggregation alters the distribution of \nionrc, indicating that the reaction proceeds through different local ionic environments. This may, in turn, contribute to the observed composition dependence of \pcross and the associated free-energy barrier. One possibility is that the larger CI-clusters observed for \conc~= 0.02 provide a greater number of potential sites for the formation of the first coordination sphere, thereby increasing the likelihood of a successful transition. Alternatively, these clusters may support additional coordination environments that modify the reaction mechanism, as discussed below, in Section~\ref{sec:water}.

While \nionrc along the second pathway decreases as \conc decreases for the larger systems, ionic aggregation becomes even more limited in the smaller systems (\nion~= 72). Consequently, the population of reactive trajectories proceeding via large \nionrc values is suppressed, making the second pathway less distinct or even completely absent. The highest probability maxima are observed at \nionrc~= 22 and 21 for \conc~= 0.01 and 0.02, respectively (see Figure~\ref{fig:cl_size_RP}). At \conc~= 0.01, no distinct pathway at larger \nionrc values is observed, reflecting the highly limited aggregation in this system. In contrast, for \conc~= 0.02, two regions of elevated probability density are present at larger \nionrc values, namely between 30 and 40 and around 57. The corresponding results obtained for smaller systems using the \ccamin OP (Figure \ref{fig:SI_cl_size_RP_cca}(a) in \SI) show a similar behaviour, with a slightly more pronounced second pathway for the system with \conc~= 0.02 ($40 \leq$~\nionrc~$\leq 60$). These findings further demonstrate that system size has a significant impact on the extent of ionic aggregation at a given \conc and, consequently, on the accessibility of the pathways for the formation of the first coordination spheres.


As already mentioned, the majority of the reactive trajectories follow one pathway. Moreover, all the identified \nionrc regions that channel the reaction overlap significantly with the \nionrc ranges associated with the free energy minima in state A (see Figure~\ref{fig:6min_cl_size}), although they may also extend to slightly different \nionrc values. This suggests that the CI-cluster formed in state A largely determines the subsequent reaction pathway. Such interpretation is particularly plausible for the first pathway, which exhibits a relatively rapid progression through the transition region, leaving little opportunity for substantial changes in \nionrc.

\begin{figure*}[h!]
    \centering
    \includegraphics{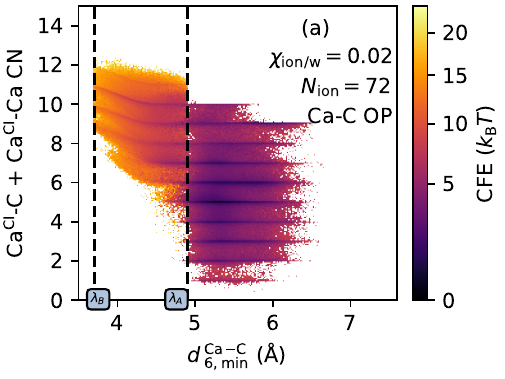}
    \includegraphics{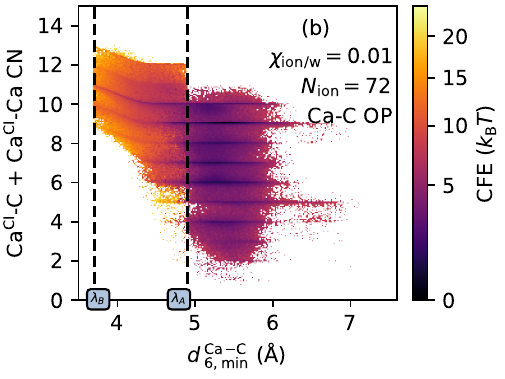}
    \includegraphics{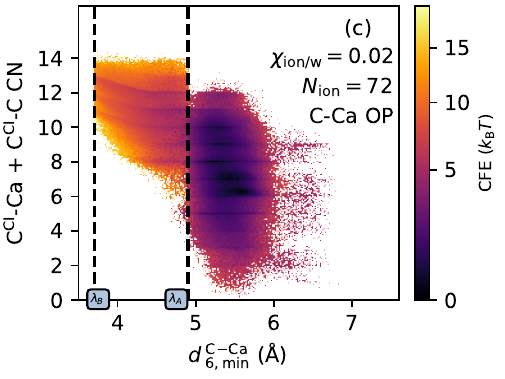}
    \includegraphics{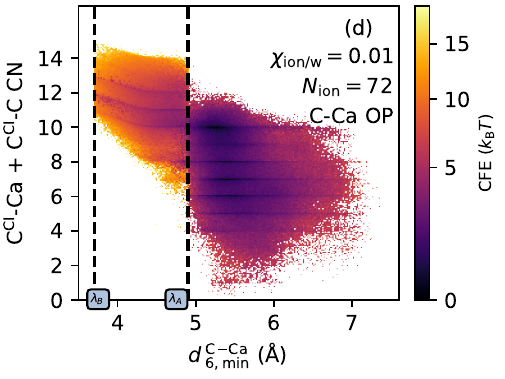}
    
    \caption{CFE surfaces as a function of (a), (b) the \cacmin OP and the \carc--C + \carc--Ca CN, and (c), (d) the \ccamin OP and the \crc--Ca + \crc--C CN. The systems compositions are indicated in the legends. Colours represent the CFE values, as shown by the colour bar.}
    \label{fig:CFE_cn_1_2_72}
\end{figure*}

To further elucidate the dynamics of \nionrc along the reactive paths, we carried out an additional analysis of \nionrc changes along the reactive trajectories, as presented in Section~\ref{sec:SI_react_traj} of the \SI. This analysis confirms the dominance of small \nionrc fluctuations along the reaction, with occasional large changes that may originate either from transitions between the two reaction pathways or from fluctuations within a pathway that, in some cases, spans a broad range of \nionrc values. No systematic trend toward either an increase or a decrease in CI-cluster size within the transition region was observed.

In summary, we have found that the formation of the first coordination sphere proceeds through two pathways that differ primarily in the size of the CI-cluster mediating the reaction. Along the first pathway, the reaction progresses rapidly within a CI-cluster of nearly minimum size (\nionrc $\approx 12$). Along the second pathway, the first coordination sphere forms within an amorphous CI-cluster, the structure of which is discussed in Section~\ref{sec:CI_struct}. The second pathway is strongly influenced by concentration, with a higher concentration enabling the formation of larger CI-clusters. Most reactive trajectories proceed along the pathway selected in state A, exhibiting fluctuations within that pathway, whereas transitions between the two pathways occur less frequently. The absence of a systematic increase in CI-cluster size indicates that the formation of the first coordination sphere is predominantly associated with structural rearrangement within pre-existing CI-clusters rather than with their progressive growth.

\subsection{Ion coordination around central ions}\label{sec:local_cn}

As a two-component system, \caco can form the first coordination sphere either around a calcium atom (\carc, monitored by \cacmin) or around a carbon atom of a carbonate ion (\crc, monitored by \ccamin). In Section~\ref{sec:results_ca_c} we found that the latter reaction proceeds with a higher rate constant.

The analysis of CI-cluster sizes and structures presented above revealed no systematic differences between the two reactions that could account for the observed differences in \kab. We therefore examined the local coordination environments around \carc and \crc. To capture the key features distinguishing the two environments, we performed CN analysis that included the entire second coordination sphere.

\carc and \crc have different types of ions as nearest neighbours. For \carc, C atoms are the nearest neighbours (excluding the oxygen atoms from \cion), and Ca atoms are the second-nearest neighbours. The second coordination sphere, comprising both C and Ca atoms, is therefore characterized by the cumulative \carc--C + \carc--Ca CN. The corresponding CFE landscapes, showing the dependencies between the \carc--C + \carc--Ca CN and the \cacmin OP, are presented in Figures~\ref{fig:CFE_cn_1_2_72}(a) and ~\ref{fig:CFE_cn_1_2_72}(b) for systems with \nion~= 72.

Likewise, Ca atoms are the nearest neighbours of \crc, followed by C atoms as the second-nearest neighbours (excluding oxygen atoms from \cion). The second coordination sphere, comprising both Ca and C atoms, is therefore characterized by the cumulative \crc--Ca + \crc--C CN. The corresponding CFE profiles, showing the dependencies between the \crc--Ca + \crc--C CN and the \ccamin OP, are presented in Figures~\ref{fig:CFE_cn_1_2_72}(c) and ~\ref{fig:CFE_cn_1_2_72}(d) for systems with \nion~= 72.

For all cases, the CFE landscapes reveal multiple parallel low CFE lines (LCFELs). These LCFELs correspond to discrete integer CN values in state A. In the transition region, some LCFELs bend such that the associated CN increases by one in state B compared to state A. This increase is consistent with the formation of the first coordination sphere. Comparing the CFE surfaces for \carc (Figure~\ref{fig:CFE_cn_1_2_72}(a),(b)) with those for \crc (Figure~\ref{fig:CFE_cn_1_2_72}(c),(d)), we observe that the LCFELs are more narrow for the former. Transitions between different LCFELs along the \carc--C + \carc--Ca CN require more energy than those along the \crc--Ca + \crc--C CN. To investigate further, we analysed the CN contributions from the first (\carc--C CN and \crc--Ca CN) and second (\carc--Ca CN and \crc--C CN) coordination layers separately. The corresponding CFE landscapes, revealing the relationships between each CN and its associated OP, are presented in Figures~\ref{fig:SI_CFE_cn_1_72} and \ref{fig:SI_CFE_cn_2_72} in \SI for the first and second coordination layer, respectively. 
Figure~\ref{fig:SI_CFE_cn_1_72} in \SI shows that for the CN component of the first coordination layer (\carc--C CN and \crc--Ca CN), the LCFELs connect minima in state A at \carc--C/\crc--Ca CN~= 3, 4, and 5 to a minimum in state B at \carc--C/\crc--Ca CN~= 6. The LCFELs connecting CN~= 5 and 6 are characterised by substantially lower CFE values than the other LCFELs. This behaviour is consistent with the formation of the first coordination sphere, which requires increase in the \carc--C/\crc--Ca CN to 6. By contrast, the CFE surfaces for the CN component of the second coordination layer (Figure~\ref{fig:SI_CFE_cn_2_72} in \SI) exhibit multiple parallel LCFECs characterised by constant integer \carc--Ca/\crc--C CN values throughout the reaction. Consequently, state B is degenerate with respect to this CN, with several CFE minima corresponding to different coordination numbers. This behaviour is consistent with the OPs not tracking Ca--Ca and C--C interactions, thereby allowing multiple values of the second-layer CN. Together, these results indicate that the bending of the LCFELs in Figure~\ref{fig:CFE_cn_1_2_72} is associated with changes in the \carc--C/\crc--Ca CN required for the formation of the first coordination sphere, while their multiplicity and parallelism arise mainly from the different discrete values of the \carc--Ca/\crc--C CN.

Despite the fact that LCFECs proceed via constant \carc--Ca/\crc--C CN, the deepest minima occur at larger CN values in state B than state A for both considered compositions. For example, in the system with \conc~= 0.01, the \carc--Ca CN reaches its deepest CFE minima at 2 and 5 in states A and B (Figure~\ref{fig:SI_CFE_cn_2_72}(b) in \SI), respectively, and the \crc--C CN shows minima at 4 and 6 (Figure~\ref{fig:SI_CFE_cn_2_72}(d) in \SI). These differences suggest an increase in the \carc--Ca and \crc--C CN as the reaction progresses. The CFE surfaces further indicate that changes in the \crc--C CN occur over lower CFE barriers than corresponding changes in the \carc--Ca CN. This observation is supported by the corresponding RDFs. The Ca--Ca RDFs shown in Figures~\ref{fig:rdf_SI_1}(c) and (d) in \SI exhibit two well-defined peaks at distances~$\simeq 4$ and 6 \AA{}, separated by a deep minimum. By contrast, the C--C RDFs shown in Figures~\ref{fig:rdf_SI_1}(e) and (f) in \SI exhibit multiple maxima that are smoothly connected, with no pronounced regions of ion depletion between them. The sharper separation of the Ca-Ca RDF peaks and the smoother C-C RDFs are consistent with the more discrete \carc--Ca CN states and the more continuous \crc--C CN states observed in the CFE landscapes, respectively. This provides a structural basis for the different free-energy barriers associated with changes in these CNs.
 
Taken together, we propose that the higher free energy barriers associated with increasing the \carc--Ca CN, relative to the \crc--C CN, may contribute to the lower \kab of the calcium-centred reaction. The role of the geometrically frustrated Ca-Ca interactions was previously highlighted by Nicholas et al., who demonstrated that they contribute to the structural complexity of the system of ACC.~\cite{geom_frustr_Ca_Ca}

For completeness, we have also carried out the coordination analysis with respect to \carc for systems with \nion~= 108. The corresponding CFE surfaces are presented in Figures~\ref{fig:SI_CFE_cn_1_2_108}-\ref{fig:SI_CFE_cn_2_108} in \SI. These surfaces reveal similar energy patterns, with parallel LCFELs connecting states A and B. 

\begin{figure*}[h!]
    \centering
    \includegraphics{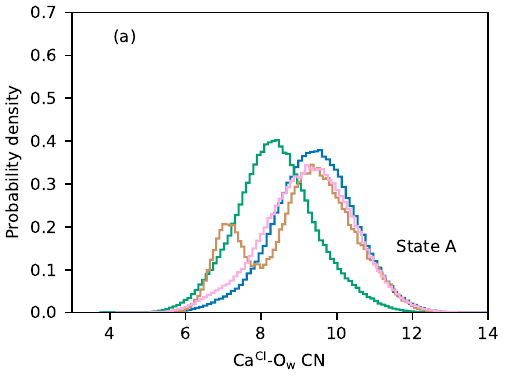}
    \includegraphics{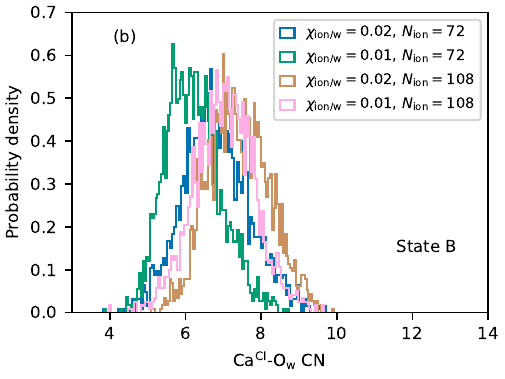}
    \caption{Probability densities of the \carc--\ow CN in (a) state A and (b) state B. The plots were obtained from trajectories generated by \infretis simulations with the \cacmin OP. System compositions are given in the legend. }
    \label{fig:cn_ow}
\end{figure*}

To evaluate the effect of system composition on the coordination environment around \carc, we calculated the distributions of the \carc--C + \carc--Ca CN in states A and B for the four systems, as shown in Figure~\ref{fig:A_full_B_cn_1_2} in \SI. For (\conc, \nion)~=~(0.01, 72), the distribution has its highest peak at a CN larger than that observed for the other systems in both states. In state B, this system is followed by the systems with (\conc, \nion)~=~(0.02, 72), and (0.01, 108), as the dominant CN decreases from 11 to 10. This trend appears to reflect increasing \nionrc associated with the energy minimum in state B.

The dependencies between the \carc--/\crc--centred CNs and the corresponding OP were also investigated for the trajectories obtained using OPES. The corresponding FES are presented in Section~\ref{sec:SI_local_cn_OPES} in \SI. These FES also show that transitions between minima occur over higher energy barriers for the \carc-C CN than for the \crc--C CN.

\subsection{Water coordination around central ions}\label{sec:water}

The role of water molecules in the formation of the \caco coordination sphere was evaluated by analysing the CN between the central atom and oxygen atoms of water molecules (\carc-\ow CN for \cacmin OP and \crc-\ow CN for \ccamin OP). The cut-off distance for computing these CNs is 5.5~\AA{} (see Section~\ref{sec:SI_CN_meth} in \SI). The distributions of the \carc--\ow CN across the four compositions are shown in Figures~\ref{fig:cn_ow}(a) and (b) for states A and B, respectively. The \crc--\ow CN distributions are presented in Figure~\ref{fig:SI_cn_ow_cca} in \SI. All presented distributions were calculated from the \infretis simulations.  

In both states A and B, the distributions for all systems are broad and overlap considerably. Nevertheless, several consistent trends are observed. In particular, the global maximum of the probability density shifts toward lower \carc--\ow CN values by $\approx 2.5$ and $3.0$ between states A and B in simulations with the \cacmin and \ccamin OPs, respectively. This indicates that water molecules are expelled from the first hydration shell of the central ion and replaced by calcium carbonate ions during the formation of the first coordination sphere. The \carc--/\crc--\ow CN never reaches zero because the chosen cut-off extends beyond the first \caco--\ow coordination sphere.

For the system with \conc~= 0.01 and \nion~= 72, the maxima are located at smaller \carc--\ow CN values compared to the other compositions in both states A and B (\carc--\ow CN~$\simeq 8.5$ and $6$ in states A and B for this composition, versus \carc--\ow~CN~$\simeq 9.5$ and 7 for the others). Therefore, the coordination environment around \carc in this system is uniquely depleted in water relative to the other compositions. This reduced hydration is compensated by the larger \carc--Ca + \carc--C CN compared to other systems (see Figure~\ref{fig:A_full_B_cn_1_2} in \SI). Interestingly, unlike the other CNs discussed above, the \crc--\ow CN distribution (Figure~\ref{fig:SI_cn_ow_cca} in \SI) shows little dependence on concentration. Moreover, for the \crc-centred reaction, the probability-density maximum is shifted toward higher CN values in both states. Such differences in the CNs of the two reactions are consistent with distinct coordination environments around C and Ca atoms.

As the differences in the \carc--\ow CNs between compositions persist in both states A and B, we infer that the water molecules responsible for these differences reside beyond the first coordination sphere of \carc, since states A and B are, by definition, characterized by markedly different first-sphere \carc--\ow CNs. Moreover, for the systems with \conc~= 0.01, although the absolute value of the \carc--\ow CN differs, \pcross remains unchanged. This suggests that \pcross depends more on the total reduction in the \carc--\ow CN between states A and B, which is the same for both \nion at this \conc, than on its absolute value.

The system with \conc~= 0.02 and \nion~= 108 uniquely exhibits a second maximum in the \carc--\ow CN probability density in state A, located at $\simeq 7$. This value also corresponds to the probability density maximum in state B. The \cacmin -- \carc--\ow CFE surfaces shown in Figure~\ref{fig:cn_ow_cac}(c) in \SI\ reveal that the two free energy minima associated with the probability density maxima in state A are directly connected to the corresponding minimum in state B by a single broad LCFEC. Thus, this composition displays a continuous LCFEC along which the first coordination sphere may form through a pure reorganisation of ions around the central ion, without requiring a reduction in the \carc-\ow CN. This behaviour may also be reflected in the complementary \carc--C CN for this system, whose CFE surface exhibits a unique LCFEL, absent from the other compositions, that smoothly connects CN values of 5 and 6 in states A and B, respectively (Figure~\ref{fig:SI_CFE_cn_1_108}(a) in \SI). Because the \carc--\ow and \carc--C CNs are correlated, with the latter increasing as the former decreases, this LCFEL for the \carc--C CN may be related to the second minimum in the \carc-\ow CN observed for this composition.

An analysis of the relationships between the \carc–\ow CN and \nionrc was carried out, as shown by the CFE surfaces in Figure~\ref{fig:cl_size_cn_ow} in \SI. Consistent with the corresponding probability distribution, the CFE surface for \conc $=0.02$ and \nion $=108$ (Figure~\ref{fig:cl_size_cn_ow}(c) in \SI) exhibits two distinct minima in state A at \carc–\ow CN $\simeq 9.5$ and $7$. Both minima align with \nionrc $\approx 71$. The remaining systems neither reach such large \nionrc values in state A nor exhibit a secondary minimum. We therefore suggest that the large CI-cluster observed for this system enables the second minimum in state A, from which the reaction can proceed. More generally, larger CI-clusters may access coordination environments that are unavailable in smaller clusters, such as the distinct hydration state identified here. These distinct environments can modify the free energy landscape and enable additional routes. The emergence of the indicated route may contribute to the higher \pcross observed for this system.

\section*{Conclusions}

This paper presents a molecular simulation study of the early stages of calcium and carbonate ions aggregation in aqueous solutions, relevant to \caco crystallisation. The organisation of calcium and carbonate ions in aqueous solution was investigated by simulating the formation of the first coordination sphere. The parallel process of ionic aggregation was analysed independently of local structural ordering and characterized in terms of the association of ions into clusters. The role of concentration and system size in these processes was evaluated by performing simulations at two ion-to-water ratio (\conc~= 0.01 and 0.02) and for different numbers of ions (\nion~= 72 and 108). In addition, as \caco ions may form a coordination sphere around either \caion or \cion, the corresponding coordination-sphere formation reactions were investigated separately. 

\infretis simulations were employed to investigate the kinetics of first-coordination-sphere formation, yielding the corresponding rate constants, \kab, while separate OPES simulations enabled exploration of the underlying free-energy landscapes. The methods were employed with the \cacmin and \ccamin OPs to study the formation of the Ca--C and C--Ca coordination spheres, respectively. 

The computed \kab values revealed different kinetics of the first-coordination-sphere formation for the two \conc values. For the larger systems (\nion~= 108), a higher \kab was observed at \conc~= 0.02. Consistently, the computed FESs revealed no free-energy barrier for the reaction at \conc~= 0.02 and \nion~= 108. Overall, both methods yielded qualitatively consistent results across all aspects of the analysis. For the smaller systems (\nion~= 72), we found that the simulations may underestimate \kab and fail to fully capture important aspects of the process, particularly ionic aggregation. Consequently, the primary conclusions regarding the impact of \conc were drawn from the larger systems. Where deviations were observed in the smaller systems, their origins were analysed and discussed. 

Through the comprehensive analysis of calcium carbonate aggregation, we found that, in addition to exhibiting distinct kinetics of first-coordination-sphere formation, the two \conc levels display different trends in ionic aggregation. The higher \conc is characterised by rapid spinodal-like aggregation, whereas aggregation remains more limited at the lower \conc. Nevertheless, ions form clusters of various sizes before the first coordination sphere forms for both investigated concentrations. The coordination sphere then forms within these pre-existing clusters rather than through the association of ions originating from distinct clusters (we count free ions and ion pairs as separate clusters). Furthermore, the extent of ionic aggregation prior to the reaction, particularly the distribution of cluster sizes, was found to contribute to the computed rate constants.

Analysis of the reactive trajectories collected with \infretis revealed two reaction pathways that differ primarily in the size of the cluster within which the formation of the first coordination sphere occurs. Along the first pathway, the reaction occurs rapidly within a cluster close to the minimum reactive size ($\gtrsim$ 12 ions). Along the second pathway, the reaction proceeds more slowly, with reactive trajectories that are 5 to 7 times longer, and takes place within a large amorphous cluster. The size of the cluster associated with the second pathway depends on the aggregation trends and therefore on \conc. Clusters of different sizes may access distinct configurational environments, thereby contributing to variations in \kab and the corresponding free-energy barriers. In particular, for \conc $=0.02$ and \nion $=108$, we identified two hydration environments of the central ion, distinguished by the coordination of surrounding water molecules.

Although the available evidence does not allow a definitive interpretation, the two observed reaction pathways may be associated with two distinct organization mechanisms. The direct formation of a compact cluster with CN 6 along the first pathway may correspond to an early stage of nucleus formation, consistent with a more compact mode of structural organization, as expected in a CNT framework. On the other hand, formation of the large amorphous cluster prior to the reaction along the second pathway is qualitatively consistent with structural organisation through an amorphous intermediate, as often considered in non-classical, multi-step crystallisation. However, analysis of structural ordering beyond the first coordination sphere and extending to larger compact clusters is required to test this hypothesis.

Another important result is that the absence of a systematic increase in  the cluster size in the transition region indicates that the formation of the first coordination sphere is predominantly associated with structural rearrangement within pre-existing clusters rather than with their progressive growth.

The amorphous cluster involved in the second pathway exhibited a DLP-like structure before the reaction, with coordination numbers between 2 and 3. Following first-coordination-sphere formation, the cluster evolved towards a denser less hydrated arrangement characterised by coordination numbers exceeding 3, indicating the emergence of higher structural organisation within the aggregate, similar to that reported for typical ACC clusters in the literature.

Finally, we compared the formation of the first coordination sphere around a \caion ion with that around a \cion ion. The rate constants obtained with \infretis were found to be slightly higher for the latter reaction. The FESs obtained with OPES indicated a similar trend. Analysis of the coordination environments around the central ions revealed that changes in the Ca--Ca CN proceed over higher energy barriers than corresponding changes in the C--C CN. This greater restriction Ca--Ca rearrangements may contribute to the lower reaction rates observed for the formation of the coordination sphere around a central \caion ion. 

This work considered only two values of \conc, with the aim of identifying trends associated with different levels of the ion-to-water ratio. Determining the specific concentration range over which ions aggregate through spinodal decomposition or nucleation would require further investigation involving a more detailed thermodynamic study with additional \conc values.

Here, we also show that combining \infretis and OPES provides insights that cannot be obtained from either method in isolation. OPES characterises the thermodynamic landscape through efficient sampling of free energy minima and barriers over a wide range of CV values, whereas \infretis resolves the dynamics of barriers crossing and identifies the reaction pathways that are kinetically accessible. Together they establish a direct connection between thermodynamics and kinetics. The limitations of both methods were also discussed, including the strong dependence on the choice of CV and the absence of dynamical description for OPES, as well as its focus on transitions between predefined states and interfaces, which limits its ability to provide a global exploration of the free-energy landscape, and the relatively complex setup for \infretis. Despite these limitations, their combined application provides a uniquely detailed picture of both the free-energy landscape and the dynamical pathways underlying early-stage \caco aggregation. 

\section*{Conflicts of interest}
There are no conflicts to declare.

\section*{Data availability}


The code used for the presented trajectory analysis together with starting configurations and input files are publicly available on GitHub at \url{https://github.com/lizasidler/ionicluster}.

\section*{Acknowledgements}
We gratefully acknowledge Titus van Erp and his group for their assistance with the \infretis software and fruitful discussion.

We acknowledge the EuroHPC Joint Undertaking for awarding project access to the EuroHPC supercomputer LUMI, hosted by CSC (Finland) and the LUMI consortium through a EuroHPC Benchmark and Development access (EHPC-BEN-2025B11-085, EHPC-DEV-2024D11-040, and EHPC-DEV-2025D12-021).

We acknowledge access to the SCARF HPC cluster provided by the STFC Computing Department staff based at Daresbury.


\bibliography{rsc} 

@article{TIS,
    author = {van Erp, Titus S. and Moroni, Daniele and Bolhuis, Peter G.},
    title = {A novel path sampling method for the calculation of rate constants},
    journal = {The Journal of Chemical Physics},
    volume = {118},
    number = {17},
    pages = {7762-7774},
    year = {2003},
    month = {05},
    issn = {0021-9606},
    doi = {10.1063/1.1562614},
    url = {https://doi.org/10.1063/1.1562614},
}

@article{
Infretis,
author = {Daniel T. Zhang  and Lukas Baldauf  and Sander Roet  and Anders Lervik  and Titus S. van Erp },
title = {Highly parallelizable path sampling with minimal rejections using asynchronous replica exchange and infinite swaps},
journal = {Proceedings of the National Academy of Sciences},
volume = {121},
number = {7},
pages = {e2318731121},
year = {2024},
doi = {10.1073/pnas.2318731121},
URL = {https://www.pnas.org/doi/abs/10.1073/pnas.2318731121},
}

@article{OPES,
author = {Invernizzi, Michele and Parrinello, Michele},
title = {Rethinking Metadynamics: From Bias Potentials to Probability Distributions},
journal = {The Journal of Physical Chemistry Letters},
volume = {11},
number = {7},
pages = {2731-2736},
year = {2020},
doi = {10.1021/acs.jpclett.0c00497},
    note ={PMID: 32191470},

URL = { 
        https://doi.org/10.1021/acs.jpclett.0c00497
},
}

@article{PLUMED,
Author = {Bonomi, Massimiliano and Bussi, Giovanni and Camilloni, Carlo and
   Tribello, Gareth A. and Banas, Pavel and Barducci, Alessandro and
   Bernetti, Mattia and Bolhuis, Peter G. and Bottaro, Sandro and
   Branduardi, Davide and Capelli, Riccardo and Carloni, Paolo and
   Ceriotti, Michele and Cesari, Andrea and Chen, Haochuan and Chen, Wei
   and Colizzi, Francesco and De, Sandip and De La Pierre, Marco and
   Donadio, Davide and Drobot, Viktor and Ensing, Bernd and Ferguson,
   Andrew L. and Filizola, Marta and Fraser, James S. and Fu, Haohao and
   Gasparotto, Piero and Gervasio, Francesco Luigi and Giberti, Federico
   and Gil-Ley, Alejandro and Giorgino, Toni and Heller, Gabriella T. and
   Hocky, Glen M. and Iannuzzi, Marcella and Invernizzi, Michele and Jelfs,
   Kim E. and Jussupow, Alexander and Kirilin, Evgeny and Laio, Alessandro
   and Limongelli, Vittorio and Lindorff-Larsen, Kresten and Lohr, Thomas
   and Marinelli, Fabrizio and Martin-Samos, Layla and Masetti, Matteo and
   Meyer, Ralf and Michaelides, Angelos and Molteni, Carla and Morishita,
   Tetsuya and Nava, Marco and Paissoni, Cristina and Papaleo, Elena and
   Parrinello, Michele and Pfaendtner, Jim and Piaggi, Pablo and Piccini,
   GiovanniMaria and Pietropaolo, Adriana and Pietrucci, Fabio and Pipolo,
   Silvio and Provasi, Davide and Quigley, David and Raiteri, Paolo and
   Raniolo, Stefano and Rydzewski, Jakub and Salvalaglio, Matteo and Sosso,
   Gabriele Cesare and Spiwok, Vojtech and Sponer, Jiri and Swenson, David
   W. H. and Tiwary, Pratyush and Valsson, Omar and Vendruscolo, Michele
   and Voth, Gregory A. and White, Andrew},
Title = {Promoting transparency and reproducibility in enhanced molecular
   simulations},
Journal = {NATURE METHODS},
Year = {2019},
Volume = {16},
Number = {8},
Pages = {670-673},
Month = {AUG},
DOI = {10.1038/s41592-019-0506-8},
ISSN = {1548-7091},
EISSN = {1548-7105},

}

@article{RETIS,
  title = {Reaction Rate Calculation by Parallel Path Swapping},
  author = {van Erp, Titus S.},
  journal = {Phys. Rev. Lett.},
  volume = {98},
  issue = {26},
  pages = {268301},
  numpages = {4},
  year = {2007},
  month = {Jun},
  publisher = {American Physical Society},
  doi = {10.1103/PhysRevLett.98.268301},
  url = {https://link.aps.org/doi/10.1103/PhysRevLett.98.268301}
}

@article{WHAM,
  title = {Optimized Monte Carlo data analysis},
  author = {Ferrenberg, Alan M. and Swendsen, Robert H.},
  journal = {Phys. Rev. Lett.},
  volume = {63},
  issue = {12},
  pages = {1195--1198},
  numpages = {0},
  year = {1989},
  month = {Sep},
  publisher = {American Physical Society},
  doi = {10.1103/PhysRevLett.63.1195},
  url = {https://link.aps.org/doi/10.1103/PhysRevLett.63.1195}
}

@article{Raiteri_2015,
Author = {Raiteri, Paolo and Demichelis, Raffaella and Gale, Julian D.},
Title = {Thermodynamically Consistent Force Field for Molecular Dynamics
   Simulations of Alkaline-Earth Carbonates and Their Aqueous Speciation},
Journal = {The Journal of Physical Chemistry C},
Year = {2015},
Volume = {119},
Number = {43},
Pages = {24447-24458},
Month = {OCT 29},
DOI = {10.1021/acs.jpcc.5b07532},
ISSN = {1932-7447},
ResearcherID-Numbers = {Demichelis, Raffaella/H-9193-2012
   Gale, Julian D/B-7987-2009
   Gale, Julian D./AAC-3668-2019
   Raiteri, Paolo/E-1465-2011},
ORCID-Numbers = {Demichelis, Raffaella/0000-0001-9741-213X
   Gale, Julian D/0000-0001-9587-9457
   Gale, Julian D./0000-0001-9587-9457
   Raiteri, Paolo/0000-0003-0692-0505},
Unique-ID = {WOS:000363994500027},
}

@article{water_pot,
    author = {Wu, Yujie and Tepper, Harald L. and Voth, Gregory A.},
    title = {Flexible simple point-charge water model with improved liquid-state properties},
    journal = {The Journal of Chemical Physics},
    volume = {124},
    number = {2},
    pages = {024503},
    year = {2006},
    month = {01},
    issn = {0021-9606},
    doi = {10.1063/1.2136877},
    url = {https://doi.org/10.1063/1.2136877},
    }

@article{LAMMPS,
  author = "A. P. Thompson and H. M. Aktulga and R. Berger and 
     D. S. Bolintineanu and W. M. Brown and P. S. Crozier and
     P. J. in 't Veld and A. Kohlmeyer and S. G. Moore and T. D. Nguyen and
     R. Shan and M. J. Stevens and J. Tranchida and C. Trott and S. J. Plimpton",
  title = "{LAMMPS} - a flexible simulation tool for
     particle-based materials modeling at the 
     atomic, meso, and continuum scales",
  journal = "Comp. Phys. Comm.",
  volume =  "271",
  pages =   "108171",
  year =    "2022",
  doi = "10.1016/j.cpc.2021.108171",
}

@article{water_is_a_key,
author = {Raiteri, Paolo and Gale, Julian D.},
title = {Water Is the Key to Nonclassical Nucleation of Amorphous Calcium Carbonate},
journal = {Journal of the American Chemical Society},
volume = {132},
number = {49},
pages = {17623-17634},
year = {2010},
doi = {10.1021/ja108508k},
    note ={PMID: 21090620},

URL = { 
    
        https://doi.org/10.1021/ja108508k
    
    

},
eprint = { 
    
        https://doi.org/10.1021/ja108508k
    
    

}

}

@article{
LLPS_MD_1,
author = {Adam F. Wallace  and Lester O. Hedges  and Alejandro Fernandez-Martinez  and Paolo Raiteri  and Julian D. Gale  and Glenn A. Waychunas  and Stephen Whitelam  and Jillian F. Banfield  and James J. De Yoreo },
title = {Microscopic Evidence for Liquid-Liquid Separation in Supersaturated CaCO<sub>3</sub> Solutions},
journal = {Science},
volume = {341},
number = {6148},
pages = {885-889},
year = {2013},
doi = {10.1126/science.1230915},
URL = {https://www.science.org/doi/abs/10.1126/science.1230915},
eprint = {https://www.science.org/doi/pdf/10.1126/science.1230915}}

@article{
PNC_first_exp,
author = {Denis Gebauer  and Antje Völkel  and Helmut Cölfen },
title = {Stable Prenucleation Calcium Carbonate Clusters},
journal = {Science},
volume = {322},
number = {5909},
pages = {1819-1822},
year = {2008},
doi = {10.1126/science.1164271},
URL = {https://www.science.org/doi/abs/10.1126/science.1164271},
eprint = {https://www.science.org/doi/pdf/10.1126/science.1164271}}

@article{
PNC_second_exp,
author = {Emilie M. Pouget  and Paul H. H. Bomans  and Jeroen A. C. M. Goos  and Peter M. Frederik  and Gijsbertus de With  and Nico A. J. M. Sommerdijk },
title = {The Initial Stages of Template-Controlled CaCO<sub>3</sub> Formation Revealed by Cryo-TEM},
journal = {Science},
volume = {323},
number = {5920},
pages = {1455-1458},
year = {2009},
doi = {10.1126/science.1169434},
URL = {https://www.science.org/doi/abs/10.1126/science.1169434},
eprint = {https://www.science.org/doi/pdf/10.1126/science.1169434}}

@article{PNC_MD_yes,
author = {Demichelis, Raffaella and Raiteri, Paolo and Gale, Julian and Quigley, David and Gebauer, Denis},
year = {2011},
month = {12},
pages = {590},
title = {Stable prenucleation mineral clusters are liquid-like ionic polymers},
volume = {2},
journal = {Nature communications},
doi = {10.1038/ncomms1604}
}

@article{
DLP_intro_PNC_MD_no,
author = {Paul J. M. Smeets  and Aaron R. Finney  and Wouter J. E. M. Habraken  and Fabio Nudelman  and Heiner Friedrich  and Jozua Laven  and James J. De Yoreo  and P. Mark Rodger  and Nico A. J. M. Sommerdijk },
title = {A classical view on nonclassical nucleation},
journal = {Proceedings of the National Academy of Sciences},
volume = {114},
number = {38},
pages = {E7882-E7890},
year = {2017},
doi = {10.1073/pnas.1700342114},
URL = {https://www.pnas.org/doi/abs/10.1073/pnas.1700342114},
eprint = {https://www.pnas.org/doi/pdf/10.1073/pnas.1700342114}}

@article{
PNC_no_1,
author = {Katja Henzler  and Evgenii O. Fetisov  and Mirza Galib  and Marcel D. Baer  and Benjamin A. Legg  and Camelia Borca  and Jacinta M. Xto  and Sonia Pin  and John L. Fulton  and Gregory K. Schenter  and Niranjan Govind  and J. Ilja Siepmann  and Christopher J. Mundy  and Thomas Huthwelker  and James J. De Yoreo },
title = {Supersaturated calcium carbonate solutions are classical},
journal = {Science Advances},
volume = {4},
number = {1},
pages = {eaao6283},
year = {2018},
doi = {10.1126/sciadv.aao6283},
URL = {https://www.science.org/doi/abs/10.1126/sciadv.aao6283},
eprint = {https://www.science.org/doi/pdf/10.1126/sciadv.aao6283}}

@article{PNC_no_2,
author = {Darkins, Robert and Duffy, Dorothy M. and Ford, Ian J.},
title = {Prenucleation Cluster Pathway is Inconsistent with CaCO3 Kinetics},
journal = {Crystal Growth \& Design},
volume = {24},
number = {10},
pages = {4013-4016},
year = {2024},
doi = {10.1021/acs.cgd.4c00092},

URL = { 
    
        https://doi.org/10.1021/acs.cgd.4c00092
    
    

},
eprint = { 
    
        https://doi.org/10.1021/acs.cgd.4c00092
}
}

@article{PNC_exp_new,
author = {Avaro, Jonathan and Moon, Ellen M. and Schulz, Kai G. and Rose, Andrew L.},
title = {Calcium Carbonate Prenucleation Cluster Pathway Observed via In Situ Small-Angle X-ray Scattering},
journal = {The Journal of Physical Chemistry Letters},
volume = {14},
number = {19},
pages = {4517-4523},
year = {2023},
doi = {10.1021/acs.jpclett.2c03192},
    note ={PMID: 37159248},

URL = { 
    
        https://doi.org/10.1021/acs.jpclett.2c03192
    
    

},
eprint = { 
    
        https://doi.org/10.1021/acs.jpclett.2c03192
    
    

}

}

@article{LLPS_1,
author = {Wolf, Stephan E. and Leiterer, Jork and Kappl, Michael and Emmerling, Franziska and Tremel, Wolfgang},
title = {Early Homogenous Amorphous Precursor Stages of Calcium Carbonate and Subsequent Crystal Growth in Levitated Droplets},
journal = {Journal of the American Chemical Society},
volume = {130},
number = {37},
pages = {12342-12347},
year = {2008},
doi = {10.1021/ja800984y},
    note ={PMID: 18717561},

URL = { 
    
        https://doi.org/10.1021/ja800984y
    
    

},
eprint = { 
    
        https://doi.org/10.1021/ja800984y
    
    

}

}

@article{LLPS_2019,
author = {Mohammed, Ahmed S. A. and Carino, Agnese and Testino, Andrea and Andalibi, Mohammad Reza and Cervellino, Antonio},
title = {In Situ Liquid SAXS Studies on the Early Stage of Calcium Carbonate Formation},
journal = {Particle \& Particle Systems Characterization},
volume = {36},
number = {6},
pages = {1800482},
doi = {https://doi.org/10.1002/ppsc.201800482},
url = {https://onlinelibrary.wiley.com/doi/abs/10.1002/ppsc.201800482},
eprint = {https://onlinelibrary.wiley.com/doi/pdf/10.1002/ppsc.201800482},
year = {2019}
}

@Article{PNC_to_ACC_LLPS,
AUTHOR = {Wang, Bing-Bing and Xiao, Yi and Xu, Zhi-Ming},
TITLE = {Variation in Properties of Pre-Nucleation Calcium Carbonate Clusters Induced by Aggregation: A Molecular Dynamics Study},
JOURNAL = {Crystals},
VOLUME = {11},
YEAR = {2021},
NUMBER = {2},
ARTICLE-NUMBER = {102},
URL = {https://www.mdpi.com/2073-4352/11/2/102},
ISSN = {2073-4352},
pages = {102},
DOI = {10.3390/cryst11020102}
}

@article{DLP_into_ACC_withMg,
author = {Seknazi, Eva and Kozachkevich, Stas and Polishchuk, Iryna and Stein, Nuphar and Villanova, Julie and Suuronen, Jussi-Petteri and Dejoie, Catherine and Zaslansky, Paul and Katsman, Alexander and Pokroy, B.},
year = {2019},
month = {10},
pages          = {4559},
article-number = {4559},
title = {From spinodal decomposition to alternating layered structure within single crystals of biogenic magnesium calcite},
volume = {10},
journal = {Nature Communications},
doi = {10.1038/s41467-019-12168-8}
}

@article{DLP_to_ACC_new,
  author       = {Jin, Biao and Chen, Ying and Pyles, Harley and Baer, Marcel D. and Legg, Benjamin A. and Wang, Zheming and Washton, Nancy M. and Mueller, Karl T. and Baker, David and Schenter, Gregory K. and others},
  title        = {Formation, chemical evolution and solidification of the dense liquid phase of calcium (bi)carbonate},
  doi          = {10.1038/s41563-024-02025-5},
  url          = {https://www.nature.com/articles/s41563-024-02025-5},
  journal      = {Nature Materials},
  issn         = {ISSN 1476-1122},
  number       = {1},
  volume       = {24},
  place        = {United States},
  publisher    = {Springer Nature},
  year         = {2025},
  pages        = {125–132},
  month        = {10}}

@article{ACC_hyd_SIM,
author = {Bushuev, Yuriy G. and Finney, Aaron R. and Rodger, P. Mark},
title = {Stability and Structure of Hydrated Amorphous Calcium Carbonate},
journal = {Crystal Growth \& Design},
volume = {15},
number = {11},
pages = {5269-5279},
year = {2015},
doi = {10.1021/acs.cgd.5b00771},

URL = { 
    
        https://doi.org/10.1021/acs.cgd.5b00771
},
eprint = { 
    
        https://doi.org/10.1021/acs.cgd.5b00771
}
}

@article{geom_frustr_Ca_Ca,
author = {Nicholas, Thomas and Stones, Adam and Patel, Adam and Michel, F. and Reeder, Richard and Aarts, Dirk and Deringer, Volker and Goodwin, Andrew},
year = {2023},
month = {09},
pages = {1-6},
title = {Geometrically frustrated interactions drive structural complexity in amorphous calcium carbonate},
volume = {16},
journal = {Nature Chemistry},
doi = {10.1038/s41557-023-01339-2}
}

@article{ACC_forms_review,
author = {Cartwright, Julyan H. E. and Checa, Antonio G. and Gale, Julian D. and Gebauer, Denis and Sainz-Díaz, C. Ignacio},
title = {Calcium Carbonate Polyamorphism and Its Role in Biomineralization: How Many Amorphous Calcium Carbonates Are There?},
journal = {Angewandte Chemie International Edition},
volume = {51},
number = {48},
pages = {11960-11970},
doi = {https://doi.org/10.1002/anie.201203125},
url = {https://onlinelibrary.wiley.com/doi/abs/10.1002/anie.201203125},
eprint = {https://onlinelibrary.wiley.com/doi/pdf/10.1002/anie.201203125},
year = {2012}
}

@article{ACC_dehydration_sim,
author = {Saharay, Moumita and Yazaydin, A. Ozgur and Kirkpatrick, R. James},
title = {Dehydration-Induced Amorphous Phases of Calcium Carbonate},
journal = {The Journal of Physical Chemistry B},
volume = {117},
number = {12},
pages = {3328-3336},
year = {2013},
doi = {10.1021/jp308353t},
    note ={PMID: 23442032},

URL = { 
    
        https://doi.org/10.1021/jp308353t
    
    

},
eprint = { 
    
        https://doi.org/10.1021/jp308353t
    
    

}

}

@article{dissol_recryst,
author = {Ihli, Johannes and Wong, Wai Ching Veronica and Noel, Elizabeth and Kim, Yi-Yeoun and Kulak, Alexander and Christenson, Hugo and Duer, Melinda and Meldrum, Fiona},
year = {2014},
month = {01},
pages = {3169},
title = {Dehydration and crystallization of amorphous calcium carbonate in solution and air},
volume = {5},
journal = {Nature communications},
doi = {10.1038/ncomms4169}
}

@article{dissol_recr_part_time,
author = {Du, Huachuan and Steinacher, Mathias and Borca, Camelia and Huthwelker, Thomas and Murello, Anna and Stellacci, Francesco and Amstad, Esther},
title = {Amorphous CaCO3: Influence of the Formation Time on Its Degree of Hydration and Stability},
journal = {Journal of the American Chemical Society},
volume = {140},
number = {43},
pages = {14289-14299},
year = {2018},
doi = {10.1021/jacs.8b08298},
    note ={PMID: 30265527},

URL = { 
    
        https://doi.org/10.1021/jacs.8b08298
    
    

},
eprint = { 
    
        https://doi.org/10.1021/jacs.8b08298
    
    

}

}

@article{dissol_recr_part_size,
author = {Zou, Zhaoyong and Bertinetti, Luca and Politi, Yael and Jensen, Anders C. S. and Weiner, Steve and Addadi, Lia and Fratzl, Peter and Habraken, Wouter J. E. M.},
title = {Opposite Particle Size Effect on Amorphous Calcium Carbonate Crystallization in Water and during Heating in Air},
journal = {Chemistry of Materials},
volume = {27},
number = {12},
pages = {4237-4246},
year = {2015},
doi = {10.1021/acs.chemmater.5b00145},
URL = { 
    
        https://doi.org/10.1021/acs.chemmater.5b00145
    
    

},
eprint = { 
    
        https://doi.org/10.1021/acs.chemmater.5b00145
    
    

}

}

@article{interface_ACC_MD,
title = {The transformation of amorphous calcium carbonate to calcite and classical nucleation theory},
journal = {Journal of Crystal Growth},
volume = {603},
pages = {126978},
year = {2023},
issn = {0022-0248},
doi = {https://doi.org/10.1016/j.jcrysgro.2022.126978},
author = {C.L. Freeman and J.H. Harding},
URL = {https://www.sciencedirect.com/science/article/pii/S0022024822004602}
}

@article{
ACC_shells,
author = {Tilman A. Grünewald  and Stefano Checchia  and Hamadou Dicko  and Gilles Le Moullac  and Manaarii Sham Koua  and Jeremie Vidal-Dupiol  and Julien Duboisset  and Julius Nouet  and Olivier Grauby  and Marco Di Michiel  and Virginie Chamard },
title = {Structure of an amorphous calcium carbonate phase involved in the formation of <i>Pinctada margaritifera</i> shells},
journal = {Proceedings of the National Academy of Sciences},
volume = {119},
number = {45},
pages = {e2212616119},
year = {2022},
doi = {10.1073/pnas.2212616119},
URL = {https://www.pnas.org/doi/abs/10.1073/pnas.2212616119},
eprint = {https://www.pnas.org/doi/pdf/10.1073/pnas.2212616119}}

@article{ACC_hydr_2010,
author = {Goodwin, Andrew L. and Michel, F. Marc and Phillips, Brian L. and Keen, David A. and Dove, Martin T. and Reeder, Richard J.},
title = {Nanoporous Structure and Medium-Range Order in Synthetic Amorphous Calcium Carbonate},
journal = {Chemistry of Materials},
volume = {22},
number = {10},
pages = {3197-3205},
year = {2010},
doi = {10.1021/cm100294d},

URL = { 
    
        https://doi.org/10.1021/cm100294d
    
    

},
eprint = { 
    
        https://doi.org/10.1021/cm100294d
    
    

}

}

@article{nanodomains_ACC,
author = {Clark, Simon and Colas, Bruno and Jacob, Dorrit and Neuefeind, Joerg and Wang, Hsiu-Wen and Page, Katherine and Soper, Alan and Schodder, Philipp and Duchstein, Patrick and Apeleo Zubiri, Benjamin and Yokosawa, Tadahiro and Pipich, Vitaliy and Zahn, Dirk and Spiecker, Erdmann and Wolf, Stephan},
year = {2022},
month = {04},
pages = {6870},
article-number = {6870},
title = {The nano‑ and meso‑scale structure of amorphous calcium carbonate},
volume = {12},
journal = {Scientific Reports},
doi = {10.1038/s41598-022-10627-9}
}

@article{ACC_bio_stable,
author = {Addadi, L. and Raz, S. and Weiner, S.},
title = {Taking Advantage of Disorder: Amorphous Calcium Carbonate and Its Roles in Biomineralization},
journal = {Advanced Materials},
volume = {15},
number = {12},
pages = {959-970},
doi = {https://doi.org/10.1002/adma.200300381},
url = {https://advanced.onlinelibrary.wiley.com/doi/abs/10.1002/adma.200300381},
eprint = {https://advanced.onlinelibrary.wiley.com/doi/pdf/10.1002/adma.200300381},
year = {2003}
}

@article{carbon_cycle,
author = {Lerman, Abraham and Mackenzie, F.},
year = {2006},
month = {11},
pages = {345-390},
title = {CO2 Air–Sea Exchange due to Calcium Carbonate and Organic Matter Storage, and its Implications for the Global Carbon Cycle},
volume = {11},
journal = {Aquatic Geochemistry},
doi = {10.1007/s10498-005-8620-x}
}

@article{carbon_cycle_ocean,
   author = "DeVries, Tim",
   title = "The Ocean Carbon Cycle", 
   journal= "Annual Review of Environment and Resources",
   year = "2022",
   volume = "47",
   number = "Volume 47, 2022",
   pages = "317-341",
   doi = "https://doi.org/10.1146/annurev-environ-120920-111307",
   url = "https://www.annualreviews.org/content/journals/10.1146/annurev-environ-120920-111307",
   publisher = "Annual Reviews",
   issn = "1545-2050",
   type = "Journal Article",
  }

@Article{soil_sequestration,
AUTHOR = {Batool, Maria and Cihacek, Larry J. and Alghamdi, Rashad S.},
TITLE = {Soil Inorganic Carbon Formation and the Sequestration of Secondary Carbonates in Global Carbon Pools: A Review},
JOURNAL = {Soil Systems},
VOLUME = {8},
YEAR = {2024},
NUMBER = {1},
ARTICLE-NUMBER = {15},
URL = {https://www.mdpi.com/2571-8789/8/1/15},
ISSN = {2571-8789},
pages = {15},
DOI = {10.3390/soilsystems8010015}
}

@article{carbon_cycle_climate,
author = {Isson, T. T. and Planavsky, N. J. and Coogan, L. A. and Stewart, E. M. and Ague, J. J. and Bolton, E. W. and Zhang, S. and McKenzie, N. R. and Kump, L. R.},
title = {Evolution of the Global Carbon Cycle and Climate Regulation on Earth},
journal = {Global Biogeochemical Cycles},
volume = {34},
number = {2},
pages = {e2018GB006061},
doi = {https://doi.org/10.1029/2018GB006061},
url = {https://agupubs.onlinelibrary.wiley.com/doi/abs/10.1029/2018GB006061},
eprint = {https://agupubs.onlinelibrary.wiley.com/doi/pdf/10.1029/2018GB006061},
note = {e2018GB006061 2018GB006061},
year = {2020}
}

@article{bacteria_biomin,
author = {Görgen, Sigrid and Benzerara, Karim and Skouri-Panet, Fériel and Gugger, Muriel and Chauvat, Franck and Cassier-Chauvat, Corinne},
year = {2020},
month = {11},
pages = {2},
title = {The diversity of molecular mechanisms of carbonate biomineralization by bacteria},
volume = {1},
journal = {Discover Materials},
doi = {10.1007/s43939-020-00001-9}
}

@article{biomin_paths,
author = {Weiner, Steve and Addadi, Lia},
year = {2011},
month = {07},
pages = {21-40},
title = {Crystallization Pathways in Biomineralization},
volume = {41},
journal = {Annual Review of Materials Research},
doi = {10.1146/annurev-matsci-062910-095803}
}

@article{
biomin_evolution,
author = {Pupa U. P. A. Gilbert  and Kristin D. Bergmann  and Nicholas Boekelheide  and Sylvie Tambutté  and Tali Mass  and Frédéric Marin  and Jess F. Adkins  and Jonathan Erez  and Benjamin Gilbert  and Vanessa Knutson  and Marjorie Cantine  and Javier Ortega Hernández  and Andrew H. Knoll },
title = {Biomineralization: Integrating mechanism and evolutionary history},
journal = {Science Advances},
volume = {8},
number = {10},
pages = {eabl9653},
year = {2022},
doi = {10.1126/sciadv.abl9653},
URL = {https://www.science.org/doi/abs/10.1126/sciadv.abl9653},
eprint = {https://www.science.org/doi/pdf/10.1126/sciadv.abl9653}}

@article{biomin_coccoliths,
url = {https://doi.org/10.1515/9781501509346-012},
title = {Biomineralization Within Vesicles: The Calcite of Coccoliths},
journal = {Reviews in Mineralogy and Geochemistry},
booktitle = {Biomineralization},
author = {Jeremy R. Young and Karen Henriksen},
editor = {Patricia M. Dove and James J. De Yoreo and Steve Weiner},
publisher = {De Gruyter},
address = {Berlin, Boston},
volume = {54},
number = {1},
pages = {189--215},
doi = {10.2113/0540189},
isbn = {9781501509346},
year = {2003},
lastchecked = {2026-04-21}
}

@article{appl_comb,
author = {Yao, Shasha and Jin, Biao and Liu, Zhaoming and Shao, Changyu and Zhao, Ruibo and Wang, Xiaoyu and Tang, Ruikang},
title = {Biomineralization: From Material Tactics to Biological Strategy},
journal = {Advanced Materials},
volume = {29},
number = {14},
pages = {1605903},
doi = {https://doi.org/10.1002/adma.201605903},
url = {https://advanced.onlinelibrary.wiley.com/doi/abs/10.1002/adma.201605903},
eprint = {https://advanced.onlinelibrary.wiley.com/doi/pdf/10.1002/adma.201605903},
year = {2017}
}

@Article{appl_biomed,
AUTHOR = {Chen, Yuanyuan and Feng, Yanmin and Deveaux, John Gregory and Masoud, Mohamed Ahmed and Chandra, Felix Sunata and Chen, Huawei and Zhang, Deyuan and Feng, Lin},
TITLE = {Biomineralization Forming Process and Bio-inspired Nanomaterials for Biomedical Application: A Review},
JOURNAL = {Minerals},
VOLUME = {9},
YEAR = {2019},
NUMBER = {2},
ARTICLE-NUMBER = {68},
URL = {https://www.mdpi.com/2075-163X/9/2/68},
ISSN = {2075-163X},
pages = {68},
DOI = {10.3390/min9020068}
}

@Article{biomed_nanoparticles,
AUTHOR = {Zhao, Pengxuan and Tian, Yu and You, Jia and Hu, Xin and Liu, Yani},
TITLE = {Recent Advances of Calcium Carbonate Nanoparticles for Biomedical Applications},
JOURNAL = {Bioengineering},
VOLUME = {9},
YEAR = {2022},
NUMBER = {11},
ARTICLE-NUMBER = {691},
URL = {https://www.mdpi.com/2306-5354/9/11/691},
PubMedID = {36421092},
ISSN = {2306-5354},
pages = {691},
DOI = {10.3390/bioengineering9110691}
}

@article{biomed_nanoparticles_new,
author = {Popova, Victoriya K. and Pyshnyi, Dmitrii V. and Dmitrienko, Elena V.},
title = {Biomedical Applications of Calcium Carbonate Nanoparticles: A Review of Recent Advances},
journal = {ACS Biomaterials Science \& Engineering},
volume = {11},
number = {11},
pages = {6359-6385},
year = {2025},
doi = {10.1021/acsbiomaterials.5c00992},
    note ={PMID: 41056498},

URL = { 
    
        https://doi.org/10.1021/acsbiomaterials.5c00992
    
    

},
eprint = { 
    
        https://doi.org/10.1021/acsbiomaterials.5c00992
    
    

}

}

@article{CAV,
author = {F. C. Meldrum},
title = {Calcium carbonate in biomineralisation and biomimetic chemistry},
journal = {International Materials Reviews},
volume = {48},
number = {3},
pages = {187--224},
year = {2003},
publisher = {Taylor \& Francis},
doi = {10.1179/095066003225005836},


URL = { 
    
        https://doi.org/10.1179/095066003225005836
    
    

},
eprint = { 
    
        https://doi.org/10.1179/095066003225005836
    
    

}

}

@article{caco3_pd,
title = {CaCO3 phase diagram studied with Raman spectroscopy at pressures up to 50 GPa and high temperatures and DFT modeling},
journal = {Physics of the Earth and Planetary Interiors},
volume = {281},
pages = {31-45},
year = {2018},
issn = {0031-9201},
doi = {https://doi.org/10.1016/j.pepi.2018.05.002},
url = {https://www.sciencedirect.com/science/article/pii/S0031920117303278},
author = {L. Bayarjargal and C.-J. Fruhner and N. Schrodt and B. Winkler}
}

@article{vaterite_str,
author = {Christy, Andrew G.},
title = {A Review of the Structures of Vaterite: The Impossible, the Possible, and the Likely},
journal = {Crystal Growth \& Design},
volume = {17},
number = {6},
pages = {3567-3578},
year = {2017},
doi = {10.1021/acs.cgd.7b00481},

URL = { 
    
        https://doi.org/10.1021/acs.cgd.7b00481
    
    

},
eprint = { 
    
        https://doi.org/10.1021/acs.cgd.7b00481
    
    

}

}

@article{CNT_NCNT_review,
author = {Karthika, S. and Radhakrishnan, T. K. and Kalaichelvi, P.},
title = {A Review of Classical and Nonclassical Nucleation Theories},
journal = {Crystal Growth \& Design},
volume = {16},
number = {11},
pages = {6663-6681},
year = {2016},
doi = {10.1021/acs.cgd.6b00794},

URL = { 
    
        https://doi.org/10.1021/acs.cgd.6b00794
    
    

},
eprint = { 
    
        https://doi.org/10.1021/acs.cgd.6b00794
    
    

}

}

@article{CNT_NCNT_both_review,
   author = "Jun, Young-Shin and Zhu, Yaguang and Wang, Ying and Ghim, Deoukchen and Wu, Xuanhao and Kim, Doyoon and Jung, Haesung",
   title = "Classical and Nonclassical Nucleation and Growth Mechanisms for Nanoparticle Formation", 
   journal= "Annual Review of Physical Chemistry",
   year = "2022",
   volume = "73",
   number = "Volume 73, 2022",
   pages = "453-477",
   doi = "https://doi.org/10.1146/annurev-physchem-082720-100947",
   url = "https://www.annualreviews.org/content/journals/10.1146/annurev-physchem-082720-100947",
   publisher = "Annual Reviews",
   issn = "1545-1593",
   type = "Journal Article",
  }

@article{
paths_coexistence,
author = {Michael H. Nielsen  and Shaul Aloni  and James J. De Yoreo },
title = {In situ TEM imaging of CaCO<sub>3</sub> nucleation reveals coexistence of direct and indirect pathways},
journal = {Science},
volume = {345},
number = {6201},
pages = {1158-1162},
year = {2014},
doi = {10.1126/science.1254051},
URL = {https://www.science.org/doi/abs/10.1126/science.1254051},
eprint = {https://www.science.org/doi/pdf/10.1126/science.1254051}}

@article{PNC_NCNT,
title = {Prenucleation clusters and non-classical nucleation},
journal = {Nano Today},
volume = {6},
number = {6},
pages = {564-584},
year = {2011},
issn = {1748-0132},
doi = {https://doi.org/10.1016/j.nantod.2011.10.005},
url = {https://www.sciencedirect.com/science/article/pii/S1748013211001174},
author = {Denis Gebauer and Helmut Cölfen}
}

@Article{ACC_to_calcite_via_vaterite,
author ="Rodriguez-Blanco, Juan Diego and Shaw, Samuel and Benning, Liane G.",
title  ="The kinetics and mechanisms of amorphous calcium carbonate (ACC) crystallization to calcite{,} via vaterite.",
journal  ="Nanoscale",
year  ="2011",
volume  ="3",
issue  ="1",
pages  ="265-271",
publisher  ="The Royal Society of Chemistry",
doi  ="10.1039/C0NR00589D",
url  ="http://dx.doi.org/10.1039/C0NR00589D"}

@article{dissol_repr_new,
    author = {Chen, Si Athena and Weber, Juliane and Starchenko, Vitalii and Eng, Peter J. and Stubbs, Joanne E. and Wang, Hsiu-Wen and Liu, Tingting and Spano, Tyler L. and Chakoumakos, Bryan C. and Stack, Andrew G.},
    title = {Real-Time Atomic-Scale
Structural Analysis Resolves
the Amorphous to Crystalline CaCO3 Mechanism Controversy},
    journal = {Crystal Growth \& Design},
    volume = {24},
    number = {12},
    pages = {5027-5038},
    year = {2024},
    month = {05},
    issn = {1528-7483},
    doi = {10.1021/acs.cgd.4c00245},
    url = {https://doi.org/10.1021/acs.cgd.4c00245},
    eprint = {https://pubs.acs.org/cgdefu/article-pdf/24/12/5027/1475575/cg4c00245.pdf},
}

@Inbook{methods_review,
author="Tiwary, Pratyush
and van de Walle, Axel",
editor="Weinberger, Christopher R.
and Tucker, Garritt J.",
title="A Review of Enhanced Sampling Approaches for Accelerated Molecular Dynamics",
bookTitle="Multiscale Materials Modeling for Nanomechanics",
year="2016",
publisher="Springer International Publishing",
address="Cham",
pages="195--221",
isbn="978-3-319-33480-6",
doi="10.1007/978-3-319-33480-6_6",
url="https://doi.org/10.1007/978-3-319-33480-6_6"
}

@article{metad,
author = {Alessandro Laio  and Michele Parrinello },
title = {Escaping free-energy minima},
journal = {Proceedings of the National Academy of Sciences},
volume = {99},
number = {20},
pages = {12562-12566},
year = {2002},
doi = {10.1073/pnas.202427399},
URL = {https://www.pnas.org/doi/abs/10.1073/pnas.202427399},
eprint = {https://www.pnas.org/doi/pdf/10.1073/pnas.202427399}}

@ARTICLE{scipy,
  author  = {Virtanen, Pauli and Gommers, Ralf and Oliphant, Travis E. and
            Haberland, Matt and Reddy, Tyler and Cournapeau, David and
            Burovski, Evgeni and Peterson, Pearu and Weckesser, Warren and
            Bright, Jonathan and {van der Walt}, St{\'e}fan J. and
            Brett, Matthew and Wilson, Joshua and Millman, K. Jarrod and
            Mayorov, Nikolay and Nelson, Andrew R. J. and Jones, Eric and
            Kern, Robert and Larson, Eric and Carey, C J and
            Polat, {\.I}lhan and Feng, Yu and Moore, Eric W. and
            {VanderPlas}, Jake and Laxalde, Denis and Perktold, Josef and
            Cimrman, Robert and Henriksen, Ian and Quintero, E. A. and
            Harris, Charles R. and Archibald, Anne M. and
            Ribeiro, Ant{\^o}nio H. and Pedregosa, Fabian and
            {van Mulbregt}, Paul and {SciPy 1.0 Contributors}},
  title   = {{{SciPy} 1.0: Fundamental Algorithms for Scientific
            Computing in Python}},
  journal = {Nature Methods},
  year    = {2020},
  volume  = {17},
  pages   = {261--272},
  adsurl  = {https://rdcu.be/b08Wh},
  doi     = {10.1038/s41592-019-0686-2},
}

@article{LLPS_ACC_Faatz,
author = {Faatz, M. and Gröhn, F. and Wegner, G.},
title = {Amorphous Calcium Carbonate: Synthesis and Potential Intermediate in Biomineralization},
journal = {Advanced Materials},
volume = {16},
number = {12},
pages = {996-1000},
doi = {https://doi.org/10.1002/adma.200306565},
url = {https://advanced.onlinelibrary.wiley.com/doi/abs/10.1002/adma.200306565},
eprint = {https://advanced.onlinelibrary.wiley.com/doi/pdf/10.1002/adma.200306565},
year = {2004}
}

@article{proto_aragonite,
author = {Farhadi-Khouzani, Masoud and Chevrier, Daniel M. and Zhang, Peng and Hedin, Niklas and Gebauer, Denis},
title = {Water as the Key to Proto-Aragonite Amorphous CaCO3},
journal = {Angewandte Chemie International Edition},
volume = {55},
number = {28},
pages = {8117-8120},
doi = {https://doi.org/10.1002/anie.201603176},
url = {https://onlinelibrary.wiley.com/doi/abs/10.1002/anie.201603176},
eprint = {https://onlinelibrary.wiley.com/doi/pdf/10.1002/anie.201603176},
year = {2016}
}

@article{ACC_stab_add,
title = {Stabilization and crystallization mechanism of amorphous calcium carbonate},
journal = {Journal of Colloid and Interface Science},
volume = {680},
pages = {24-35},
year = {2025},
issn = {0021-9797},
doi = {https://doi.org/10.1016/j.jcis.2024.11.076},
url = {https://www.sciencedirect.com/science/article/pii/S0021979724026523},
author = {Qihang Wang and Wenyang Huang and Jilin Wang and Fei Long and Zhengyi Fu and Jingjing Xie and Zhaoyong Zou}
}

@article{polymorph_T_phase_diag,
    author = {Ma, Ming and Wang, Yihu and Cao, Xiaofeng and Lu, Weipeng and Guo, Yanchuan},
    title = {Temperature and Supersaturation as Key Parameters
Controlling the Spontaneous Precipitation of Calcium Carbonate with
Distinct Physicochemical Properties from Pure Aqueous Solutions},
    journal = {Crystal Growth \& Design},
    volume = {19},
    number = {12},
    pages = {6972-6988},
    year = {2019},
    month = {10},
    issn = {1528-7483},
    doi = {10.1021/acs.cgd.9b00758},
    url = {https://doi.org/10.1021/acs.cgd.9b00758},
    eprint = {https://pubs.acs.org/cgdefu/article-pdf/19/12/6972/14437459/cg9b00758.pdf},
}

@article{ACC_stab_add_ions,
    author = {Jung, Gwan Yeong and Shin, Eunhye and Park, Ju Hyun and Choi, Byoung-Young and Lee, Seung-Woo and Kwak, Sang Kyu},
    title = {Thermodynamic Control of Amorphous Precursor Phases
for Calcium Carbonate via Additive Ions},
    journal = {Chemistry of Materials},
    volume = {31},
    number = {18},
    pages = {7547-7557},
    year = {2019},
    month = {08},
    issn = {0897-4756},
    doi = {10.1021/acs.chemmater.9b02346},
    url = {https://doi.org/10.1021/acs.chemmater.9b02346},
    eprint = {https://pubs.acs.org/cmatex/article-pdf/31/18/7547/30970999/cm9b02346.pdf},
}

@article{CFE,
    author = {van Erp, Titus S. and Zhang, Daniel T. and Wils, Elias and Safaei, Sina and Ghysels, An},
    title = {Generalized path reweighting and history-dependent free energies},
    journal = {The Journal of Chemical Physics},
    volume = {164},
    number = {21},
    pages = {214107},
    year = {2026},
    month = {06},
    issn = {0021-9606},
    doi = {10.1063/5.0326023},
    url = {https://doi.org/10.1063/5.0326023},
    eprint = {https://pubs.aip.org/aip/jcp/article-pdf/doi/10.1063/5.0326023/21021628/214107_1_5.0326023.pdf},
}

@article{10.1063/1.4989844,
    author = {Cabriolu, Raffaela and Skjelbred Refsnes, Kristin M. and Bolhuis, Peter G. and van Erp, Titus S.},
    title = {Foundations and latest advances in replica exchange transition interface sampling},
    journal = {The Journal of Chemical Physics},
    volume = {147},
    number = {15},
    pages = {152722},
    year = {2017},
    month = {10},
    issn = {0021-9606},
    doi = {10.1063/1.4989844},
    url = {https://doi.org/10.1063/1.4989844},
    eprint = {https://pubs.aip.org/aip/jcp/article-pdf/doi/10.1063/1.4989844/15536514/152722_1_online.pdf},
}

@article{str_trans,
    author = {Sidler, Elizaveta and Cabriolu, Raffaela},
    title = {Structural transitions of calcium carbonate by molecular dynamics simulation},
    journal = {The Journal of Chemical Physics},
    volume = {161},
    number = {21},
    pages = {214707},
    year = {2024},
    month = {12},
    issn = {0021-9606},
    doi = {10.1063/5.0233713},
    url = {https://doi.org/10.1063/5.0233713},
    eprint = {https://pubs.aip.org/aip/jcp/article-pdf/doi/10.1063/5.0233713/20280252/214707_1_5.0233713.pdf},
}

@article{Bayarjargal_phase_diagr_VII,
title = {CaCO3 phase diagram studied with Raman spectroscopy at pressures up to 50 GPa and high temperatures and DFT modeling},
journal = {Physics of the Earth and Planetary Interiors},
volume = {281},
pages = {31-45},
year = {2018},
issn = {0031-9201},
doi = {https://doi.org/10.1016/j.pepi.2018.05.002},
url = {https://www.sciencedirect.com/science/article/pii/S0031920117303278},
author = {L. Bayarjargal and C.-J. Fruhner and N. Schrodt and B. Winkler}
}
\bibliographystyle{rsc}

\clearpage
\onecolumn 

\begin{center}
  {\Large \bfseries Supplementary Information:\\[0.3em]
  Organisation and aggregation of calcium carbonate ions in water via molecular simulation \par}
  \vspace{1em}
  {\large Elizaveta Sidler, Rita S. Dias, and Raffaela Cabriolu \par}
  \vspace{2em}
\end{center}

\setcounter{section}{0}
\setcounter{figure}{0}
\setcounter{table}{0}
\setcounter{equation}{0}

\renewcommand{\thesection}{S\arabic{section}}
\renewcommand{\thefigure}{S\arabic{figure}}
\renewcommand{\thetable}{S\arabic{table}}
\renewcommand{\theequation}{S\arabic{equation}}

\renewcommand{\theHsection}{S\arabic{section}}
\renewcommand{\theHfigure}{S\arabic{figure}}
\renewcommand{\theHtable}{S\arabic{table}}
\renewcommand{\theHequation}{S\arabic{equation}}


\newcommand*{\addFileDependency}[1]{
\typeout{(#1)}
\@addtofilelist{#1}
\IfFileExists{#1}{}{\typeout{No file #1.}}
}\makeatother

\newcommand*{\myexternaldocument}[1]{
\externaldocument{#1}
\addFileDependency{#1.tex}
\addFileDependency{#1.aux}
}



\section{Methods}

\subsection{\texorpdfstring{Interfaces $\lambda_A$ and $\lambda_B$}{Interfaces lambda A and lambda B}}\label{sec:Interfaces}

\begin{figure}[h!]
    \centering
    \includegraphics{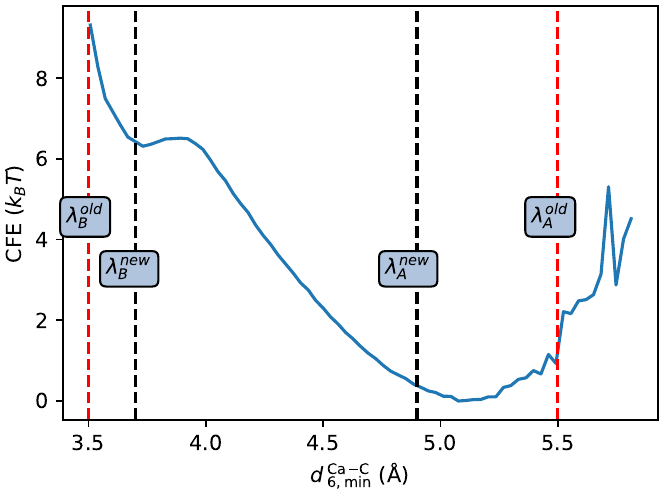}
    \caption{CFE as a function of the \cacmin for the system with \conc~= 0.01 and \nion~= 72. The dashed lines indicate the first (red, $\lambda^{\rm old}$) and final (black, $\lambda^{\rm new}$) positions of \la and \lb interfaces.}
    \label{fig:fes_OP}
\end{figure}

Interfaces \la and \lb were initially placed at 3.5 and 5.5 \AA{}, respectively (see $\lambda^{\rm old}$ in Figure~\ref{fig:fes_OP}). As described in the main article, this choice was based on the Ca--C radial distribution function and the free energy profile for the Ca--C distance in a two-ion system.

For the system with \conc~= 0.01 and \nion~= 72, an $\infty$init simulation (based on \infretis and provided by \texttt{$\infty$tools}) was first conducted to establish an initial positioning of the intermediate interfaces between \la and \lb. Subsequently, \infretis simulation was performed for 2000 steps. Based on this preliminary \infretis simulation, a conditional free energy (CFE) profile was evaluated as a function of the \cacmin order parameter, as presented in Figure~\ref{fig:fes_OP}. It revealed two minima at the \cacmin values of $\simeq 3.7$ and $\simeq 5.1$ \AA{}, both of which lying between the initial placement of \la and \lb (between $\lambda^{\rm old}$ in Figure~\ref{fig:fes_OP}). Consequently, such placement of \la and \lb can lead to \infretis trajectories becoming trapped in CFE minima instead of undergoing the reaction, increasing trajectory lengths and computational cost. To mitigate this issue, the interfaces \la and \lb were repositioned closer to the CFE minima, at 3.7 and 4.9 \AA{}, respectively (see $\lambda^{\rm new}$ in Figure~\ref{fig:fes_OP}). In this way, each interface lies between the corresponding CFE minimum and the CFE barrier maximum. The same \la and \lb values were used in all \infretis simulations across different system compositions and order parameters. After completion of the simulations, \la and \lb were consistently found to lie between the CFE barrier maximum and the corresponding CFE minima for all systems, thereby justifying their placement (see Section~\ref{sec:SI_ln_P}).

\subsection{\texorpdfstring{$\infty$RETIS protocol}{infRETIS protocol}}\label{sec:SI_protocol}

First, a trajectory connecting state A to state B was generated using the OPES method available in \texttt{PLUMED}\cite{PLUMED}. Along this trajectory, 10 interfaces were uniformly placed between \la and \lb. The \texttt{$\infty$init} algorithm was used to generate MD trajectories to obtain an initial estimate of the reaction rate and to update the intermediate interfaces between \la and \lb (both number and values).

Starting from the subset of trajectories produced by $\infty$init, we conducted \infretis simulations for 3000-5000 \infretis steps using the updated interfaces. The resulting trajectories provided an updated evaluation of \kab and of the intermediate interfaces. If the number or locations of these intermediate interfaces changed significantly compared to the previous set, we repeated the \infretis simulation for 3000-5000 steps up to two additional times. The intermediate interfaces were updated after each iteration. Each restart was initialized from the most recent trajectories of the preceding \infretis simulation, thereby providing additional equilibration of the system with respect to the order parameter (OP). Finally, we conducted a production \infretis simulation for 30000-34000 steps, terminating it once the relative errors of \kab, \pcross, and $\Phi_A$ fell below 0.5.

To assess and update the positions of the intermediate interfaces after the \infretis simulation, we used the \texttt{estimate\_interfaces*} routine provided by \texttt{$\infty$tools}. For this routine, we imposed the condition that the local crossing probability (\texttt{ploc}) for each intermediate interface is equal to 0.4. Finally, during the $\infty$RETIS production simulations, each system had 8 to 12 interfaces in total. The errors for \kab, \pcross, and $\Phi_A$ were calculated with a block-averaging procedure available in \texttt{$\infty$tools}. All MD simulations initiated by \infretis were carried out in the NVE ensemble. Since \infretis requires an OP that increases from \la to \lb, we used \cacmin and \ccamin multiplied by --1 as an OP in the actual \infretis calculations. 

\subsection{Cluster analysis}\label{sec:SI_cluster_analysis}

The analysis of ionic clusters composed of \caion and \cion ions was performed using Compressed Sparse Graph Routines implemented in \texttt{scipy} Python library \cite{scipy}. The clusters were identified with respect to the Ca and C atomic coordinates, with a cut-off distance of 4~\AA{}. This distance lies near the barrier maximum that separates the water-mediated and contact ion-pair configurations, with the exact position of the barrier maximum differing between systems. A cluster is then defined as a set of ions, each of which is `in contact' with at least one other ion.

\subsection{Radial distribution function (RDF)}\label{sec:SI_RDF}

The observed ionic aggregates consist of a dynamic and finite number of ions, and are non-uniform in both density and shape. Consequently, the standard expression for the RDF, which is based on the bulk density, cannot be applied. Instead, the RDFs between atom types $a$ and $b$, $g_{ab}(r)$, were calculated using the following expression:

\begin{equation}
g_{ab}(r)= \frac{1}{N_{a}}\frac{\delta N_{ab}(r)}{\delta V(r)}, \\
\label{eq:rdf}
\end{equation}

\noindent where $\delta N_{ab}(r)$ is the number of $a$-$b$ atom pairs located within a spherical shell volume $\delta V(r)$ at a distance $r$, and $N_{a}$ is the number of atoms of type $a$ in relation to which the distribution function is calculated.

\subsection{Coordination number (CN)}\label{sec:SI_CN_meth}

A cosine switching function, decaying between $r_1$ and $r_2$, was used to calculate the CN between different types of atoms. The values of $r_1$ and $r_2$ were determined from the positions of the minimum in the corresponding RDF (see Section~\ref{sec:SI_RDF}). Specifically, they were set 0.5 \AA{} on either side of the minimum, which defines the first coordination sphere. The selected values are reported in Table~\ref{tab:CN_r1_r2}. 

To calculate the CN for interactions between Ca/C and water (\ow), a larger difference between $r_1$ and $r_2$ was used to monitor the correlation between hydration and dehydration of the coordination sphere more smoothly. In addition. $r_2$ extends beyond the first coordination sphere, enabling contributions from more distant water molecules to be included.

\begin{table}[h!]
\caption{Cut-off distances used in the calculations of coordination number for each pair of atomic types.}
\centering
\begin{tabular}{|c c c|}
\hline
atom pair & $r_1$ (\AA{}) & $r_2$ (\AA{}) \\
\hline
Ca--C/C--Ca &  3.5  &  4.5    \\
Ca--Ca  &  4.5 &  5.5  \\
C--C  &  5.1  &  6.1   \\
Ca/C--\ow  &  3.5 &  5.5    \\
\hline
\end{tabular}
\label{tab:CN_r1_r2}
\end{table}

\subsection{Shape parameters $K_1$ and $K_2$}\label{sec:SI_K1_K2_meth}

To analyse the shapes of ionic clusters, we introduce the shape parameters $K_1$ and $K_2$. These parameters are derived from the eigenvalues $L_1, L_2, L_3$ of the geometric radius of gyration (without mass weighting), which are sorted in descending order, as follows:

\begin{align}
{K_1} = \frac{(L_2 + L_3)}{(L_1 + L_2)} \\
{K_2} = \frac{(L_1 + L_3)}{(L_1 + L_2)}
\label{eq:rdf}
\end{align}

By using $K_1$ and $K_2$, we can discern different morphologies of the cluster, as listed in Table~\ref{tab:K1_K2}.

\begin{table}[h!]
\caption{The correspondence between the values of $K_1$ and $K_2$, and the shape of the cluster.}
\centering
\begin{tabular}{|c c c|}
\hline
shape & $K_1$ & $K_2$ \\
\hline
sphere &  1  &  1    \\
rod  &  0 &  1  \\
disk  &  0.5  &  0.5   \\
\hline
\end{tabular}
\label{tab:K1_K2}
\end{table}

\newpage
\section{Results}\label{sec:SI_res}

\subsection{Crossing probability and barrier height}\label{sec:SI_ln_P}

According to the transition state theory, the crossing probability is proportional to the height of the free energy barrier as follows:

\begin{gather}
P(\lambda) \propto \exp{(-\frac{G(\lambda)}{k_{\rm B}T})}, \\
\frac{d\ln(P(\lambda))}{d\lambda}) \propto -\beta\frac{d(G(\lambda))}{d\lambda},
\label{eq:arrh}
\end{gather}

\noindent where $\lambda$ is an OP. Therefore, the maximum of the energy barrier ($\frac{d(G(\lambda))}{d\lambda} \approx 0$) should correspond to a point along $\ln(P_{\rm cross})$ where the slope approaches zero. 

\begin{figure}[h!]
    \centering
    \includegraphics{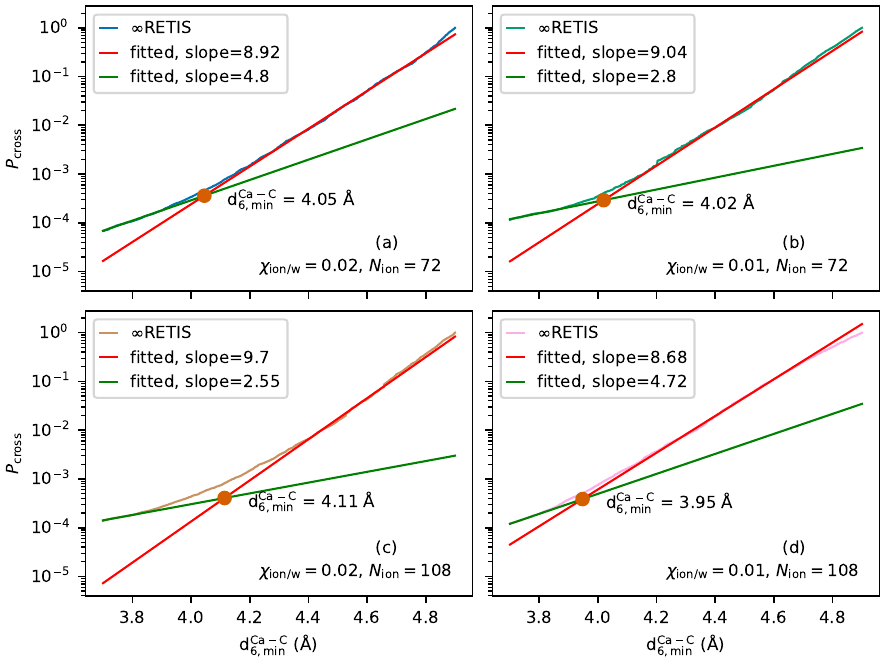}
    \includegraphics{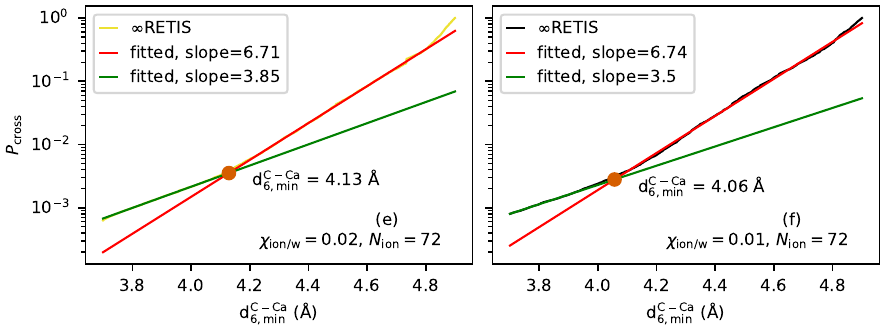}
    \caption{Dependence of \pcross on \cacmin ((a)--(d)) and \ccamin ((e)--(f)) OPs for the systems with \conc and \nion values indicated in the plots. The red and green lines correspond to the slopes of \pcross before and after the barrier, respectively. They were fitted to two regions where \pcross changes approximately linearly. Their intersection is marked by an orange dot, and the corresponding $d_{6, \rm min}$ values are annotated.} 
    \label{fig:P_cross}
\end{figure}

In our systems, all barriers are relatively low, allowing trajectories to recross after passing the maximum. As a result, the slope decreases significantly beyond the maximum, but does not approach zero. The change in the slope allowed us to identify the location of the energy maximum for each system, as illustrated in Figure~\ref{fig:P_cross}. These data indicate that the maximum of the energy barrier for the formation of the Ca--C/C--Ca first coordination sphere is located at $d_{6,\rm min}\simeq 4.05$~\AA\ on average for all considered systems. The location of the barrier maximum within the configurational space boarded by the chosen $\lambda_A$ and $\lambda_B$ confirms the validity of their placement.

\newpage
\clearpage

\subsection{Energy barriers with OPES}\label{sec:SI_FE_1d}

\begin{figure}[h!]
    \centering
    \includegraphics{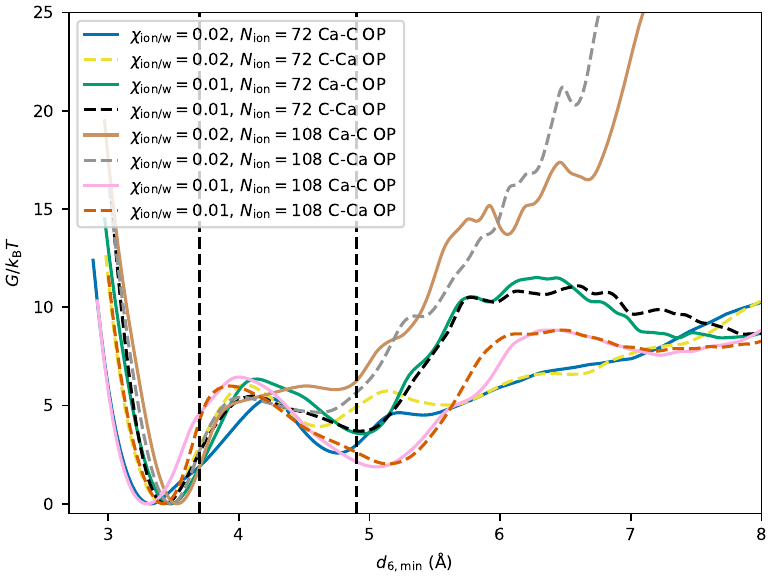}
    \caption{One-dimensional free energy profiles as functions of \cacmin (solid lines) or \ccamin (dashed lines) for systems with different compositions, as indicated in the legend.}
    \label{fig:SI_FE_1d}
\end{figure}

\newpage
\subsection{CFE surfaces projected onto \cacmin and \ccamin}\label{sec:SI_CFE_OPs}

\begin{figure}[h!]
    \centering
    \includegraphics{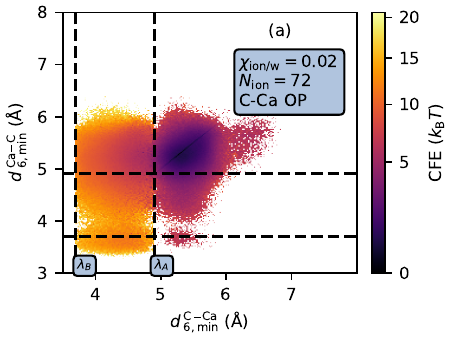}
    \includegraphics{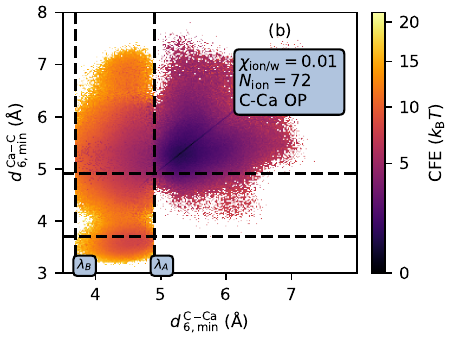}
    \includegraphics{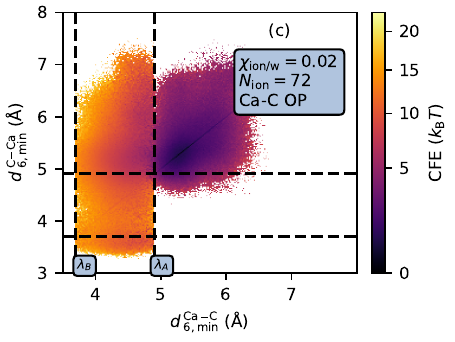}
    \includegraphics{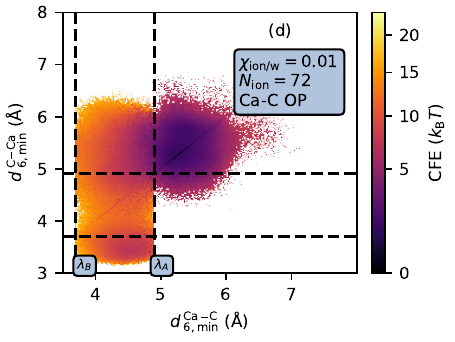}
    \includegraphics{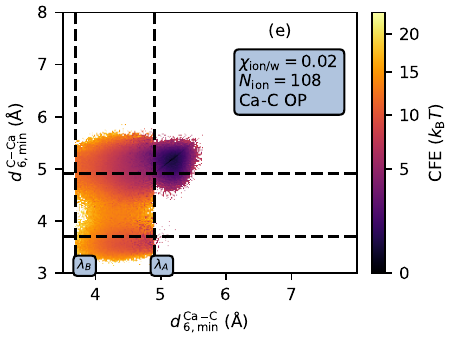}
    \includegraphics{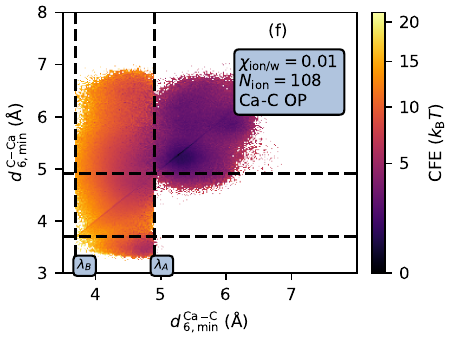}
    \caption{CFE surfaces projected onto \cacmin and \ccamin for the systems with \conc, \nion and OP indicated in the legends.}
    \label{fig:cfe_ops}
\end{figure}

All CFE landscapes in Figure~\ref{fig:cfe_ops} exhibit multiple minima, with one of them located at either \cacmin or \ccamin corresponding to state B (along the y-axis), whereas the \infretis simulations were carried out using the complementary OP (\ccamin or \cacmin along the x-axis, respectively). Therefore, the \infretis simulations inherently include pathways in which, for a reaction defined by one OP (\cacmin or \ccamin), the complementary coordinate (\ccamin or \cacmin, respectively) may also evolve toward or reach state B.

\newpage
\clearpage

\subsection{Cluster analysis}\label{sec:SI_CA}

\subsubsection{\infretis vs. OPES}\label{sec:SI_infretis_vs_OPES} ~\\

The energy surfaces computed with OPES and \infretis as functions of \nionrc (Figures~\ref{fig:6min_cl_size} and \ref{fig:6min_cac_RC_size_OPES} in \mainp) demonstrate a noticeable mismatch in the location of the minima with respect to \nionrc for the systems with \nion $=72$. One of the possible reasons for this discrepancy may lie in the limited suitability of \cacmin and \ccamin CVs for the accurate sampling of \nionrc with OPES, as these CVs do not explicitly track ionic clustering. 

As a biasing technique, OPES is inherently sensitive to the choice of CVs. Accurately reproducing FES as a function of cluster size -- whose evolution is slow at the low concentrations considered here -- would likely require additional CVs that explicitly depend on the CI-cluster size. In their current form, \cacmin and \ccamin do not capture orthogonal degrees of freedom associated with ionic aggregation, which limits their ability to describe this slow process. 

Another possible reason for the observed mismatch in the energy minima along \nionrc in state A is that, unlike \infretis, OPES does not explicitly account for dynamical and kinetic factors. The minimum at \nionrc $\approx 50$ observed in the OPES simulations but not highlighted in $\infty$RETIS for the system with \conc $=0.01$ and \nion $=72$, would require substantial ionic aggregation prior to the reaction. Such aggregation appears kinetically unlikely in a system this dilute, because the minima associated with state B correspond to smaller cluster sizes. This limitation in sampling different cluster sizes with the chosen CVs might be evident for this particular system the most, because it has the lowest effective concentration of free calcium carbonate ions (or units) outside the CI-cluster, which limits changes in cluster size.

This under-representation of aggregation within the chosen CVs may also contribute to the difference in between the apparent energy barriers obtained with OPES and \infretis (see Section~\ref{sec:results_barr_OPES} in the main paper).

Based on this discussion, we conclude that the free-energy valley near \nionrc~$\approx 50$ observed in the OPES simulations but not highlighted in \infretis for the system with \conc~= 0.01 and \nion~= 72, does not correspond to a primary reaction pathway, despite being energetically feasible. 

\newpage
\clearpage
\subsubsection{Flux}\label{sec:SI_flux}

\flux is determined from the average lengths of the MD trajectories within state A (see Section~\ref{sec:method_infretis} of \mainp for the exact definition). Shorter trajectories correspond to a larger flux (see Section~\ref{sec:method_infretis} in \mainp). The same set of trajectories is used to construct CFE surfaces for state A. Consequently, CFE surfaces may help rationalise the computed flux values. 

Among all CFE surfaces investigated in this work, we found that Figure~\ref{fig:6min_cl_size} in \mainp, showing the CFE projected onto \cacmin and the corresponding \nionrc, most clearly highlights the origin of the observed fluxes. The surface reveals how \flux (reported in Table~\ref{tab:rates} in \mainp) is related to the extent of configurational diversity in state A, expressed through the accessible combinations of \nionrc and \cacmin.

The largest difference in \flux is found for the systems with \conc~= 0.02 and different \nion. For \nion~= 72 (Figure~\ref{fig:6min_cl_size}(a) in \mainp), state A spans \nionrc values from 0 to 57 and extends up to $\approx$ 6.5~\AA{} along \cacmin. As such, trajectories explore a large configurational region before returning to \la, in line with a smaller flux.
By contrast, for \nion~= 108 (Figure~\ref{fig:6min_cl_size}(c) in \mainp), the accessible region is substantially more confined, with \nionrc ranging mostly between 60 and 80 and \cacmin extending only up to $\approx$ 5.5~\AA{}. This reduced configurational space is consistent with shorter trajectory lengths and, consequently, a larger \flux. For \conc~= 0.01, the opposite trend is observed: the region preceding $\lambda_A$ is more extensive for \nion~= 108, which correlates with a smaller flux in this system. 

Overall, for the systems considered here, differences in \flux appear to correlate with the extent of state A in the corresponding CFE landscapes, as quantified by the range of dynamically accessible \cacmin and \nionrc values.

\newpage
\clearpage
\subsubsection{FES and CFE landscapes dependent on \ccamin and the corresponding \nionrc.}\label{sec:SI_FE_CFE}

\begin{figure}[h!]
    \centering
    \includegraphics{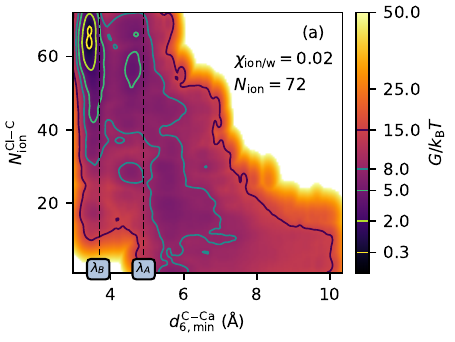}
    \includegraphics{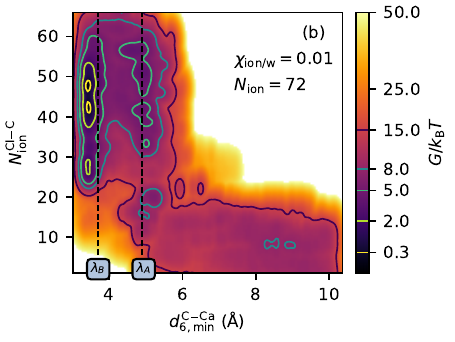}
    \includegraphics{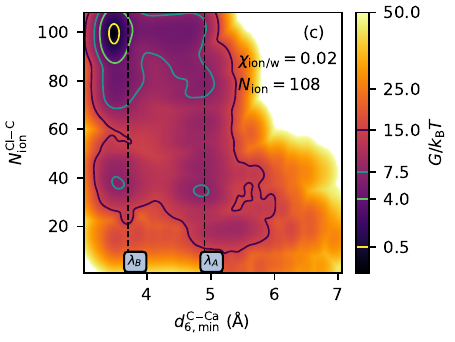}
    \includegraphics{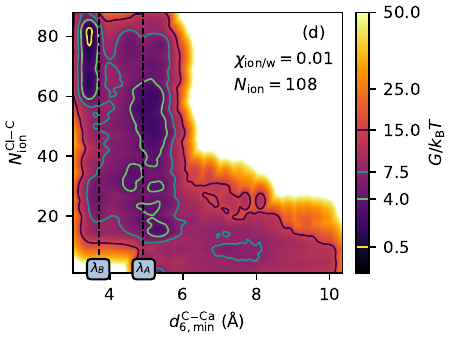}
    \caption{FES projected onto \ccamin and the corresponding \nionrc for the systems with compositions indicated in the legends.}
    \label{fig:fes_cca_RC_size}
\end{figure}

\begin{figure}[h!]
    \centering
    \includegraphics{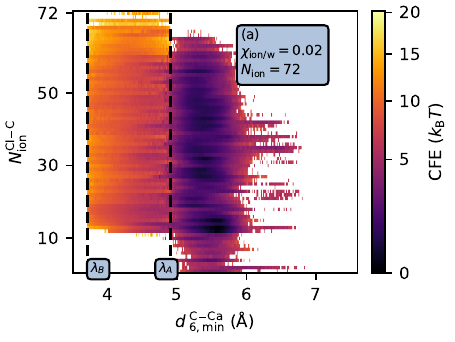}
    \includegraphics{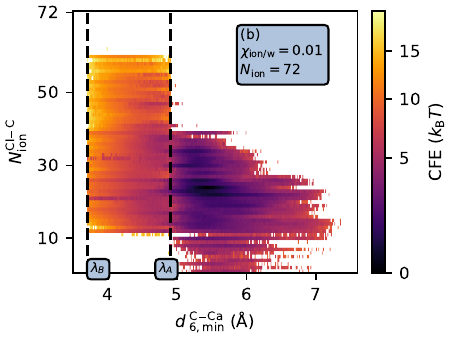}
    \caption{CFE surfaces as a function of the \nionrc and \ccamin for the systems with compositions indicated in the legends.}
    \label{fig:6min_cl_cize_cca}
\end{figure}

\newpage
\clearpage
\subsection{Structural analysis}\label{sec:SI_CI_struct}

\subsubsection{RDFs}\label{sec:SI_RDFs}

\begin{figure}[h!]
    \centering
\includegraphics{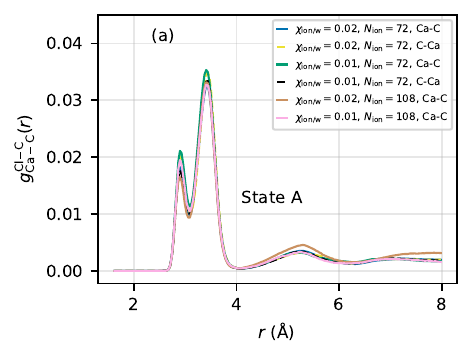}
\includegraphics{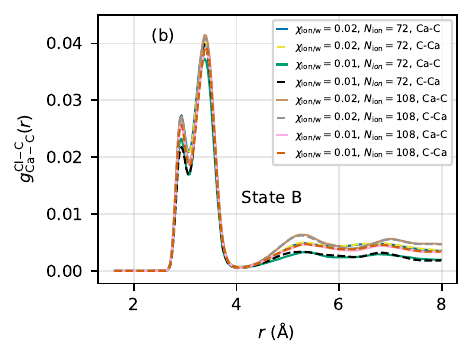}
\includegraphics{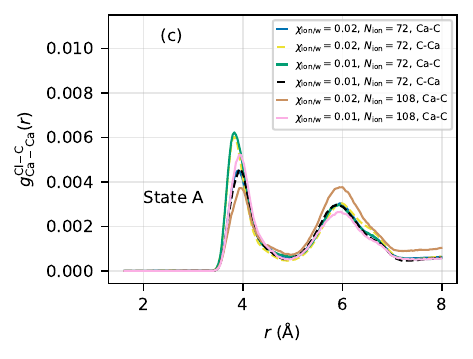}
\includegraphics{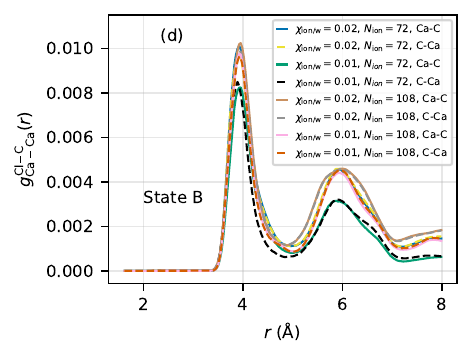}
\includegraphics{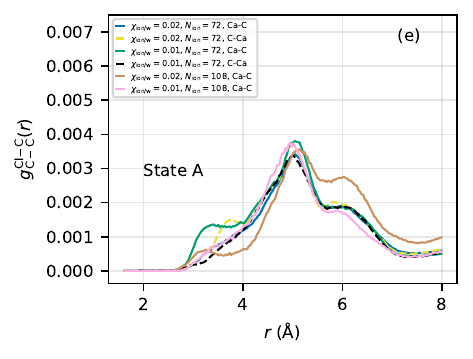}
\includegraphics{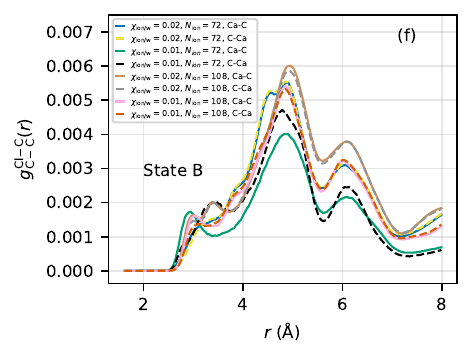}
\caption{RDF, $g(r)$, computed between (a,b)Ca--C, (c,d) Ca--Ca, and (e,f) C--C atom pairs. Left column: state A (a,c,e); right column: state B (b,d,f). All RDFs are computed for the systems indicated in the legends and characterise the CI-cluster associated with either the \cacmin or \ccamin OP, as specified in the legends.}
 \label{fig:rdf_SI_1}
\end{figure}

\newpage

\begin{figure}[h!]
    \centering
\includegraphics{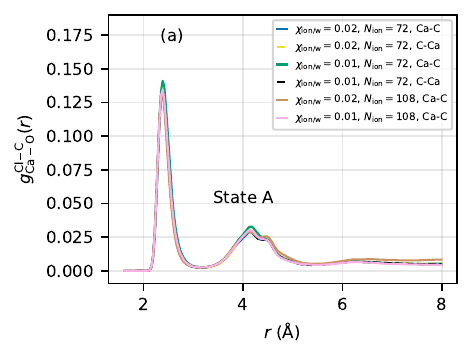}
\includegraphics{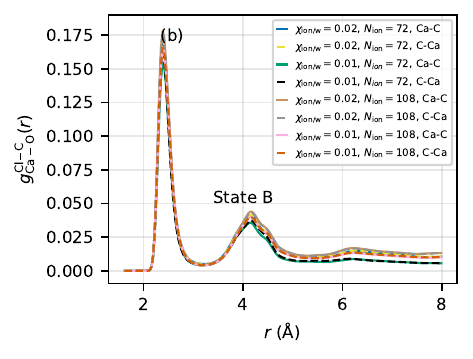}
\includegraphics{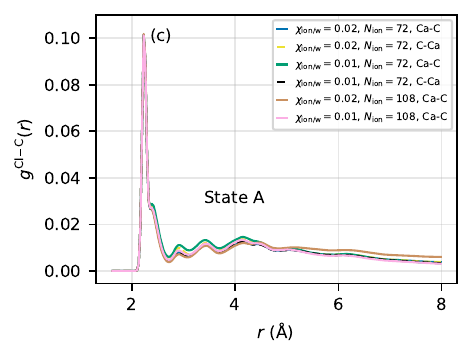}
\includegraphics{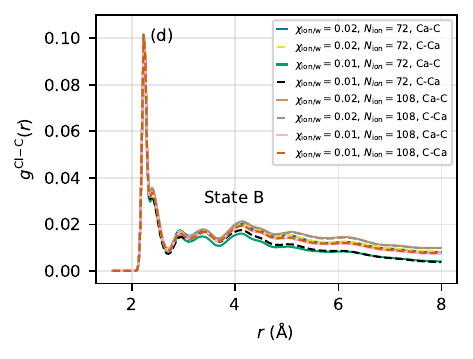}
    
    \caption{RDF, $g(r)$, computed between (a,b)Ca--O and (c,d) all atom pairs. Left column: state A (a,c,e); right column: state B (b,d,f). All RDFs are computed for the systems indicated in the legends and characterise the CI-cluster associated with either the \cacmin or \ccamin OP, as specified in the legends.}
    \label{fig:rdf_SI_2}
\end{figure}

\subsubsection{Ca-C/C-Ca CN in different systems}\label{sec:SI_CN}

\begin{figure}[h!]
    \centering
    \includegraphics{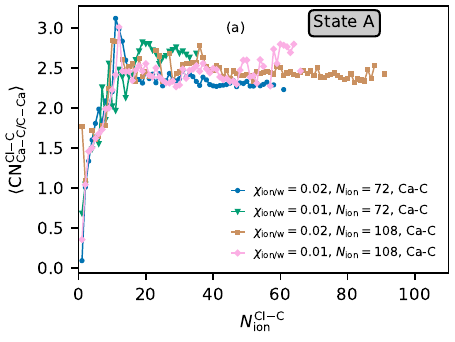}
    \includegraphics{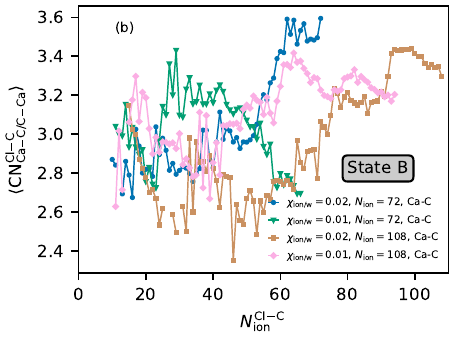}
    \includegraphics{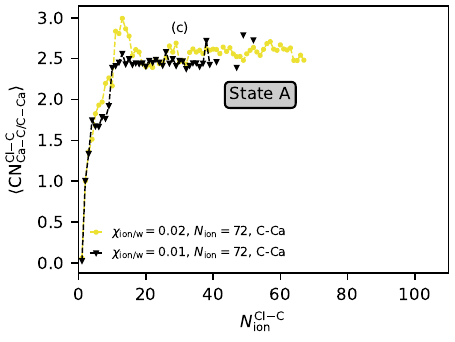}
    \includegraphics{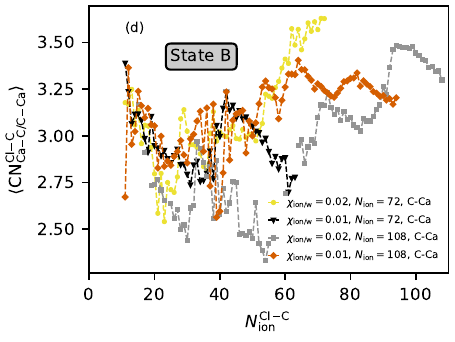}
    \caption{The average Ca--C/C--Ca CN in the CI-cluster associated with the OP indicated in the legends (\cacmin or \ccamin) as a function of the corresponding \nionrc. Left column (a, c): state A from \infretis trajectories; right column (b, d): state B from OPES trajectories. Top (a, b): \cacmin; bottom (c, d): \ccamin. System compositions are given in the figure. Error bars along the $y$-axis are below 0.5.}
    \label{fig:CN_per_cl_size_SI}
\end{figure}

We computed the average Ca-C/C-Ca CN per CI-cluster as a function of \nionrc, $\langle \rm CN^{\rm CI-C}_{\rm Ca-C/C-Ca} \rangle$, in state A and B, as shown in Figure~\ref{fig:CN_per_cl_size_SI}. These data indicate that, in state B, the CN reaches a maximum near the \nionrc values corresponding to the free energy minimum for each system (see Section~\ref{sec:cluster_analysis} in \mainp). For example, for \conc~= 0.02 and \nion~= 108, the CN reaches a maximum of $\simeq 3.4$ at \nionrc $\approx 100$. This maximum CN value is nearly identical across all systems, which indicates a similar degree of local order in the energy minimum. Furthermore, most systems, including all systems with \nion~= 108, exhibit relatively large CN values of approximately 2.8 and 3.2 in states A and B, respectively, at \nionrc~$\approx 11$--17.

\begin{figure}[h!]
    \centering
    \includegraphics{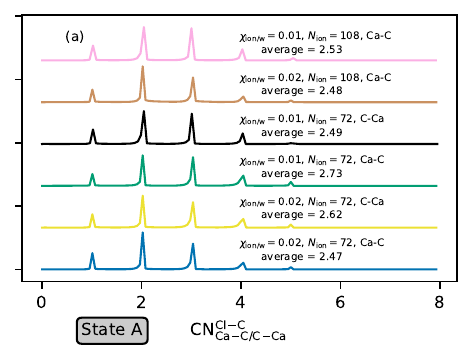}
    \includegraphics{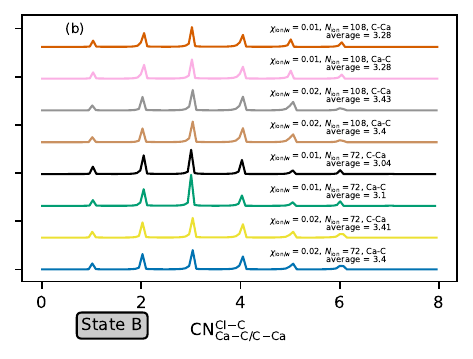}
    \caption{Probability densities of the Ca--C/C--Ca CN in the CI-cluster associated with the OP indicated in the legends (\cacmin or \ccamin). System compositions are given in the figure. (a) state A: \infretis trajectories; (b) state B: OPES trajectories.}
    \label{fig:CN_hist_SI}
\end{figure}

\newpage
\clearpage
\subsubsection{Global cluster distribution}\label{sec:SI_GC}

\begin{figure}[h!]
    \centering
    \includegraphics{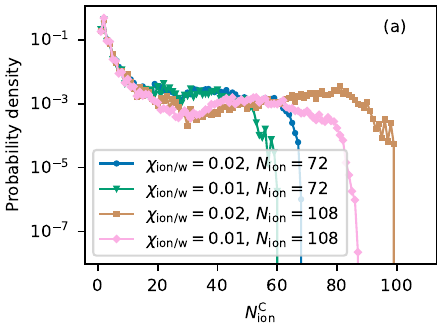}
    \includegraphics{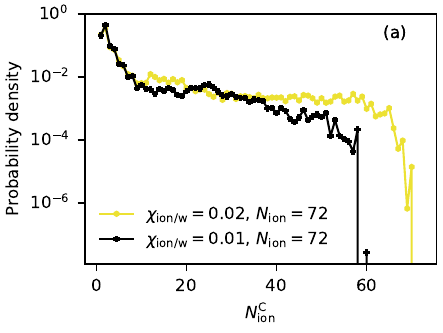}
    
    \caption{Probability densities of \nionc computed using the \infretis simulations with the (a) \cacmin and (b) \ccamin OP. The system compositions are indicated in the legends.}
    \label{fig:SI_cl_hist_cca}
\end{figure}

To characterise overall ionic clustering, the cluster distribution over all \caco aggregates in the simulation box was evaluated. All trajectories generated with \infretis were used in this analysis. The resulting probability density as a function of the cluster size \nionc is shown in Figures~\ref{fig:SI_cl_hist_cca}(a) and ~\ref{fig:SI_cl_hist_cca}(b) for the \infretis simulations employing the \cacmin and \ccamin OPs, respectively.

All curves exhibit the highest maxima at \nionc $=1$ and 2, indicating that isolated ions and ion pairs are the dominant cluster species. Furthermore, because \nionc reaches values of up to 100, the systems also contain oligomeric species and large \caco aggregates. The latter are likely associated with CI-clusters. This is particularly evident from the second maximum observed at \nionc $> 40$ for the larger systems (\nion $=108$), whose range of cluster sizes substantially overlaps with that corresponding to the free energy minimum in state A (see Figure~\ref{fig:6min_cl_size} in \mainp) and to the second reaction pathway identified in Section~\ref{sec:trans_region} in \mainp. The structure of the CI-cluster is discussed in detail in Section~\ref{sec:CI_struct} in \mainp.

Oligomeric aggregates with sizes \nionc~$< 20$ also exhibit a significant probability of formation. Previous work by Demichelis et al., investigating the association of \caion and \cion ions using the same force field, showed that PNCs can be formed with the DOLLOP structure \cite{PNC_MD_yes}. DOLLOPs are characterised by large \rg, which are attainable only for highly hydrated configurations. Consequently, \rg provides a useful descriptor for distinguishing DOLLOPs from other \caco aggregates, such as DLP and ACC-like clusters, which are less hydrated and less dynamic \cite{PNC_MD_yes}. To determine whether the oligomeric and larger aggregates observed in our simulations exhibit a DOLLOP-like structure, we computed \rg for clusters of different sizes.

\begin{figure}[h!]
    \centering
    \includegraphics{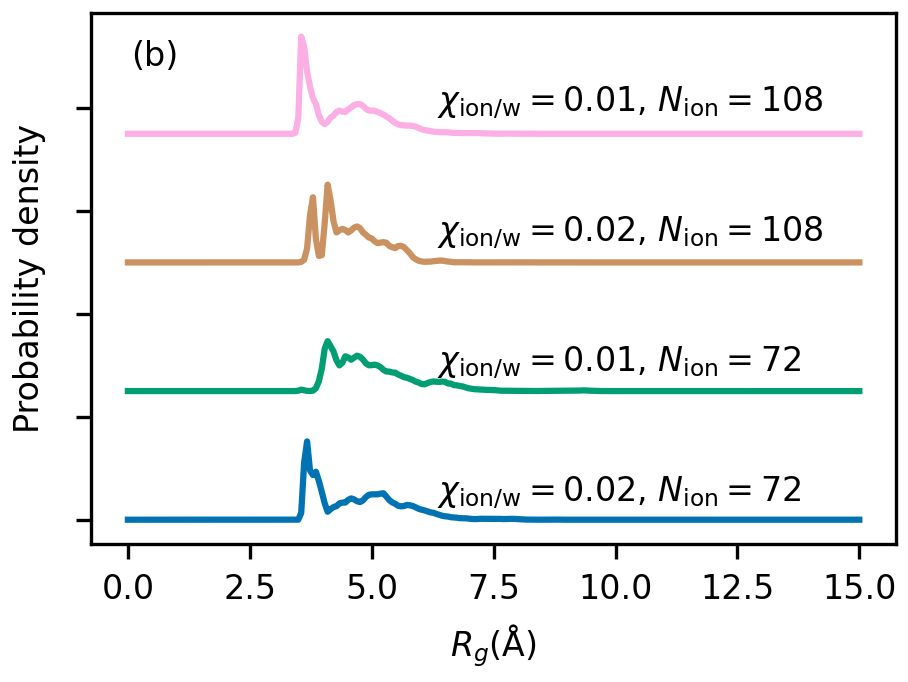}
    \includegraphics{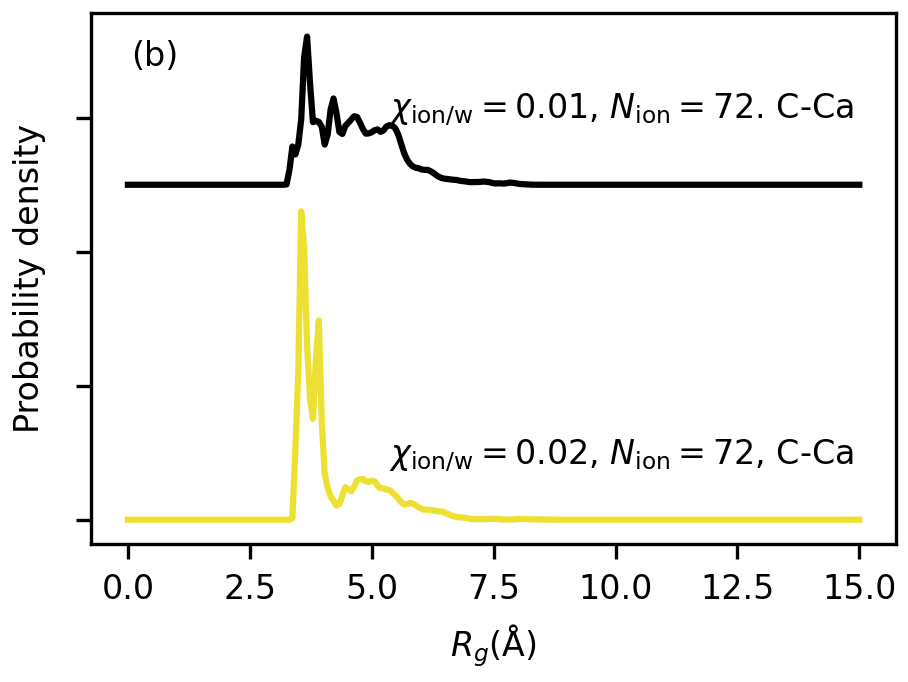}
    
    \caption{Probability densities of \rg computed using the \infretis simulations with the (a) \cacmin and (b) \ccamin OPs. The system compositions are indicated in the legends.}
    \label{fig:SI_rg_hist}
\end{figure}

The \rg distributions for clusters with \nionc~= 12 are presented in Figures~\ref{fig:SI_rg_hist}(a) and \ref{fig:SI_rg_hist}(b) for the \infretis simulations using the \cacmin and \ccamin OPs, respectively. DOLLOP clusters of this size are characterized by \rg values exceeding 6 \AA{} \cite{PNC_MD_yes}. In contrast, the \rg distributions obtained here span values from approximately 3.1 to 7 \AA{}, with maxima at $\approx 3.6{-}4$ \AA{} for all studied systems. These relatively small \rg values are consistent with highly coordinated, less hydrated aggregates rather than DOLLOPs. This observation can be rationalised by the fact that some of these clusters act as CI-clusters mediating the formation of the first coordination sphere, which requires a high CN of 6 and is therefore incompatible with a DOLLOP structure. It should be noted that, although \infretis does not directly bias the structural properties of the system, it preferentially samples trajectories associated with the transition under investigation. Consequently, the cluster statistics obtained from the \infretis ensembles are expected to be enriched in structures relevant to the reaction mechanism rather than representative of equilibrium states.

\begin{figure}[h!]
     \centering
     \includegraphics{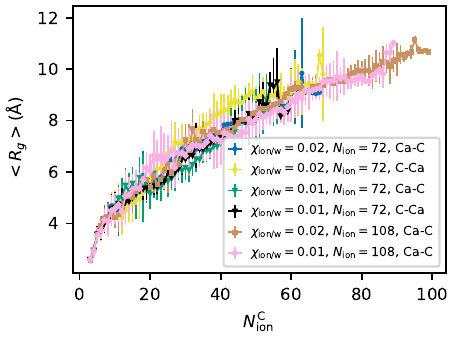}
     \caption{Average radius of gyration $\langle$\rg$\rangle$ as a function of \nionrc obtained from the \infretis simulation with the OP and composition indicated in the legend.}
     \label{fig:SI_av_rg}
 \end{figure}

In addition, we computed the radius of gyration  averaged over all clusters of a given size, $\langle$\rg$\rangle$, as a function of \nionc, as presented in Figure~\ref{fig:SI_av_rg}. For all considered cluster sizes, the observed \rg values were found to be smaller than those expected for DOLLOP structures. Specifically, for \nionc $= 72$, DOLLOP clusters have been reported to exhibit \rg values between 9 and 12 \AA{}, whereas in our simulations \rg fluctuates around 9 \AA{} \cite{PNC_MD_yes}. The same study reported \rg values strictly $< 9$ \AA{} for ACC clusters of the same size. Thus, the exhibited \rg $\simeq 9$ \AA{} lies at the boundary between the characteristic ranges of ACC and DOLLOP structures. This finding is consistent with the DLP-like cluster structure proposed for CI-clusters in Section~\ref{sec:CI_struct}, which is expected to be structurally intermediate between ACC and DOLLOP.

Across all compositions and OPs, \rg exhibits a similar dependence on \nionrc, with no systematic differences observed between the systems.

Overall, no statistically significant population of DOLLOP-structured PNCs, as reported in ref.~\cite{PNC_MD_yes}, was observed in our simulations.

\newpage
\clearpage
\subsubsection{Shapes of the CI-clusters}\label{sec:SI_K1_K2}

\begin{figure}[h!]
    \centering
    \includegraphics{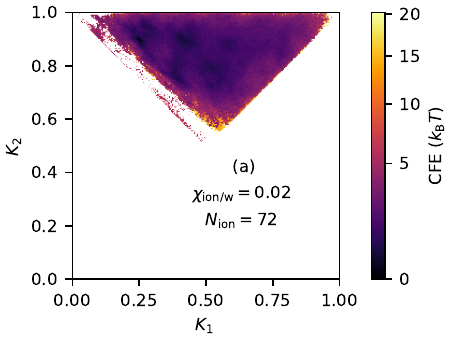}
    \includegraphics{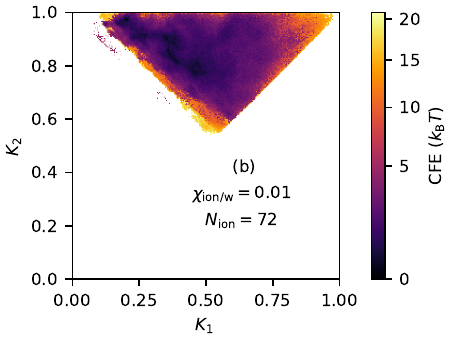}
    \includegraphics{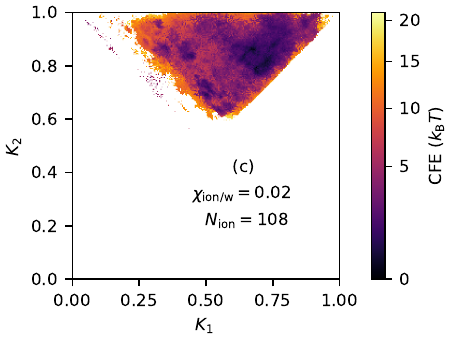}
    \includegraphics{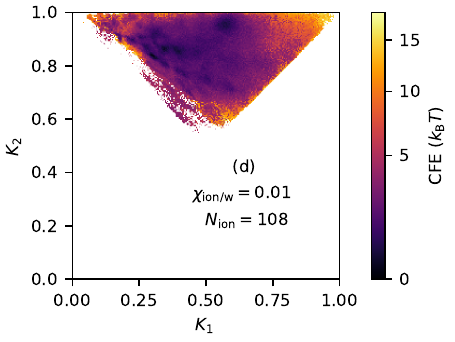}
    
    \caption{CFE surfaces as functions of $K_1$ and $K_2$ shape parameters for the CI-cluster associated with the \cacmin OP. System compositions are indicated in the legends.}
    \label{fig:SI_K1_K2_A}
\end{figure}

\begin{figure}[h!]
    \centering
    \includegraphics{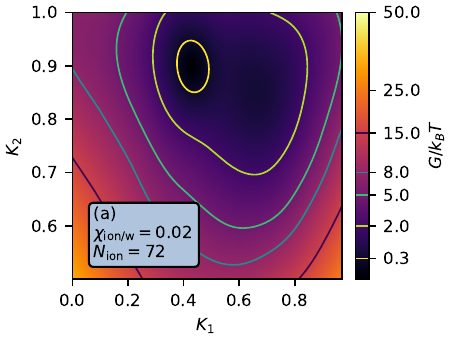}
    \includegraphics{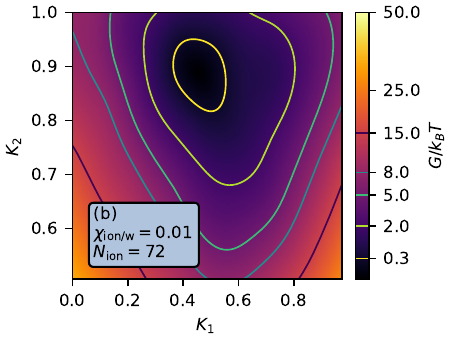}
    \includegraphics{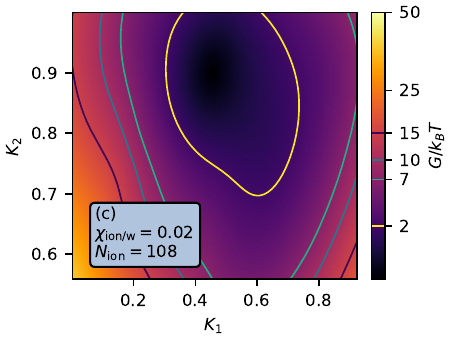}
    \includegraphics{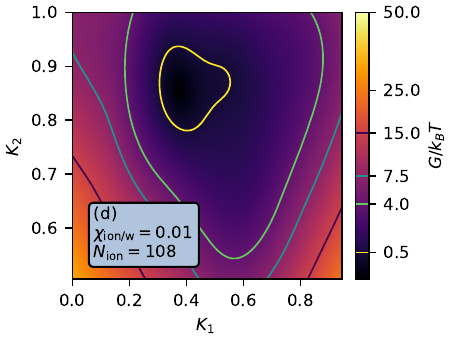}
    
    \caption{FES as functions of $K_1$ and $K_2$ shape parameters for the CI-cluster associated with the \cacmin OP. System compositions are indicated in the legends.} 
    \label{fig:SI_K1_K2_B}
\end{figure}

We investigated the shape of the CI-clusters using $K_1$ and $K_2$ shape parameters calculated from the positions of the Ca and C atoms. The corresponding $K_1-K_2$ CFE surfaces, obtained with \infretis (whose minima represent state A), and the FES, computed from reweighing of OPES trajectories (whose minima represent state B), are presented in Figures~\ref{fig:SI_K1_K2_A} and \ref{fig:SI_K1_K2_B}, respectively. The provided energy landscapes correspond to the CI-cluster of the \cacmin OP; those computed for the CI-cluster of the \ccamin OP show very similar results. 

In state A (Figure~\ref{fig:SI_K1_K2_A}), the free energy basin is spread over the broad range of $K_1$ and $K_2$ values, with the lowest energy regions approaching a rod-like shape characterized by $K_1$ $\leq 0.5$ and $K_2$ $\geq 0.75$ for all compositions except \conc~= 0.02 with \nion~= 108. The latter system exhibits its lowest energies at $K_1$ $\geq 0.5$ and $K_2$ $\geq 0.7$, which corresponds to a sphere-like shape. In state B, all four compositions are characterised by rod-like CI-clusters, as indicated by the free energy minima at $K_1$ $\simeq 0.4$ and $K_2$ $\simeq 0.9$ (Figure~\ref{fig:SI_K1_K2_B}). In OPES simulations, CI-clusters do not exhibit a sphere-like shape for the studied systems.

\newpage
\clearpage
\subsubsection{CI-cluster snapshots}\label{sec:SI_snapshots}

\begin{figure}[h!]
    \centering
    \includegraphics[width=1\textwidth]{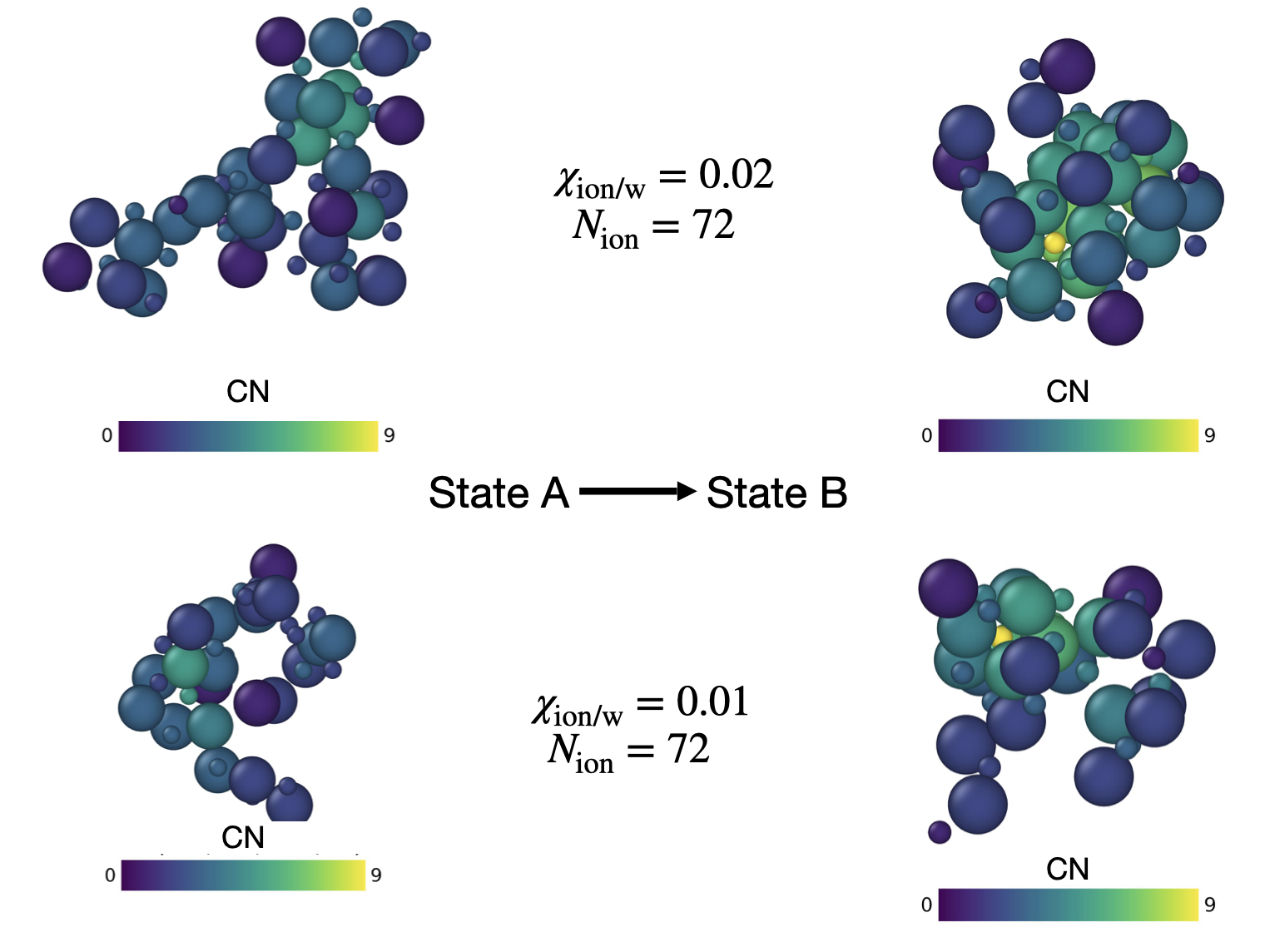}
    \caption{Simulation snapshots of the CI-cluster in state A (left) and state B (right), for the indicated compositions. Larger spheres correspond to calcium ions, and smaller spheres correspond to carbon atoms. Oxygen atoms and water molecules are not shown. Colours represent the CN, computed using a cutoff distance of 3.7 \AA{} between atom pairs (C--Ca, C--C, Ca--C, and Ca--Ca), as indicated in the colour bar.}
    \label{fig:photo_72_SI}
\end{figure}

\begin{figure}[h!]
    \centering
    \includegraphics[width=1\textwidth]{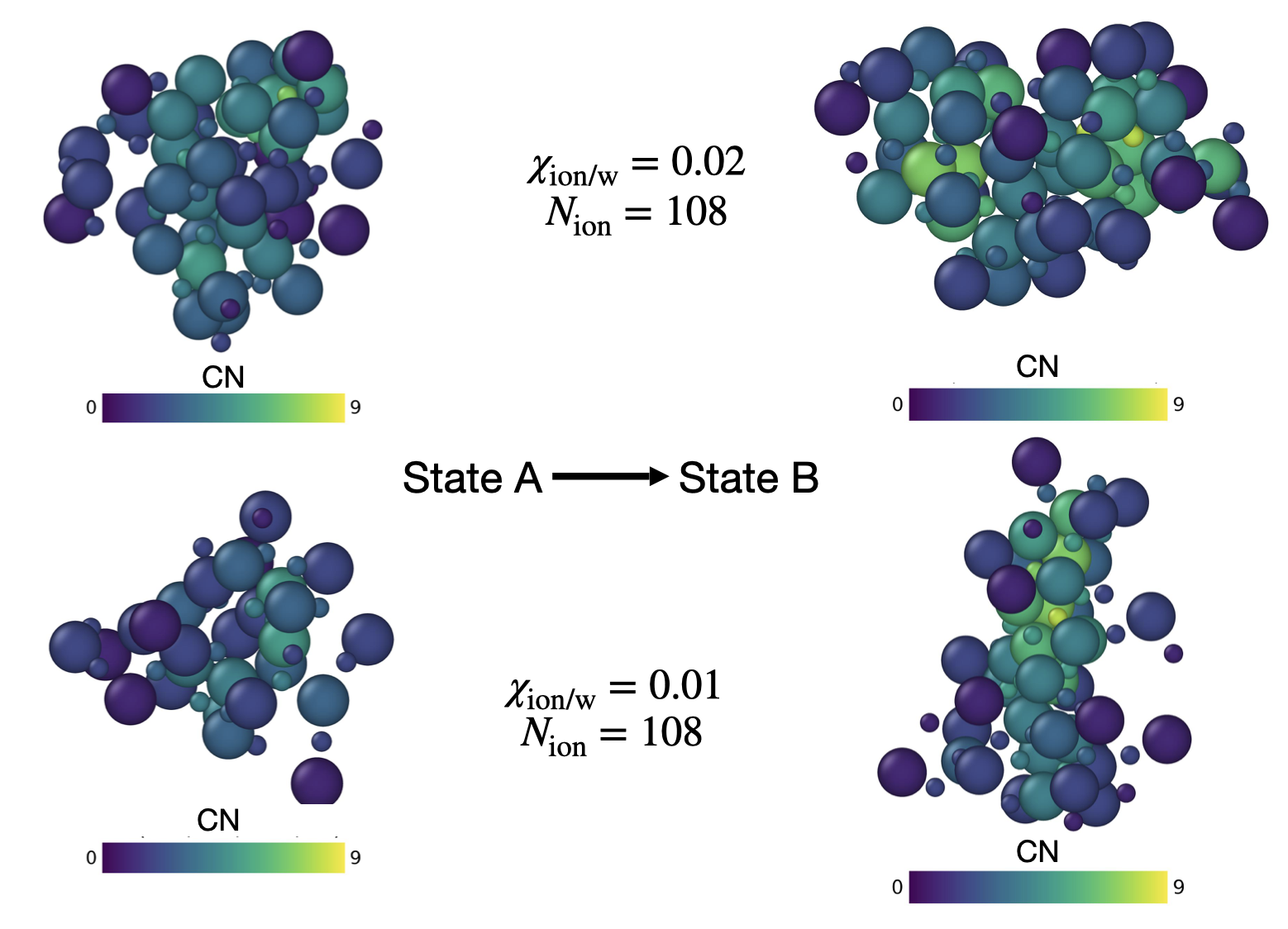}
    \caption{Simulation snapshots of the CI-cluster in state A (left) and state B (right) for the indicated compositions. Larger spheres correspond to calcium ions, and smaller spheres correspond to carbon atoms. Oxygen atoms and water molecules are not shown. Colours represent the CN, computed using a cutoff distance of 3.7 \AA{} between atom pairs (C--Ca, C--C, Ca--C, and Ca--Ca), as indicated in the colour bar.}
    \label{fig:photo_108_SI}
\end{figure}

\newpage
\clearpage 

\subsection{Analysis of reactive paths}\label{sec:SI_react_traj}

\begin{figure}[h!]
    \centering
    \includegraphics{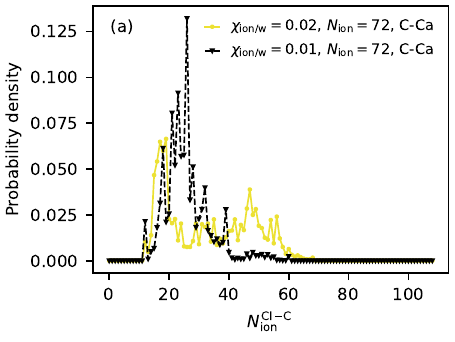}
    \includegraphics{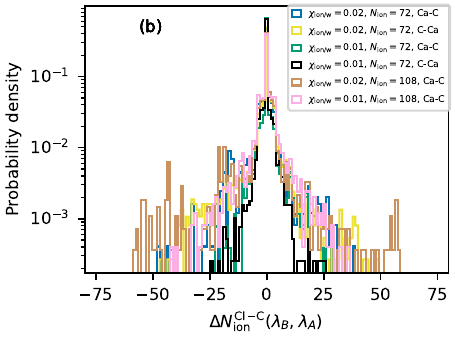}
    
    \caption{(a) Probability densities of \nionrc along the reactive paths, normalised by the trajectory length obtained from the \infretis simulations using the \ccamin OP; (b) probability densities of the total change of the CI-cluster size between $\lambda_B$ and $\lambda_A$, defined as \nionrc($\lambda_B$) - \nionrc($\lambda_A$) = $\Delta$\nionrc($\lambda_B,\lambda_A$)), obtained from the \infretis\ simulations with the OPs given in the legend. The system compositions are indicated in the legends.}
    \label{fig:SI_cl_size_RP_cca}
\end{figure}

Figure~\ref{fig:SI_cl_size_RP_cca}(b) shows the probability density of $\Delta$\nionrc($\lambda_B,\lambda_A$) = \nionrc($\lambda_B$) -- \nionrc($\lambda_A$) for the reactive paths obtained from the \infretis simulations. No significant differences are observed between the distributions obtained using the \cacmin and \ccamin OPs.

For all compositions, small values of $\Delta$\nionrc($\lambda_B,\lambda_A$) are more probable than large values. The CI-cluster size does not change between \la and \lb ($\Delta$\nionrc($\lambda_B,\lambda_A$) = 0) in approximately 50\% of the reactive trajectories. The only exception is the system with \conc~= 0.01 and \nion $= 72$, for which this probability increases to approximately 75\%. However, this does not imply that the CI-cluster size remains constant throughout the reaction. Even in trajectories with $\Delta$\nionrc($\lambda_B,\lambda_A$) = 0, transient fluctuations in \nionrc of up to about 6 can occur before the cluster ultimately returns to its initial size. Nevertheless, these results show that a large fraction of reactive trajectories cross the barrier at nearly constant \nionrc, corresponding to the value established in state A.

\begin{figure}[h!]
    \centering
    
    \includegraphics{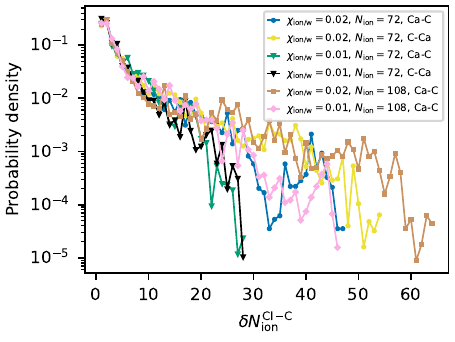}
    \caption{The probability density of instantaneous changes in \nionrc, $\delta$\nionrc, computed from the \infretis simulation with the OPs and compositions indicated in the legend.}
    \label{fig:SI_d_RC_instant}
\end{figure}

The distribution of instantaneous changes in \nionrc along the reactive paths, $\delta$\nionrc, is presented in Figure~\ref{fig:SI_d_RC_instant}; $\delta$\nionrc denotes the absolute change in \nionrc each time it changes. The distribution is dominated by small changes, with the highest probabilities observed for $\delta$\nionrc~= 1 and 2. The probability density decreases by approximately an order of magnitude by $\delta$\nionrc~$\simeq 6$. This indicates that changes in \nionrc arise primarily from the attachment and detachment of individual ions, ion pairs, and small ionic oligomers. Such behaviour is consistent with the predominance of small ionic clusters in solution, as shown by the cluster size distribution in Figure~\ref{fig:SI_cl_hist_cca}. 

Although small changes dominate the statistics, both $\Delta$\nionrc($\lambda_B,\lambda_A$) and $\delta$\nionrc can occasionally reach large values. Maximum $\Delta$\nionrc($\lambda_B,\lambda_A$) ranges from $\approx 25$ for the system with \conc~= 0.01 and \nion $=72$ to $\approx 60$ for the system with \conc~= 0.02 and \nion $=108$, while $\delta$\nionrc can reach values as high as 60 within a single 0.1 ps sampling interval. Such large changes in \nionrc may originate from reactive trajectories switching between the two pathways identified in \mainp in Section~\ref{sec:trans_region}. 
This interpretation is consistent with the largest changes occurring for \conc~= 0.02 and \nion~= 108, where the two pathways are most strongly separated in \nionrc, and the smallest changes occurring for \conc $= 0.01$ and \nion $= 72$, where only a single pathway is observed. Additional support comes from the approximately symmetric distribution of $\Delta$\nionrc($\lambda_B,\lambda_A$): because the two pathways span both low and high \nionrc values, transitions between them can produce either increases or decreases in \nionrc. Some large changes may also arise from fluctuations within a single pathway, particularly for the second pathway in the systems with \nion $=108$, which extends over a broad range of \nionrc values (approximately 30-40).

It is important to note that changes in \nionrc do not arise solely from ion exchange between the CI-cluster and the surrounding solution. Due to the definitions of the OP and the CI-cluster, the identity of the CI-cluster itself may change. For example, consider a configuration containing an 80-ion cluster identified as the CI-cluster together with a smaller 12-ion cluster. The latter becomes the new CI-cluster if one of its ions exhibits a smaller sixth-nearest-neighbour distance than any ion within the larger cluster, i.e. if it becomes the CI associated with \cacmin/\ccamin. Such changes in cluster identity may contribute significantly to large jumps in \nionrc. In contrast, this mechanism is unlikely to explain small changes ($\leq 11$), since values of \nionrc$\leq 11$ occur only in a negligible fraction of configurations within the transition region.

\newpage
\clearpage

\subsection{Ion coordination around central ions}\label{sec:SI_local_cn}
\subsubsection{The \infretis results}\label{sec:SI_local_cn_infretis}

\begin{figure}[h!]
    \centering
    \includegraphics{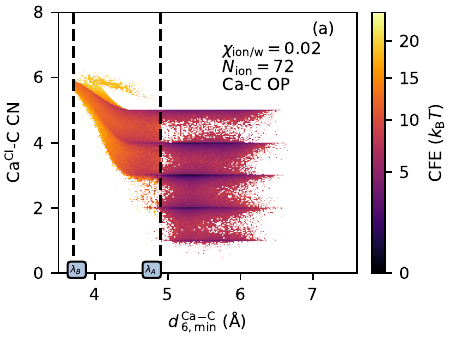}
    \includegraphics{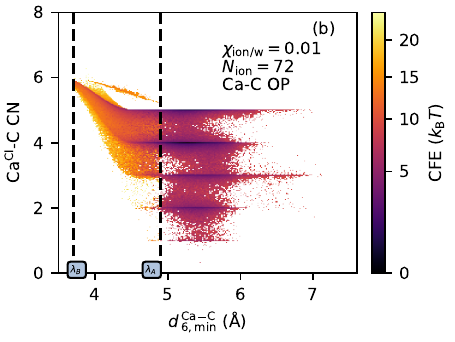}
    \includegraphics{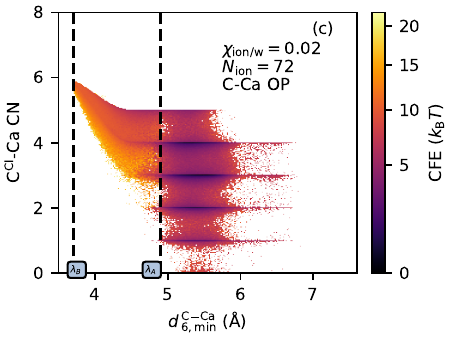}
    \includegraphics{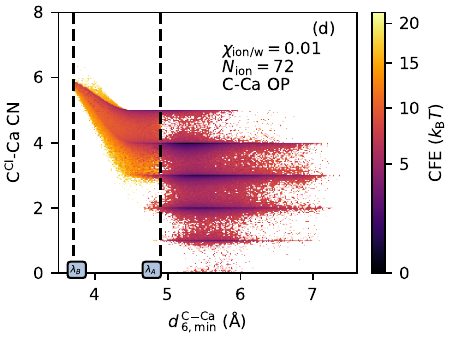}
    
    \caption{CFE surfaces as a function of (a), (b) the \cacmin OP and the \carc--C CN, and (c), (d) the \ccamin OP and the \crc--Ca CN. The systems compositions are indicated in the legends.} 
    \label{fig:SI_CFE_cn_1_72}
\end{figure}

\begin{figure}[h!]
    \centering
    \includegraphics{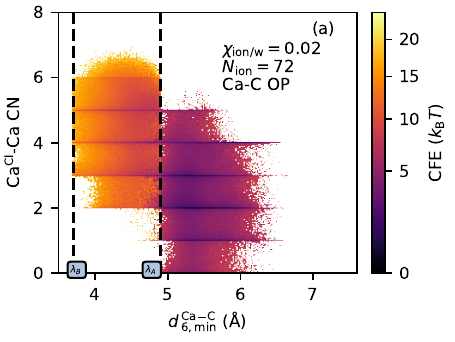}
    \includegraphics{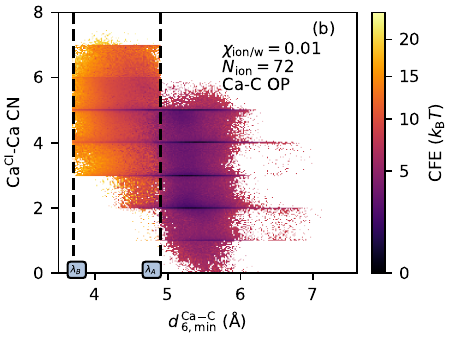}
    \includegraphics{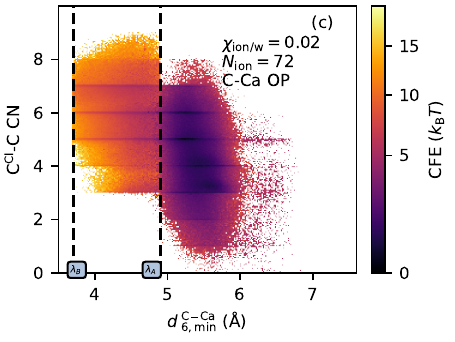}
    \includegraphics{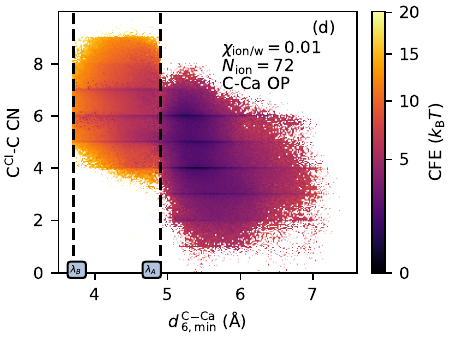}
    
    \caption{CFE surfaces as a function of (a), (b) the \cacmin OP and the \carc--Ca CN, and (c), (d) the \ccamin OP and the \crc--C CN. The systems compositions are indicated in the legends.}
    \label{fig:SI_CFE_cn_2_72}
\end{figure}

\begin{figure}[h!]
    \centering
    \includegraphics{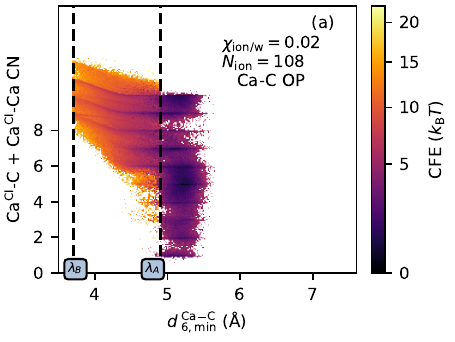}
    \includegraphics{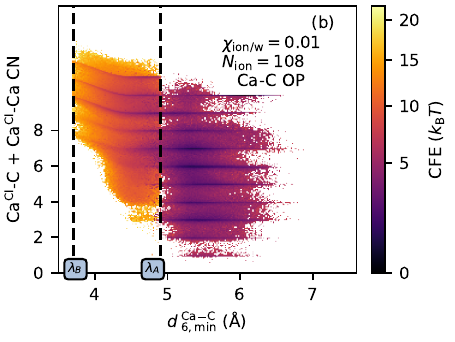}
    \caption{CFE surfaces as a function of the \cacmin OP and the \carc--C + \carc--Ca CN. The systems compositions are indicated in the legends.}
    \label{fig:SI_CFE_cn_1_2_108}
\end{figure}

\begin{figure}[h!]
    \centering
    \includegraphics{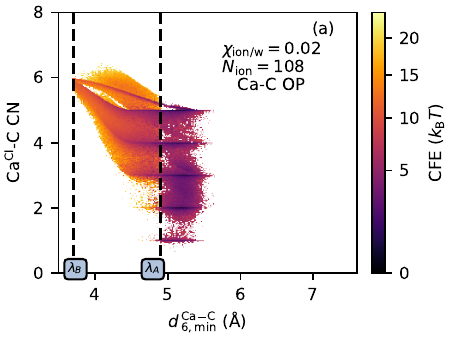}
    \includegraphics{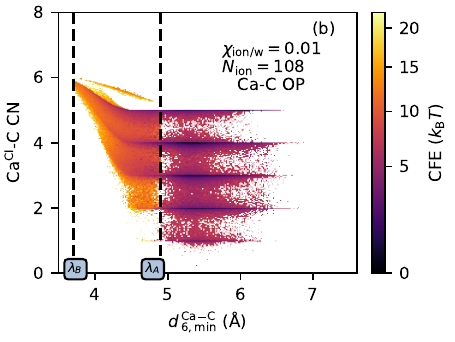}
    \caption{CFE surfaces as a function of the \cacmin OP and the \carc--C CN. The systems compositions are indicated in the legends.}
    \label{fig:SI_CFE_cn_1_108}
\end{figure}

\begin{figure}[h!]
    \centering
    \includegraphics{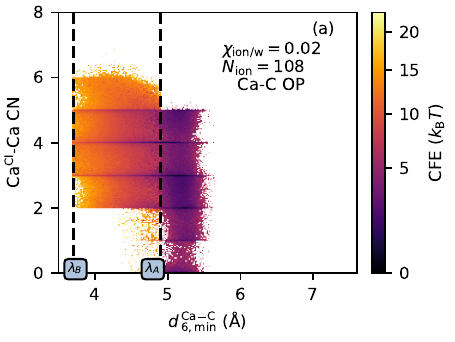}
    \includegraphics{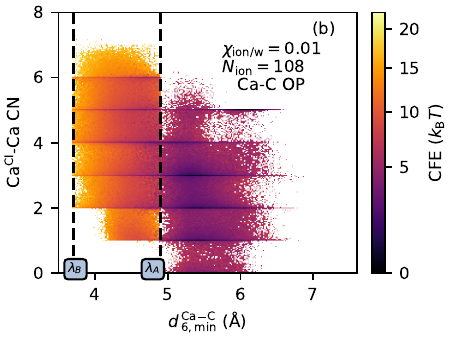}
    \caption{CFE surfaces as a function of the \cacmin OP and the \carc--Ca CN. The systems compositions are indicated in the legends.}
    \label{fig:SI_CFE_cn_2_108}
\end{figure}

\begin{figure}[h!]
    \centering
    \includegraphics{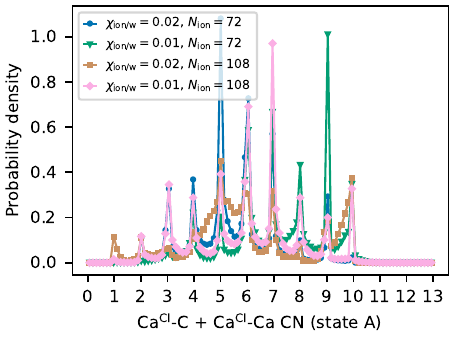}
    \includegraphics{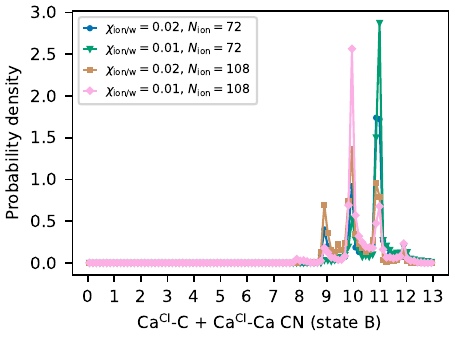}
    \caption{The probability densities of \carc--C + \carc--Ca CN in (a) state A and (b) state B for the trajectories from the \infretis simulations with the \cacmin OP. The systems compositions are indicated in the legends.}
    \label{fig:A_full_B_cn_1_2}
\end{figure}

\newpage
\clearpage
\subsubsection{The OPES results}\label{sec:SI_local_cn_OPES}

\begin{figure}[h!]
    \centering
    \includegraphics{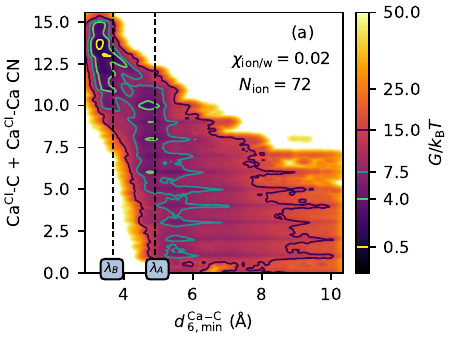}
    \includegraphics{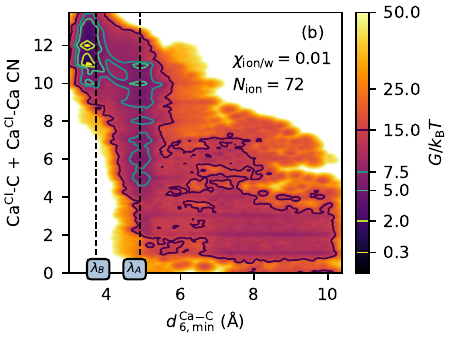}
    \includegraphics{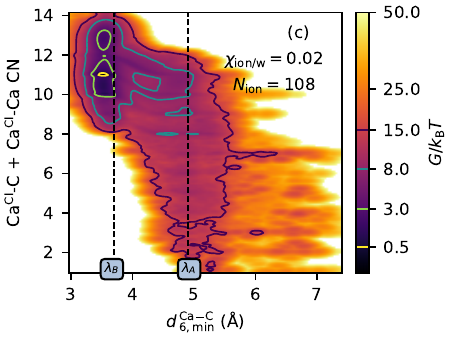}
    \includegraphics{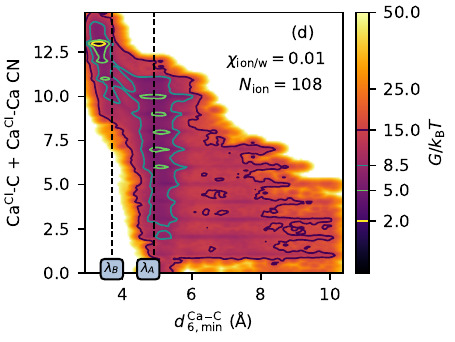}
    \caption{FES projected onto the \cacmin OP and the \carc--C + \carc--Ca CN. System compositions are given in the legends.}
    \label{fig:fes_cac_R_Ca_CN}
\end{figure}

\begin{figure}[h!]
    \centering
    \includegraphics{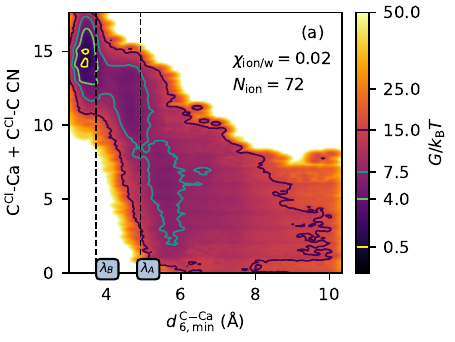}
    \includegraphics{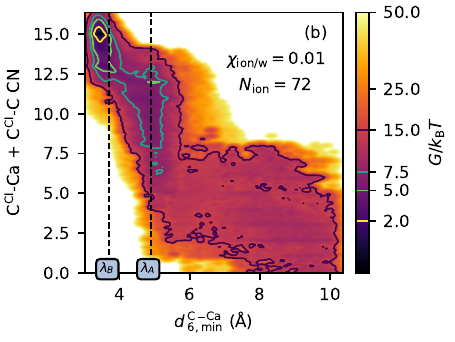}
    \includegraphics{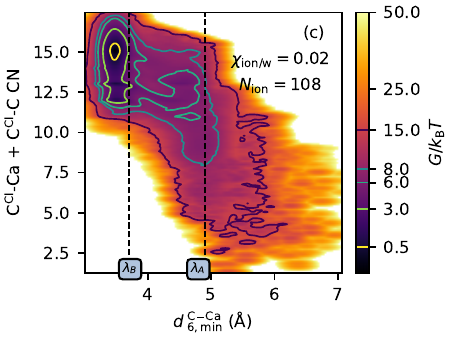}
    \includegraphics{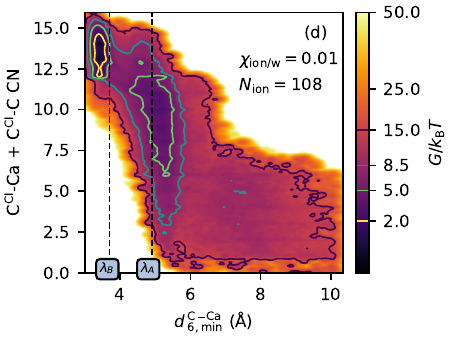}
    \caption{FES projected onto the \ccamin OP and the \crc--Ca + \crc--C CN. System compositions are given in the legends.}
    \label{fig:fes_cca_R_C_CN}
\end{figure}

\begin{figure}[h!]
    \centering
    \includegraphics{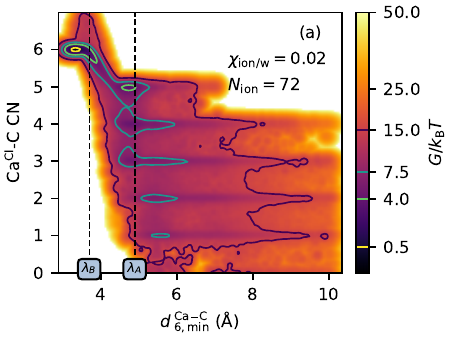}
    \includegraphics{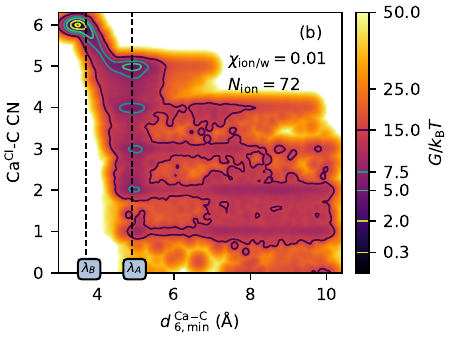}
    \includegraphics{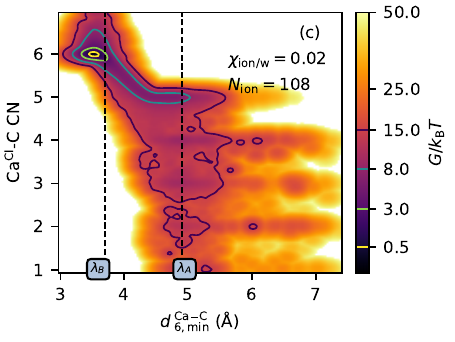}
    \includegraphics{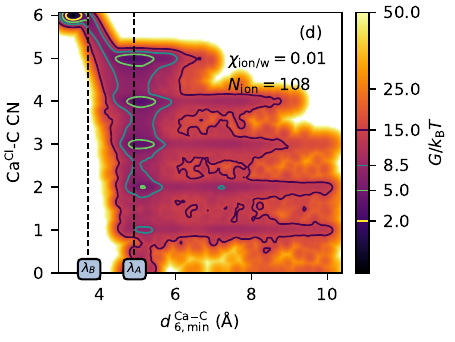}
    \caption{FES projected onto the \cacmin OP and the \carc--C CN. System compositions are given in the legends.}
    \label{fig:fes_cac_R_Ca_CN_1}
\end{figure}

\begin{figure}[h!]
    \centering
    \includegraphics{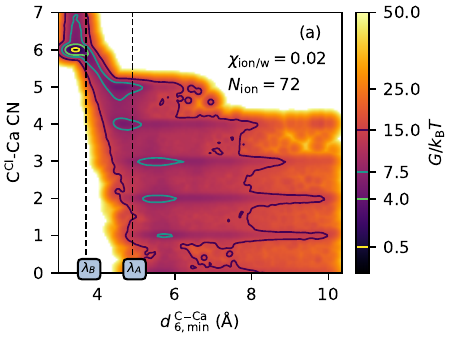}
    \includegraphics{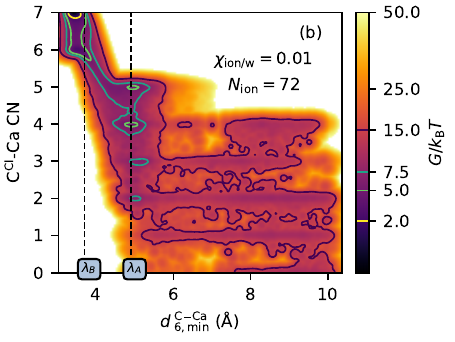}
    \includegraphics{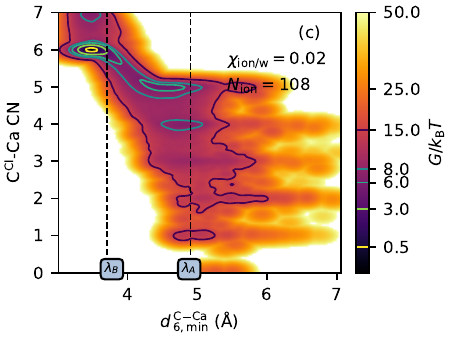}
    \includegraphics{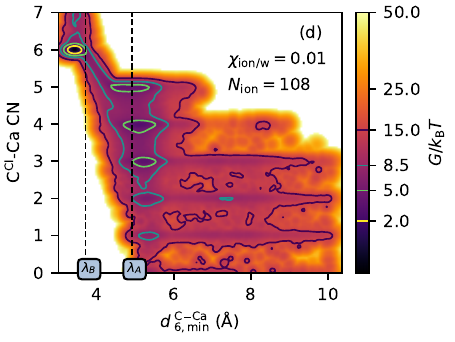}
    \caption{FES projected onto the \ccamin OP and the \crc--Ca CN. System compositions are given in the legends.}
    \label{fig:fes_cca_R_C_CN_1}
\end{figure}

\begin{figure}[h!]
    \centering
    \includegraphics{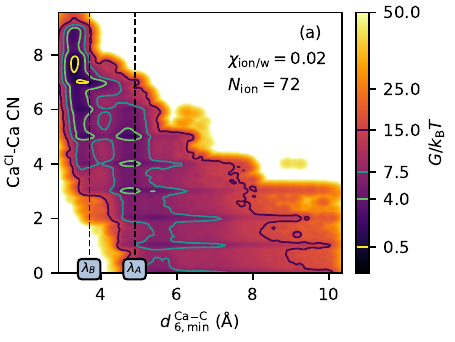}
    \includegraphics{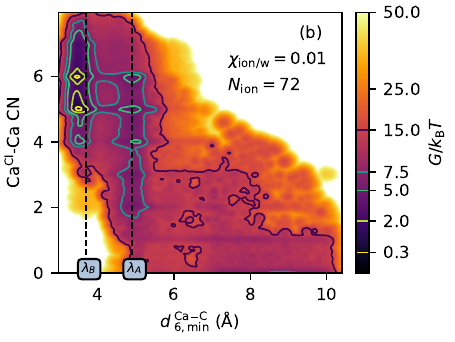}
    \includegraphics{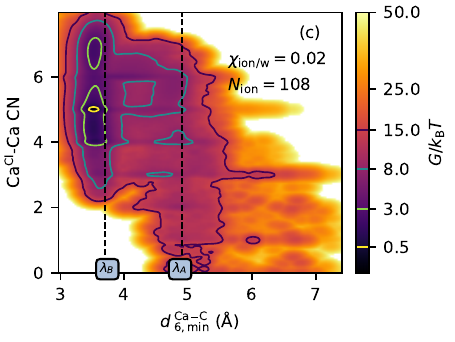}
    \includegraphics{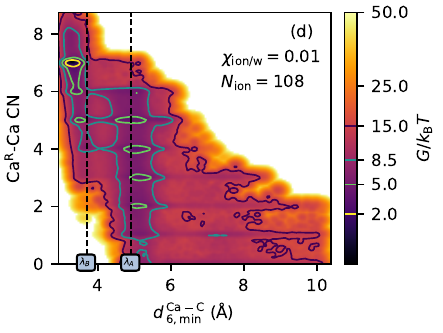}
    \caption{FES projected onto the \cacmin OP and the \carc--Ca CN. System compositions are given in the legends.}
    \label{fig:fes_cac_R_Ca_CN_2}
\end{figure}

\begin{figure}[h!]
    \centering
    \includegraphics{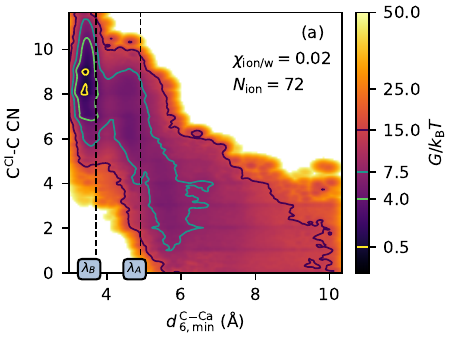}
    \includegraphics{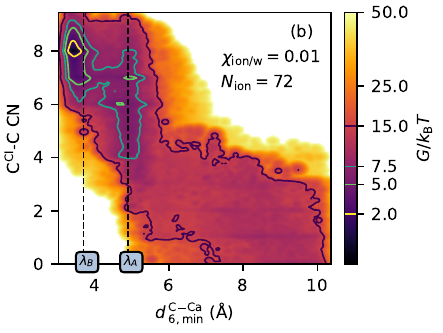}
    \includegraphics{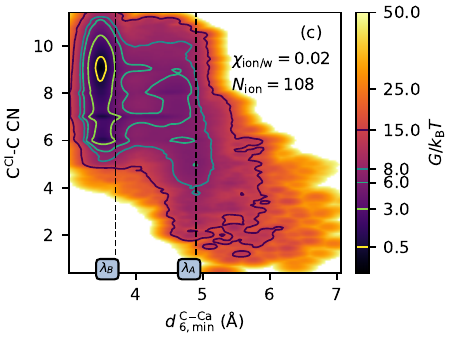}
    \includegraphics{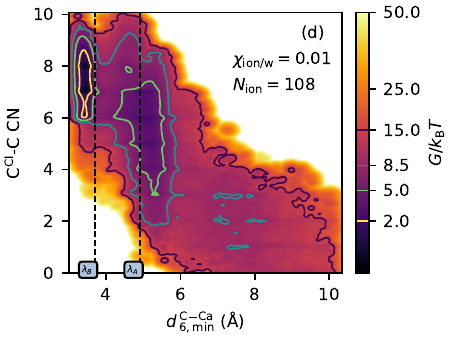}
    \caption{FES projected onto the \ccamin OP and the \crc--C CN. System compositions are given in the legends.}
    \label{fig:fes_cca_R_C_CN_2}
\end{figure}

\newpage
\clearpage
\subsection{Water coordination around central ions}\label{sec:SI_water}

\begin{figure}[h!]
    \centering
    \includegraphics{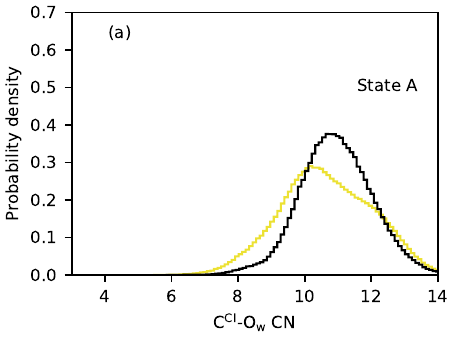}
    \includegraphics{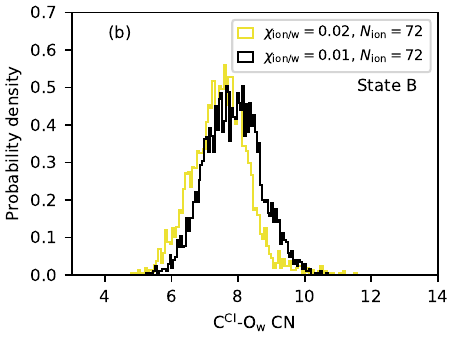}
    \caption{Probability densities of the \crc--\ow CN in (a) state A and (b) state B. The plots were obtained from trajectories generated by the \infretis simulations with the \ccamin OP. System compositions are given in the legends.}
    \label{fig:SI_cn_ow_cca}
\end{figure}

\begin{figure}[h!]
    \centering
    \includegraphics{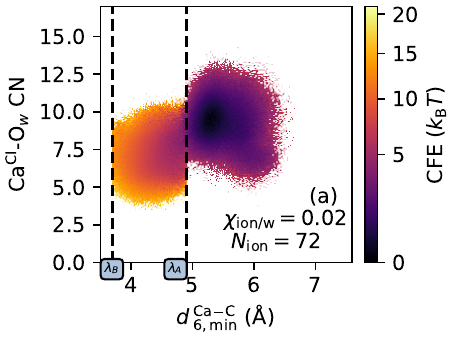}
    \includegraphics{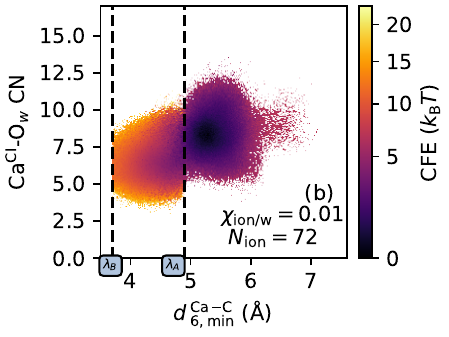}
    \includegraphics{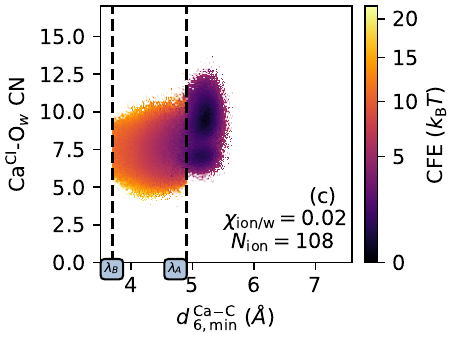}
    \includegraphics{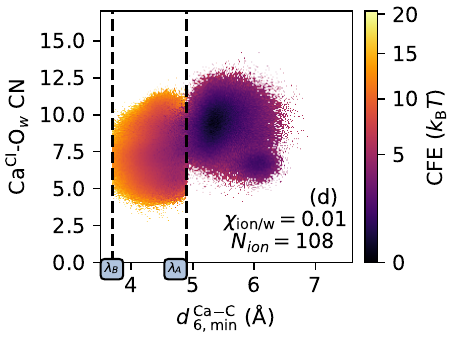}
    \caption{CFE surfaces as a function of \cacmin OP and \carc-O$_w$ CN. The systems compositions are indicated in the legends.} 
    \label{fig:cn_ow_cac}
\end{figure}

\begin{figure}[h!]
    \centering
    \includegraphics{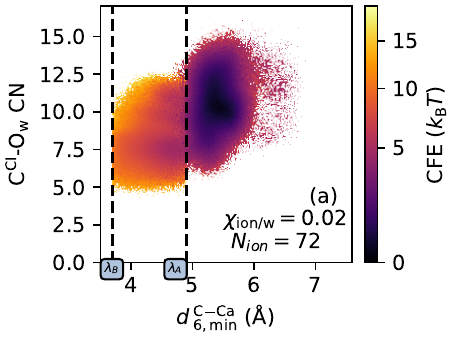}
    \includegraphics{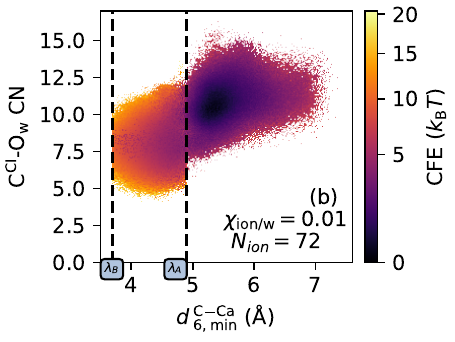}
    \caption{CFE surfaces as a function of \ccamin OP and \crc-O$_w$ CN. The systems compositions are indicated in the legends.}
    \label{fig:cn_ow_cca}
\end{figure}

\begin{figure}[h!]
    \centering
    \includegraphics{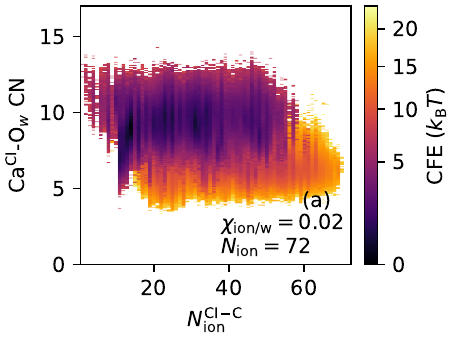}
    \includegraphics{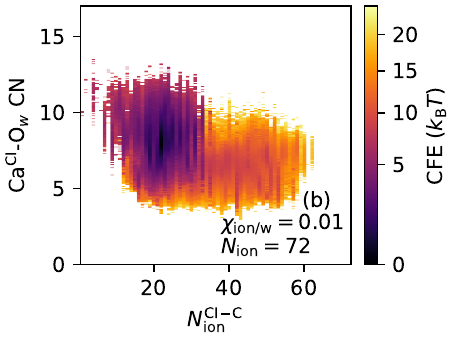}
    \includegraphics{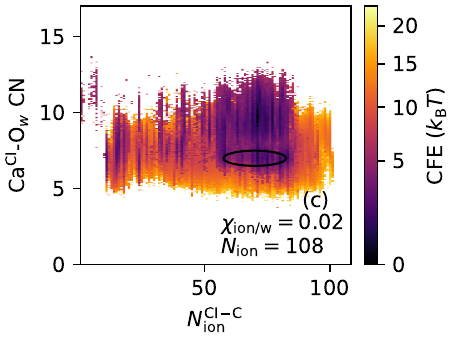}
    \includegraphics{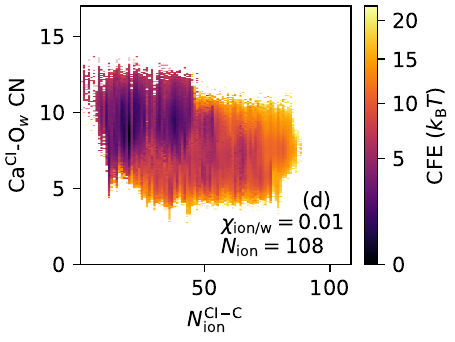}
    \caption{CFE surfaces as a function of \nionrc and the \carc--\ow CN. The systems compositions are indicated in the legends.} 
    \label{fig:cl_size_cn_ow}
\end{figure}


\end{document}